\documentclass[journal]{IEEEtran}

\usepackage{graphicx}
\graphicspath{{./figures/}}
\usepackage[utf8x]{inputenc}
\usepackage[T1]{fontenc}
\usepackage[american]{babel}
\usepackage[cmex10]{amsmath}
\usepackage{amsthm,amssymb,bm,bbm,ucs,cite,booktabs}
\usepackage{balance}
\theoremstyle{remark}
\newtheorem{proposition}{Proposition}
\newtheorem{remark}{Remark}
\newtheorem{example}{Example}

\newcommand*{\herm}{^{\mathsf{H}}}
\newcommand*{\transp}{^{\mathsf{T}}}
\DeclareMathOperator*{\argmin}{\arg\min}
\DeclareMathOperator*{\argmax}{\arg\max}
\newcommand{\e}{\mathrm{e}}
\renewcommand{\i}{\mathrm{i}}

\IEEEoverridecommandlockouts
\title{Parameter estimation of range-migrating targets using OTFS signals from LEO satellites}
\author{Tong~Ding, Luca~Venturino,~\IEEEmembership{Senior~Member,~IEEE}, and Emanuele~Grossi,~\IEEEmembership{Senior~Member,~IEEE}
	\thanks{Copyright (c) 20xx IEEE. Personal use of this material is permitted. However, permission to use this material for any other purposes must be obtained from the IEEE by sending a request to pubs-permissions@ieee.org.}
    \thanks{Tong~Ding is with the Key Laboratory of Civil Aviation Data Governance and Decision Optimization, Civil Aviation Management Institute of China, Beijing, 100102, China (e-mail: dingtong@camic.cn). L.~Venturino and E.~Grossi are with the Department of Electrical and Information Engineering, University of Cassino and Southern Lazio (UNICAS), 03043 Cassino, Italy, with the People Oriented Smart Technology Lab (POSTLab), European University of Technology (EUt+), European Union, and with the National Inter-University Consortium for Telecommunications (CNIT), 43124 Parma, Italy (e-mail: l.venturino@unicas.it, e.grossi@unicas.it).}
    \thanks{The work of L.~Venturino is supported by the project CIRCE, CUP H53D23000420006, funded by the European Union under the Italian National Recovery and Resilience Plan of NextGenerationEU. The work of E.~Grossi is supported by the project FLARE, CUP E63C22002040007, funded by the European Union under the Italian National Recovery and Resilience Plan of NextGenerationEU, partnership on ``Telecommunications of the Future'' (Project PE00000001, program ``RESTART'').}
    \thanks{A preliminary version of this work was presented at the 2025 IEEE International Workshop on Technologies for Defense and Security.}
}

\begin{document}
\bstctlcite{BSTcontrol}
\maketitle
\IEEEpeerreviewmaketitle

\begin{abstract}
    This study investigates a communication-centric integrated sensing and communication system that utilizes orthogonal time-frequency space (OTFS) modulated signals emitted by low Earth orbit (LEO) satellites to estimate the initial-range, range-rate, and amplitude of point-like targets experiencing range migration under a first-order range model over the OTFS frame. Leveraging the signal samples produced by off-the-shelf OTFS demodulators, we derive the delay-Doppler-domain input-output relationship governing the target echo under both ideal and rectangular shaping filters. The target response exhibits a sparse structure in the delay-Doppler domain, whose support is determined by the target initial-range and range-rate. Exploiting this sparse model, we derive a lower-complexity approximation of the maximum-likelihood estimator for the target parameters. The proposed method first obtains coarse estimates of the target initial-range and range-rate using block orthogonal matching pursuit; then, a bank of matched filters focused on a smaller initial-range/range-rate region refines these two estimates and computes the amplitude estimate. The single-target procedure is extended to multiple targets via iterative estimation, reconstruction, and cancellation of dominant echoes. Finally, numerical examples are provided to evaluate the estimation performance.
\end{abstract}


\section{Introduction}
Future wireless networks and Internet-of-Things (IoT) systems will utilize non-terrestrial stations to ensure seamless and ubiquitous connectivity~\cite{Mojtaba-2022}. Among space transceivers, low Earth orbit (LEO) satellites are attracting significant attention for their low latency, high data rates, and capability to support integrated sensing and communication (ISAC)~\cite{Clerckx-2023,You-2024,Wang-2025}, enabling applications such as environmental monitoring, global navigation, and space exploration.

Orthogonal time-frequency space (OTFS) modulation is a promising transmission scheme for LEO constellations due to its robustness in doubly-dispersive channels~\cite{Viterbo-2018,Buzzi-2023,Shi-2024}. Its delay-Doppler-domain representation is also well suited to radar sensing, as the underlying processing resembles that of pulsed-Doppler radars~\cite{Bondre-2022-A}; in particular, the range and range-rate parameters of detected objects can be inferred from the estimated delay-Doppler channel response.\footnote{In this work, the range is defined as the propagation distance of the signal path, while the range-rate is defined as the negative of its time derivative, so that positive range-rate corresponds to a decreasing propagation distance. The range and range-rate provide a convenient physical parameterization of the delay and Doppler shifts along the propagation path, respectively.} These features enable OTFS-based LEO satellites to support ISAC in various forms~\cite{Wu-2025,Yuan-2025}. Communication-centric ISAC is promising in this context, as existing signals are reused for sensing at low cost~\cite{Raviteja-2019,Venturino-2021,Piotr-2022,Singh-2022,Stock-2024}, and this approach is considered in this study.

A large body of work has addressed signal detection and channel estimation for OTFS systems by exploiting the sparsity of the delay-Doppler representation under the standard approximation that range and range-rate variation are negligible over the OTFS frame, an assumption that is usually valid for terrestrial scenarios; see~\cite{Aldababsa-2024} and references therein. For example,~\cite{Viterbo-2018} derived an input-output signal model under the assumption of integer range and fractional range-rate for both ideal and rectangular shaping filters, along with a message passing algorithm for data detection.\footnote{Integer (fractional) range or range-rate means that the range or range-rate is an integer (fractional) multiple of the respective resolution.} Based on this model, pilot-aided channel estimation schemes have been proposed in~\cite{Raviteja-2019-pilot}, where guard symbols are introduced to avoid interference between pilot and data symbols at the receiver. For multiple-input multiple-output OTFS systems, \cite{Wenqian-2019} exploits three-dimensional structured sparsity for downlink estimation. Sparse Bayesian learning approaches have been developed in~\cite{LiuFei-2021} and extended in~\cite{Zhiqiang-2022} to handle both fractional range and range-rate. On the sensing side,~\cite{Raviteja-2019} exploited OTFS signals for radar sensing within a communication-centric ISAC framework. This idea has been developed in several directions, including the derivation of Cramér–Rao lower bounds for range and range-rate estimation~\cite{Colavolpe-2020}, the analysis of pulse shaping effects~\cite{Vani-2022}, the design of sparse recovery algorithms with fractional refinement~\cite{Zacharia-2023-ISAC}, and atomic norm-based formulations for target parameter estimation~\cite{LIU-2024}. In addition, OTFS signals have been employed for joint localization and communication in~\cite{Zijun-2023} and for target tracking in~\cite{Gao-2025}.

Assuming the range of each propagation path to be constant within an OTFS frame may no longer be valid in sensing scenarios based on LEO satellite signals. For instance, non-cooperative space targets, such as debris, asteroids, spacecraft, or missiles, may exhibit range-rates of several km/s. For example, when considering a system with $M=512$ subcarriers, $N=128$ symbol intervals, a subcarrier spacing of $\Delta=15$~kHz, a symbol interval of $T=1/\Delta$, and a range-rate of $\bar v=15$~km/s, the range variation over the OTFS frame is $\bar vNT=128$~m, which greatly exceeds the range resolution $c/(M\Delta)\approx 39$~m, where $c$ is the speed of light.

This work investigates the use of OTFS communication signals for estimating the initial-range, range-rate, and amplitude of point-like targets undergoing range migration within an OTFS frame, hereafter referred to as high-speed targets. Specifically, we relax the standard constant-range approximation while retaining the assumption of negligible range-rate variation over the frame. We adopt a communication-centric approach in which the emitted waveform is not modified. Sensing high-speed targets is challenging, and prior radar studies have primarily relied on trains of unmodulated or linear frequency-modulated pulses and developed motion-compensation strategies in the time or frequency domain to coherently integrate consecutive pulses~\cite{Penghui-2016,Tong-2022,Mingxing-2023,Xiong-2023}. In the OTFS setting considered here, however, properly accounting for range migration requires characterizing its impact on the delay-Doppler-domain signal model. Accordingly, we focus on deriving a range-migration-aware echo model, designing the corresponding radar receiver, and analyzing its estimation performance. The major contributions are summarized below.
\begin{itemize}
    \item We derive a model for the echo produced by a high-speed point-like target in the delay-Doppler domain, considering both ideal and rectangular shaping filters; the target range is assumed constant over a block of a few symbols, but changes from block to block consistently with the range-rate. Ideal filters yield a circular convolution model, whereas rectangular filters break this convolution structure because of inter-carrier interference (ICI) and inter-symbol interference (ISI). This model subsumes the one in~\cite{Viterbo-2018} when target-range migration is negligible and the target range lies on the delay grid.

    \item We demonstrate that the response of a high-speed target maintains a sparse structure in the delay-Doppler domain, determined by its initial-range (i.e., the target range at the beginning of the OTFS frame) and range-rate. Specifically, range migration results in an expanded support of this response not captured by existing models.

    \item For the single-target case, we exploit this sparse structure to derive a lower-complexity approximation of the maximum likelihood (ML) estimator. Block orthogonal matching pursuit (BOMP)~\cite{eldar2009block} is first used to obtain coarse estimates of the target initial-range and range-rate; a local matched-filter search then refines these two estimates and computes the target amplitude estimate. For multiple targets, this two-step estimator is combined with the CLEAN algorithm~\cite{Colone-2016,Bose-2011,Misiurewicz-2012,Bosse-2018,Venturino-2021}, which iteratively estimates, reconstructs, and cancels the target echoes.

    \item Numerical simulations are conducted to validate the echo model and assess the performance of the estimation procedures under different system configurations, also in comparison with the OTFS-based estimators in~\cite{Raviteja-2019-pilot} and~\cite{Wenqian-2019} and a motion-aware matched-filter benchmark directly operating in the time domain.
\end{itemize}

\paragraph*{Organization}
The remainder of the paper is organized as follows. Sec.~\ref{SEC:system} describes the system model; Sec.~\ref{SEC_IO} derives the delay-Doppler input-output model; Sec.~\ref{SEC_Detector_Estimator} presents the proposed target-parameter estimation procedure; Sec.~\ref{SEC_Analysis} reports numerical results; and Sec.~\ref{SEC_Conclusions} concludes the paper. The appendices provide the supporting derivations.

\paragraph*{Notation}
In the following, $\mathbb R$ and $\mathbb C$ denote the sets of real and complex numbers, respectively. Lowercase and uppercase boldface letters denote column vectors and matrices. The symbols $(\,\cdot\,)^{*}$, $(\,\cdot\,)\transp$, and $(\,\cdot\,)\herm$ denote conjugate, transpose, and conjugate-transpose, respectively, while $\bm{I}_{M}$ is the $M\times M$ identity matrix. The $i$-th entry of $\bm a$ and the $(i,j)$-th entry of $\bm A$ are denoted by $(\bm a)_i$ and $(\bm A)_{i,j}$, respectively. The notations $(\bm{A}_{1} \ \cdots \ \bm{A}_{n})$ and $(\bm{A}_{1}; \ \cdots \ ; \bm{A}_{n})$ indicate horizontal and vertical matrix concatenation. $\mathbbm 1_{\mathcal A}$ is the indicator of condition $\mathcal A$, $[\,\cdot\,]_{M}$ denotes reduction modulo $M$, and $\delta(t)$ and $\delta[n]$ are the Dirac delta and discrete unit sample functions, respectively. Finally, $\Re\{\cdot\}$, $\i$, $\odot$, and $\mathrm{E}[\,\cdot\,]$ denote the real part, imaginary unit, Schur product, and statistical expectation, respectively.

\section{System description}\label{SEC:system}

\begin{figure}[!t]
    \centerline{\includegraphics[width=0.65\columnwidth]{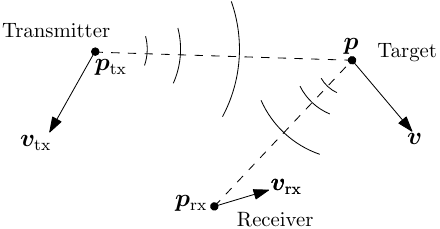}}
    \caption{System geometry.}
    \label{Fig_01_system_geometry}
\end{figure}

Consider a system comprising a communication transmitter and a radar receiver. The transmitter is mounted on a LEO satellite, uses OTFS modulation, and operates at carrier frequency $c/\lambda$, where $\lambda$ is the carrier wavelength; its waveform is specified by the underlying communication protocol and is not modified here. The receiver senses targets illuminated by the transmitter and may be co-located with it (monostatic configuration) or mounted on a separate platform (bistatic configuration), e.g., another LEO satellite or a high-altitude platform. The receiver is synchronized with the transmitter and knows the emitted symbols, which is readily ensured in the monostatic case~\cite{Venturino-2021}, while the bistatic case requires coordination or a separate reference channel~\cite{Malanowski_2019,Cristallini-2019,Cristallini-2022}. For illustration, we consider the bistatic configuration in Fig.~\ref{Fig_01_system_geometry}.

We assume point-like targets, each generating a single line-of-sight propagation path. This model is appropriate when each target is much smaller than the range resolution and its response is dominated by a single scattering center. A larger non-rotating object consisting of multiple resolvable scattering centers can also be approximately represented within the present framework as multiple point-like targets.\footnote{The treatment of more complex scenarios involving extended target structures with rotational micro-Doppler effects and multipath propagation is outside the scope of this contribution.} For clarity of exposition, we initially focus on a single target, while multiple targets are addressed in Sec.~\ref{SEC_multitarget}.

\subsection{Transmit signal}
Consider an OTFS frame with bandwidth $M\Delta$ and duration $NT$, where $M$ and $N$ are the number of subcarriers and symbol intervals, respectively, $\Delta $ is the subcarrier spacing, and $T=1/\Delta$ is the symbol interval. Accordingly, the corresponding range and range-rate resolutions are~\cite{Malanowski_2019}
\begin{equation}\label{resolutions}
    R_{\rm r}=c/(M \Delta),\quad
    R_{\rm rr}=\lambda/(NT),
\end{equation}
respectively.\footnote{$1/(M \Delta)$ and $1/(NT)$ are the delay and Doppler resolutions, respectively.} Up to a positive scaling factor accounting for the radiated energy, the baseband signal emitted by the communication transmitter is
\begin{equation}\label{tx_signal}
    x(t)=\sum_{n=0}^{N-1}\underbrace{\sum_{m=0}^{M-1} X_{\rm TF}[n,m] g_{\rm tx}\!\left(t-nT\right)\e^{\i 2\pi m \Delta (t-nT)}}_{x_{n}(t)},
\end{equation}
where $x_{n}(t)$ is the signal emitted in the $n$-th symbol interval, $X_{\rm TF}[n,m]$ is the time-frequency symbol sent in the $n$-th symbol interval on subcarrier $m$, and $g_{\rm tx}(t)$ is the transmit shaping filter with unit energy, effective temporal duration $T$, and effective spectral width $\Delta$. The time-frequency sequence $X_{\rm TF}[n,m]$ is obtained as the inverse Symplectic Finite Fourier Transform of the delay-Doppler sequence $X_{\rm DD}[k,l]$, i.e.,
\begin{equation}\label{ISFFT_TX}
    X_{\rm TF}[n,m]=\frac{1}{\sqrt{NM}}\sum_{k=0}^{N-1}\sum_{l=0}^{M-1} X_{\rm DD}[k,l]\e^{\i 2\pi\left(\frac{n k}{N}-\frac{m l}{M}\right)},
\end{equation}
for $n=0,\ldots,N-1$ and $m=0,\ldots,M-1$. By considering $g_{\rm tx}(t)$ time limited in $[0, T]$, then $x(t)$ has support $[0,NT]$. Also, by assuming the spectrum of $g_{\rm tx}(t)$ is contained in $[0, \Delta]$, then the spectrum of $x(t)$ is contained in $[0,M\Delta]$.

\subsection{Target kinematics and design assumptions}\label{Sec:design-assumptions}
Let $\bm p(t)$, $\bm p_{\rm tx}(t)$, and $\bm p_{\rm rx}(t)$ denote the sufficiently smooth trajectories of the target, communication transmitter, and radar receiver, respectively. Accordingly, the target range is
\begin{equation}\label{true-range}
    \bar{r}(t)=\left \| \bm{p}(t)-\bm{p}_{\rm tx}(t) \right \|+\left \|\bm{p}(t)-\bm{p}_{\rm rx}(t)\right\|.
\end{equation}
Hereafter, we make the following assumptions.

\subsubsection{First-order range model}\label{Sec:First-order-range-model}
Let $\bar{v}(t)=-\dot{\bar{r}}(t)$ be the instantaneous range-rate and define $\bar{d}=\bar{r}(0)$ and $\bar{v}=\bar{v}(0)$. Upon denoting by $\bm p=\bm p(0)$, $\bm p_{\rm tx}=\bm p_{\rm tx}(0)$, and $\bm p_{\rm rx}=\bm p_{\rm rx}(0)$, the initial positions of the target, communication transmitter, and radar receiver, respectively, and by $\bm v=\dot{\bm p}(0)$, $\bm v_{\rm tx}=\dot{\bm p}_{\rm tx}(0)$, and $\bm v_{\rm rx}=\dot{\bm p}_{\rm rx}(0)$ their initial velocities, we have~\cite{Tong-2022}
\begin{subequations}
    \begin{align}
        \bar{d}&=\left\|\bm{p}-\bm{p}_{\rm tx}\right\|+\left\|\bm{p}-\bm{p}_{\rm rx}\right\|,\\
        \bar{v}&=\frac{\left(\bm{p}-\bm{p}_{\rm tx}\right)\transp\!\left(\bm{v}_{\rm tx}-\bm{v}\right)}{\left\|\bm{p}-\bm{p}_{\rm tx}\right\|}
        +\frac{\left(\bm{p}-\bm{p}_{\rm rx}\right)\transp\!\left(\bm{v}_{\rm rx}-\bm{v}\right)}{\left\|\bm{p}-\bm{p}_{\rm rx}\right\|}.
    \end{align}
\end{subequations}
Then, the target range can be written exactly as
\begin{equation}
    \bar{r}(t)=\bar{d}-\bar{v}t+\varepsilon(t),
    \qquad
    \varepsilon(t)=-\int_{0}^{t}\left[\bar{v}(\tau)-\bar{v}\right]d\tau,
\end{equation}
where $\varepsilon(t)$ is the deviation from the first-order range model. To proceed, define the peak magnitude of the derivative of the instantaneous range-rate within the OTFS frame as
\begin{equation}
    \bar{a}_{\rm pk}=\max_{t\in[0,NT]}\left|\dot{\bar{v}}(t)\right|.
\end{equation}
Since $|\dot{\bar{v}}(t)|\leq\bar{a}_{\rm pk}$ for $t\in[0,NT]$, we have
\begin{equation}
    \left|\bar{v}(t)-\bar{v}\right|
    \leq\bar{a}_{\rm pk}t,
    \qquad
    \left|\varepsilon(t)\right|
    \leq\frac{1}{2}\bar{a}_{\rm pk}t^2,
\end{equation}
for $t\in[0,NT]$. In this study, we assume
\begin{equation}\label{cond_range_rate_migration}
    \bar{a}_{\rm pk}NT\ll R_{\rm rr},
    \qquad
    \frac{1}{2}\bar{a}_{\rm pk}(NT)^2\ll R_{\rm r}.
\end{equation}
The first condition ensures that the instantaneous range-rate can be regarded as constant over the OTFS frame at the range-rate resolution, while the second ensures that the target range can be regarded as linear in time at the range resolution.\footnote{The quantity $\bar a_{\rm pk}$ has units of m/s$^2$ and can equivalently be interpreted as the peak magnitude of the instantaneous range acceleration. For each propagation leg, the instantaneous range acceleration includes two contributions: one due to the relative acceleration projected along the line of sight and another due to transverse relative motion, which changes the line-of-sight direction and hence the radial component of the relative velocity~\cite{Hennessy-2025}. Under the system setup considered in Sec.~\ref{SEC_Analysis}, a hypothetical value $\bar a_{\rm pk}=200$~m/s$^2$ would correspond to a maximum range-rate variation of about $1.7$~m/s, i.e., about $20\%$ of $R_{\rm rr}$, and to a maximum deviation from the first-order range model of about $7.3$~mm, i.e., about $0.02\%$ of $R_{\rm r}$.} Under this assumption, we adopt the first-order range model
\begin{equation}
    \bar{r}(t)=\bar{d}-\bar{v}t, \quad \forall t\in[0,NT], \label{equivalent_range_specific}
\end{equation}
within the OTFS frame. If~\eqref{cond_range_rate_migration} is not satisfied, a higher-order range model and corresponding receiver processing are required, which are left for future investigation.\footnote{The transmitter and receiver trajectories may be known from navigation or ephemeris data. If a nominal target trajectory is also available from a preceding tracking stage, the predicted range evolution can be compensated for in the receiver processing. In this case, an analogous condition to~\eqref{cond_range_rate_migration} only needs to hold for the residual range evolution.}

\subsubsection{Unambiguous initial-range and range-rate intervals}
Denote by $\bar{r}_{\min}$ and $\bar{r}_{\max}$ the minimum and maximum initial-range values, respectively, and by $\bar{v}_{\min}$ and $\bar{v}_{\max}$ the minimum and maximum range-rates, respectively. These values are tied to the prior uncertainty on the target location and mobility, as resulting from the previous detection phase, and are assumed to be known. Finally, let $r_{\max}=\bar{r}_{\max}-\bar{r}_{\min}$ and $v_{\max}=\bar{v}_{\max}-\bar{v}_{\min}$ be the lengths of the inspected initial-range and range-rate intervals, respectively. Then, we assume
\begin{equation}\label{nec_cond_tau_nu}
    \tau_{\max}=\frac{r_{\max}}{c}< T, \qquad
    \nu_{\max}=\frac{v_{\max}}{\lambda}< \Delta,
\end{equation}
so as to avoid ambiguity in the estimation of the initial-range and the range-rate of the target, respectively.\footnote{For the parameters adopted in Sec.~\ref{SEC_Analysis}, we have $c/\Delta\approx 19.9$~km and $\lambda\Delta\approx 1.12$~km/s, respectively. By varying $\bar r_{\min}$ and $\bar v_{\min}$, any initial-range interval of length $r_{\max}<19.9$~km and any range-rate interval of length $v_{\max}<1.12$~km/s can be jointly inspected without ambiguity.}

\subsubsection{Blockwise constant-range approximation}
Let $B$ be a divisor of $N$, and partition the OTFS frame into $B$ blocks of $N/B$ consecutive symbol intervals. For receiver design, we assume that the target range in~\eqref{equivalent_range_specific} is approximately constant within each block, i.e.,
\begin{equation}\label{stop-and-go-approx}
    \bar{r}(t)\approx\bar{r}(bT_{B}),\quad \forall t\in[bT_B,(b+1)T_B),
\end{equation}
where $T_{B}=NT/B$ and $b=0,\ldots,B-1$. Under a first-order range model, prior studies effectively neglect target-range migration over the OTFS frame, which corresponds to $B=1$ in the present formulation. When target-range migration is non-negligible over the OTFS frame, increasing $B$ improves the accuracy of~\eqref{stop-and-go-approx}. Specifically, define the peak magnitude of the range-rate within the inspected interval as
\begin{equation}
    \bar{v}_{\rm pk}=\max\left\{|\bar{v}_{\min}|,|\bar{v}_{\max}|\right\}.
\end{equation}
Within each block, the range variation is then bounded by
\begin{equation}
    |\bar{r}(t)-\bar{r}(bT_{B})|
    \leq \bar v_{\rm pk}(t-bT_{B}),
    \quad \forall t\in[bT_B,(b+1)T_B).
\end{equation}
Accordingly, the maximum range variation normalized to the range resolution is
\begin{equation}\label{ratio_range_migration}
    \eta_B=\frac{\bar v_{\rm pk}T_{B}}{R_{\rm r}}
    =\frac{\bar v_{\rm pk}MN}{cB},
\end{equation}
and a direct design criterion is to select $B$ such that $\eta_B\ll 1$ (more on this in Sec.~\ref{SEC_Analysis}).

\subsection{Received signal}
Upon neglecting the noise contribution, the radio-frequency target echo at the radar receiver is
\begin{equation}
    \bar{y}_{\rm RF}(t)= \Re\left\{\bar{\alpha} x\left(t-\frac{\bar{r}(t)}{c}\right)\e^{\i 2\pi\frac{c}{\lambda}\left(t-\frac{\bar{r}(t)}{c}\right)}\right\},
\end{equation}
where $\bar{\alpha}\in\mathbb{C}$ is the target amplitude (accounting for the radiated energy, the transmit and receive antenna gains, the target radar cross-section, and any signal attenuation). For convenience, we advance the time axis by $\bar{r}_{\min}/c$, thus obtaining
\begin{equation}
    \bar{y}_{\rm RF}\left(t+\frac{\bar{r}_{\min}}{c}\right)=\Re\left\{y(t)\e^{\i 2\pi\frac{c+\bar{v}_{\min}}{\lambda}t}\right\},
\end{equation}
where
\begin{equation}\label{baseband_echo}
    y(t)=\bar{\alpha} x\left(t-\frac{r(t)}{c}\right)\e^{-\i 2\pi \frac{r(t)}{\lambda}} \e^{-\i 2\pi\frac{\bar{v}_{\min}}{\lambda}t}
\end{equation}
is the baseband representation of the target echo with respect to the reference frequency $\frac{c+\bar{v}_{\min}}{\lambda}$ and $r(t)= \bar{r}(t)-\bar{r}_{\min}$ is the excess target range (i.e., the target range in excess with respect to $\bar{r}_{\min}$). Note that the support of $y(t)$ is contained in the time interval $[0,NT+r_{\max}/c]$, and its spectrum is negligible outside the frequency interval $[0,M\Delta+v_{\max}/\lambda]$.

Define the excess initial-range $d=\bar{d}-\bar{r}_{\min}\in[0,r_{\max}]$, the excess range-rate $v=\bar{v}-\bar{v}_{\min}\in[0,v_{\max}]$, and $\alpha=\bar{\alpha}\e^{-\i 2\pi d/\lambda}$. If the target-range variation within each symbol interval is neglected, then~\eqref{baseband_echo} becomes
\begin{equation}
    y(t)=\alpha\sum_{n=0}^{N-1}
    x_n\left(t-\frac{r(nT)}{c}\right)
    \e^{\i 2\pi\frac{v}{\lambda}t},
    \label{rx_signal-fast}
\end{equation}
where $r(nT)=d-(v+\bar v_{\min})nT$ under a first-order range model. Note that each symbol interval experiences a target delay $r(nT)/c$ which may vary across the OTFS frame, while the Doppler shift $v/\lambda$ remains constant. The range evolution is retained here at the symbol level; instead, the coarser blockwise constant-range approximation in~\eqref{stop-and-go-approx} is invoked only in Sec.~\ref{SEC_IO} to derive the delay-Doppler-domain models.

\subsection{OTFS demodulation}
The radar receiver relies on an OTFS demodulator to obtain discrete samples of the received signal in the delay-Doppler domain. Such samples are then processed to estimate the target parameters as shown in Sec.~\ref{SEC_Detector_Estimator}.

First, the OTFS demodulator transforms $y(t)$ into the time-frequency domain via the Wigner Transform; in particular, the following measurements are obtained
\begin{equation}\label{rx_signal_TF}
    Y_{\rm TF}[n,m]=\int_{-\infty}^{\infty}y(t)g_{\rm rx}^{*}(t-nT)\e^{-\i 2 \pi m \Delta (t-nT)}dt,
\end{equation}
for $n=0,\ldots,N-1$ and $m=0,\ldots,M-1$, where $g_{\rm rx}(t)$ is the receive shaping filter with unit energy, effective temporal duration $T$ and effective spectral width $\Delta$. Denote by
\begin{equation}
    \gamma(\tau,\nu)=\int_{-\infty}^{\infty}g_{\rm tx}(\beta) g^{*}_{\rm rx}(\beta-\tau)\e^{-\i 2\pi \nu (\beta-\tau)}d\beta
\end{equation}
the cross-ambiguity function of $g_{\rm tx}(t)$ and $g_{\rm rx}(t)$; then, the following proposition establishes a connection between $\{Y_{\rm TF}[n,m]\}$ and $\{X_{\rm TF}[n,m]\}$ that generalizes the previous result in~\cite[Theorem~1]{Viterbo-2018}. The proof is reported in Appendix~\ref{Appendix_TF}.
\begin{proposition}\label{Proposition_TF}
    If the variation of the target range within each symbol interval can be ignored, then
    \begin{equation}
        Y_{\rm TF}[n,m]=\sum_{n'=0}^{N-1}\sum_{m'=0}^{M-1} H_{n,m}[n',m']X_{\rm TF}[n',m'],\label{rx_signal_TF_fast}
    \end{equation}
    for $n=0,\ldots,N-1$ and $m=0,\ldots,M-1$, where
    \begin{multline}
        H_{n,m}[n',m']= \alpha \e^{\i 2\pi\frac{v}{\lambda}nT}
        \e^{-\i 2\pi m' \Delta\frac{r\left(n'T\right)}{c}} \\
        \times \gamma\left((n-n')T-\frac{r\left(n'T\right)}{c},(m-m') \Delta -\frac{v}{\lambda}\right). \label{H_signal_TF_fast}
    \end{multline}
\end{proposition}

Next, the OTFS demodulator transforms $Y_{\rm TF}[n,m]$ into the delay-Doppler domain via the Symplectic Finite Fourier Transform; in particular, we have
\begin{equation}\label{rx_signal_DD}
    Y_{\rm DD}[k,l]=\frac{1}{\sqrt{NM}}\sum_{n=0}^{N-1}\sum_{m=0}^{M-1} Y_{\rm TF}[n,m]\e^{-\i 2\pi\left(\frac{n k}{N}-\frac{m l}{M}\right)},
\end{equation}
for $k=0,\ldots,N-1$ and $l=0,\ldots,M-1$.

\section{Delay–Doppler Input–Output Model} \label{SEC_IO}

We establish here a relationship between $\{Y_{\rm DD}[k,l]\}$ and $\{X_{\rm DD}[k,l]\}$ for two representative shaping filters, namely ideal and rectangular filters. Ideal shaping filters lead to a simplified input--output relationship, and the resulting analysis provides useful insights into the impact of range migration and serves as a benchmark. In contrast, the more practical rectangular shaping filters yield a more involved relationship.

\subsection{Ideal shaping filters} \label{SEC_IO_ideal}
Ideal shaping filters satisfy the bi-orthogonality condition~\cite[Sec.~III]{Viterbo-2018}. In this case, the ambiguity function satisfies $\gamma(\tau,\nu)=1$ for $(\tau,\nu)\in[-\tau_{\max},\tau_{\max}]\times[-\nu_{\max},\nu_{\max}]$, and $\gamma(\tau,\nu)=0$ for $(\tau,\nu)\in[nT-\tau_{\max},nT+\tau_{\max}]\times[m\Delta-\nu_{\max},m\Delta+\nu_{\max}]$ for all integer pairs $(m,n)\neq(0,0)$. Consequently, \eqref{rx_signal_TF_fast} becomes
\begin{equation}\label{rx_signal_TF_ideal_fast}
    Y_{\rm TF}[n,m]=H_{n,m}[n,m] X_{\rm TF}[n,m],
\end{equation}
for $n=0,\ldots,N-1$ and $m=0,\ldots,M-1$, where
\begin{equation}\label{H_fast_ideal}
    H_{n,m}[n,m]=\alpha\e^{\i 2\pi\frac{v}{\lambda}nT}
    \e^{-\i 2\pi \frac{r\left(nT\right)}{c} m \Delta}.
\end{equation}

To proceed, denote by
\begin{equation}
    \mathcal{D}_{Q}(\nu)=\frac{1}{Q}\sum_{q=0}^{Q-1}\e^{-\i 2\pi \nu q}=\frac{\e^{-\i 2\pi \nu Q}-1}{Q\left(\e^{-\i 2\pi \nu }-1\right)}
\end{equation}
the periodic $\mathrm{sinc}$ function (also called Dirichlet function)\footnote{ Recall that $\mathcal{D}_{Q}(\nu)$ is a periodic function of $\nu$ with period one and is zero at multiples of $1/Q$, except for $\nu=0,\pm 1, \pm 2, \ldots$ where it is one; also, $|\mathcal{D}_{Q}(\nu)|$ has a main-lobe centered at $0$ of width $2/Q$ (between zero crossings) and $Q-2$ sidelobes of width $1/Q$ (between zero crossings) in $[-0.5,0.5]$. The first sidelobe is approximately $13$~dB down from the main-lobe, and the subsequent sidelobes fall off at about $6$~dB per octave~\cite{Harris-1978}.} of degree $Q$. Then, the following result extends~\cite[Proposition~2]{Viterbo-2018} to the case of target range variations every $N/B$ symbol intervals. The proof is reported in Appendix~\ref{Appendix_DD_ideal}.
\begin{proposition}\label{Proposition_DD_ideal}
    Assume that ideal shaping filters are employed. Under the blockwise constant-range approximation, we have
    \begin{equation}
        Y_{\rm DD}[k,l]= \alpha \sum_{k'=0}^{N-1}\sum_{l'=0}^{M-1} \! X_{\rm DD}[k',l']\Phi[k-k',l-l'], \label{rx_signal_DD_ideal_fast}
    \end{equation}
    for $k=0,\ldots,N-1$ and $l=0,\ldots,M-1$, where
    \begin{align}
        \Phi[k,l]&= \frac{1}{B}\sum_{b=0}^{B-1} \e^{-\i 2 \pi\left(\frac{k}{N}-\frac{vT}{\lambda}\right)\frac{bN}{B}} \mathcal{D}_{\frac{N}{B}}\!\left(\frac{k}{N}-\frac{vT}{\lambda}\right)\notag\\
        &\quad \times \mathcal{D}_{M}\!\left(-\left(\frac{l}{M }-\frac{r\left(bT_{B}\right)\Delta}{c}\right)\right).\label{hw_fast}
    \end{align}
\end{proposition}

Next, we provide further insight into Proposition~\ref{Proposition_DD_ideal}.
\begin{remark}\label{Remark-periodicity}
    $\Phi[k,l]$ is periodic in $k$ and $l$ with period $N$ and $M$, respectively. This implies that~\eqref{rx_signal_DD_ideal_fast} is a two-dimensional circular convolution, which can be rewritten as
    \begin{subequations}\label{rx_signal_DD_ideal_fast-circ}
        \begin{align}
            Y_{\rm DD}[k,l]&=\alpha \sum_{k'=0}^{N-1}\sum_{l'=0}^{M-1} X_{\rm DD}[k',l']\Phi\Big[[k-k']_{N},[l-l']_{M}\Big]\label{rx_signal_DD_ideal_fast-circ1} \\
            &=\alpha \sum_{k'=0}^{N-1}\sum_{l'=0}^{M-1} X_{\rm DD}\Big[[k-k']_{N},[l-l']_{M}\Big]\Phi[k',l'].
            \label{rx_signal_DD_ideal_fast-circ2}
        \end{align}
    \end{subequations}
    Here, $\Phi[k,l]$ is the convolution kernel relating the transmitted and received delay-Doppler symbols. Accordingly, $\alpha \Phi[k,l]$ represents the \emph{target response} in the delay-Doppler domain, and $\Phi[k,l]$ can be interpreted as the associated \emph{point-spread function}, specified by the target initial-range and range-rate.
\end{remark}

\begin{remark}
    Proposition~\ref{Proposition_DD_ideal} subsumes~\cite[Proposition~2]{Viterbo-2018} as a special case. To illustrate this point, first use~\eqref{resolutions} to write the range at time $bT_{B}$ and the range-rate as
    \begin{equation}\label{tau_nu_int_fra_fast}
        r\left(bT_{B}\right)=(l_{b,{\rm int}}+l_{b,{\rm fra}})R_{\rm r}, \quad
        v=(k_{\rm int}+k_{\rm fra})R_{\rm rr},
    \end{equation}
    where $l_{b,{\rm int}}\in\{0,1,\ldots,M-1\}$, $k_{\rm int}\in\{0,1,\ldots,N-1\}$, and $l_{b,{\rm fra}},\, k_{\rm fra}\in(-0.5,0.5]$. Next, assume no range migration, whereby $l_{b,{\rm int}}=l_{{\rm int}}$ and $l_{b,{\rm fra}}=l_{{\rm fra}}$ for $b=0,\ldots,B-1$. Upon plugging~\eqref{tau_nu_int_fra_fast} in~\eqref{rx_signal_DD_ideal_fast-circ2}, we obtain
    \begin{multline}
        Y_{\rm DD}[k,l]=\alpha \sum_{\bar{k}=-\lfloor N/2\rfloor}^{\lceil N/2\rceil-1}\sum_{\bar{l}=-\lfloor M/2\rfloor}^{\lceil M/2\rceil-1}
        \mathcal{D}_{N}\!\left(-\frac{\bar{k}+k_{\rm fra}}{N}\right)
        \\ \times \mathcal{D}_{M}\!\left(\frac{\bar{l}+l_{\rm fra}}{M}\right)
        X_{\rm DD}\Big[[k-k_{\rm int}+\bar{k}]_{N},[l-l_{\rm int}+\bar{l}]_{M}\Big],\label{rx_signal_DD_ideal_fast_05}
    \end{multline}
    consistently with~\cite[Proposition~2]{Viterbo-2018}. In addition, if $k_{\rm fra}=l_{\rm fra}=0$, then~\eqref{rx_signal_DD_ideal_fast_05} becomes
    \begin{align}
        Y_{\rm DD}[k,l]&=\alpha X_{\rm DD}\Big[[k-k_{\rm int}]_{N},[l-l_{{\rm int}}]_{M}\Big].\label{rx_signal_DD_ideal_fast_07}
    \end{align}
\end{remark}

\begin{remark}\label{Remark-ideal-sparse}
    By substituting~\eqref{tau_nu_int_fra_fast} into~\eqref{hw_fast}, we obtain
    \begin{align}
        \Phi[k,l]
        &= \frac{1}{B}\sum_{b=0}^{B-1}
        \e^{-\i 2 \pi\left(\frac{k-k_{\rm int}-k_{\rm fra}}{B}\right)b}
        \mathcal{D}_{\frac{N}{B}}\!\left(
        \frac{k-k_{\rm int}-k_{\rm fra}}{N}\right)\notag\\
        &\quad\times
        \mathcal{D}_{M}\!\left(
        -\frac{l-l_{b,{\rm int}}-l_{b,{\rm fra}}}{M}\right).
        \label{hw_fast_new}
    \end{align}
    Consider first the delay domain. For large $M$, $\big|\mathcal{D}_{M}\!\big(-(l-l_{b,{\rm int}}-l_{b,{\rm fra}})/M\big)\big|$ is appreciable over approximately $2D_{\tau}+1$ delay indices around $l_{b,{\rm int}}$, where $D_{\tau}$ is a non-negative integer controlling the approximation error and satisfying $0<2D_{\tau}+1\ll M$~\cite{Viterbo-2018}. Hence, the number of delay indices over which $|\Phi[k,l]|$ is appreciable is approximately bounded by
    \begin{equation}\label{delay_support_phi}
        \max_{b\in\{0,\ldots,B-1\}} l_{b,{\rm int}}-\min_{b\in\{0,\ldots,B-1\}} l_{b,{\rm int}}+2D_{\tau}+1,
    \end{equation}
    and range migration enlarges the dominant delay-domain support according to the number of range-resolution bins traversed by the target. In the considered applications, the bound in~\eqref{delay_support_phi} still remains much smaller than $M$. Consider now the Doppler domain. In the absence of range migration,~\eqref{hw_fast_new} simplifies to
    \begin{equation}
        \!\Phi[k,l]
        =
        \mathcal{D}_{N}\!\left(\!
        \frac{k-k_{\rm int}-k_{\rm fra}}{N}\!\right)
        \mathcal{D}_{M}\!\left(\!
        -\frac{l-l_{\rm int}-l_{\rm fra}}{M}\!\right),
    \end{equation}
    consistently with~\eqref{rx_signal_DD_ideal_fast_05}. In this case, for large $N$, $|\Phi[k,l]|$ is appreciable over approximately $2D_{\nu}+1$ Doppler indices around $k_{\rm int}$, where $D_{\nu}$ is a non-negative integer controlling the approximation error and satisfying $0<2D_{\nu}+1\ll N$. With range migration, the block-dependent delay factors in~\eqref{hw_fast_new} may induce additional Doppler spreading. The following example further illustrates these effects.
\end{remark}

\begin{figure*}[!t]
    \centerline{
        \includegraphics[width=0.3\textwidth,trim=0cm 0.8cm 0cm 1cm, clip]{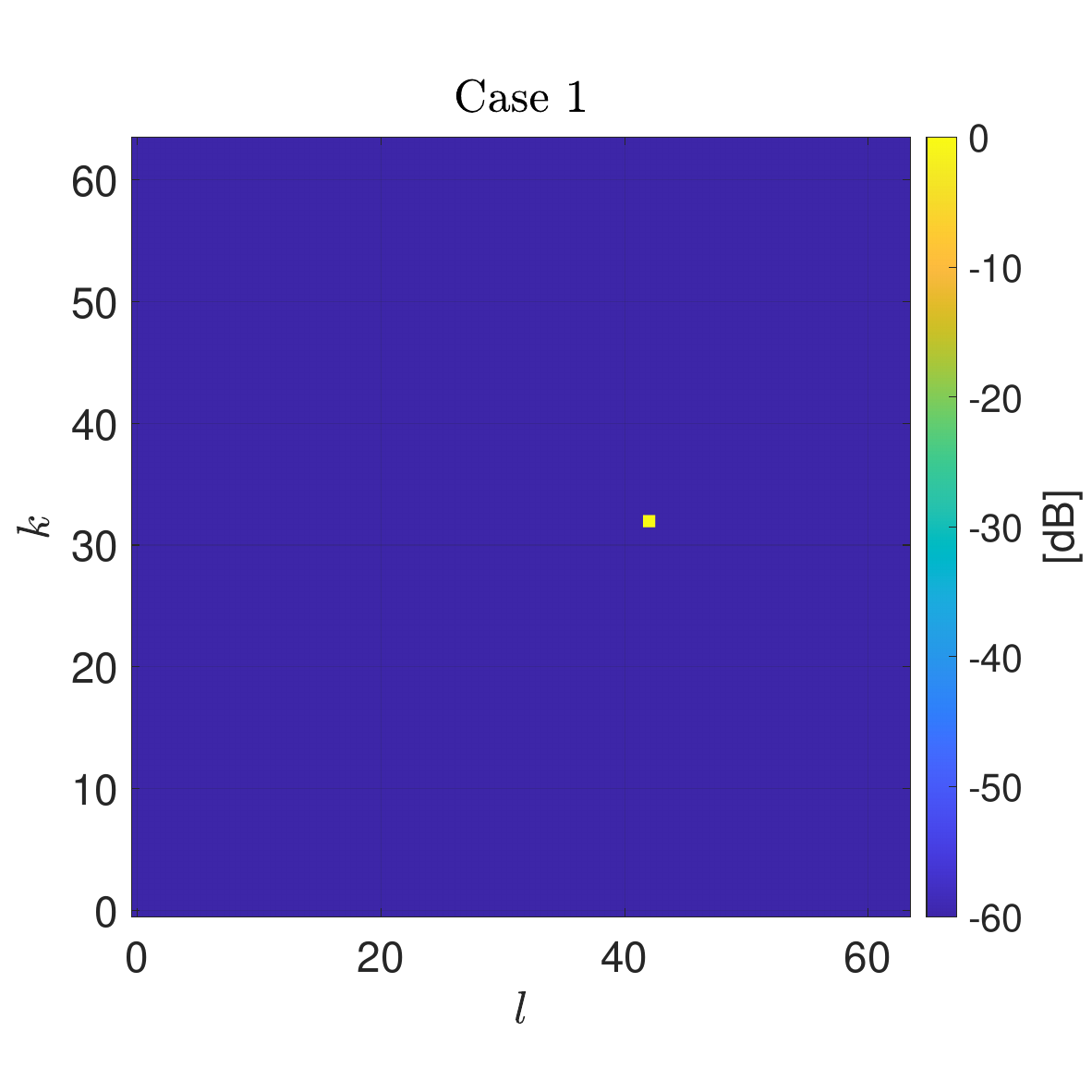}\hspace{0.5cm}
        \includegraphics[width=0.3\textwidth,trim=0cm 0.8cm 0cm 1cm, clip]{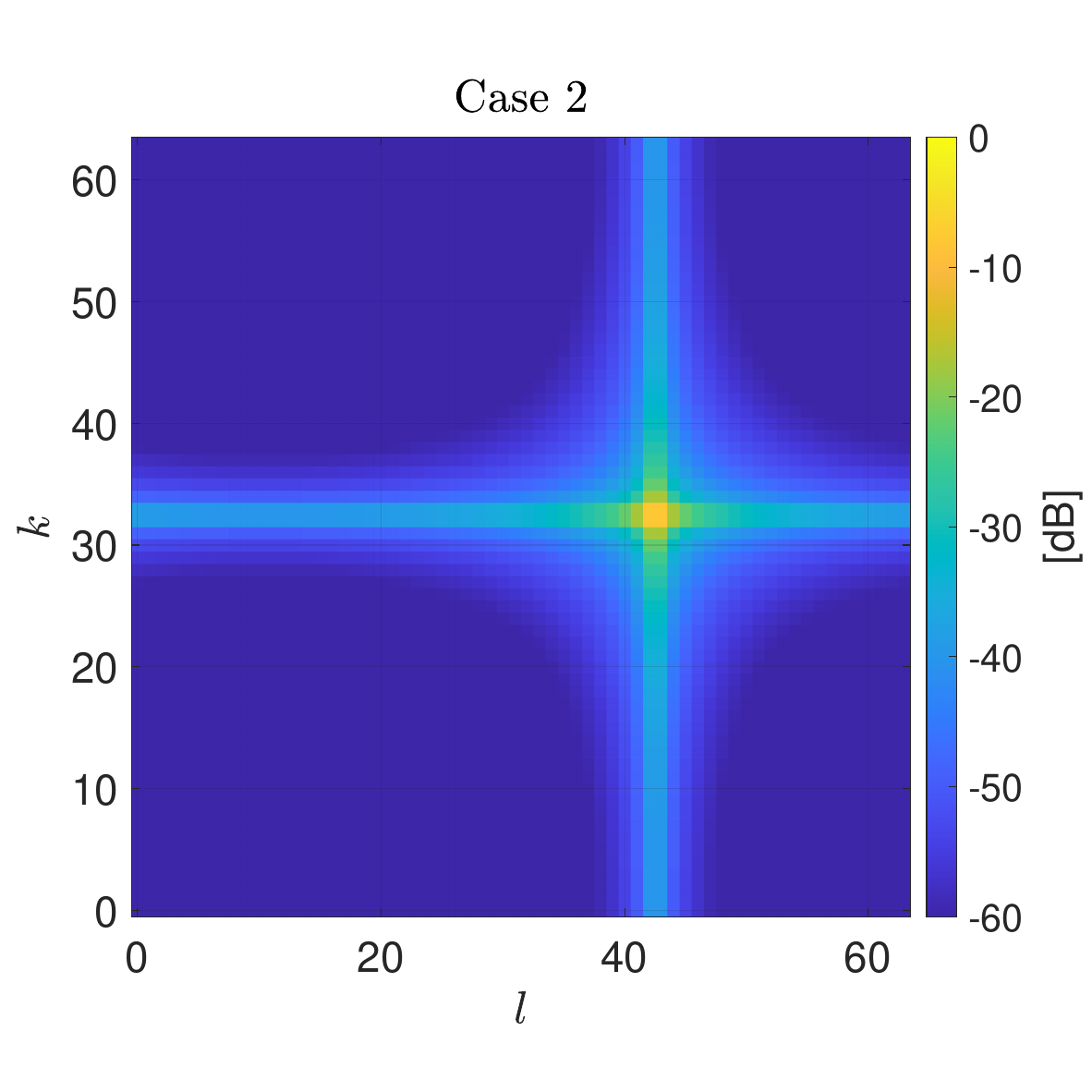}\hspace{0.5cm}
        \includegraphics[width=0.3\textwidth,trim=0cm 0.8cm 0cm 1cm, clip]{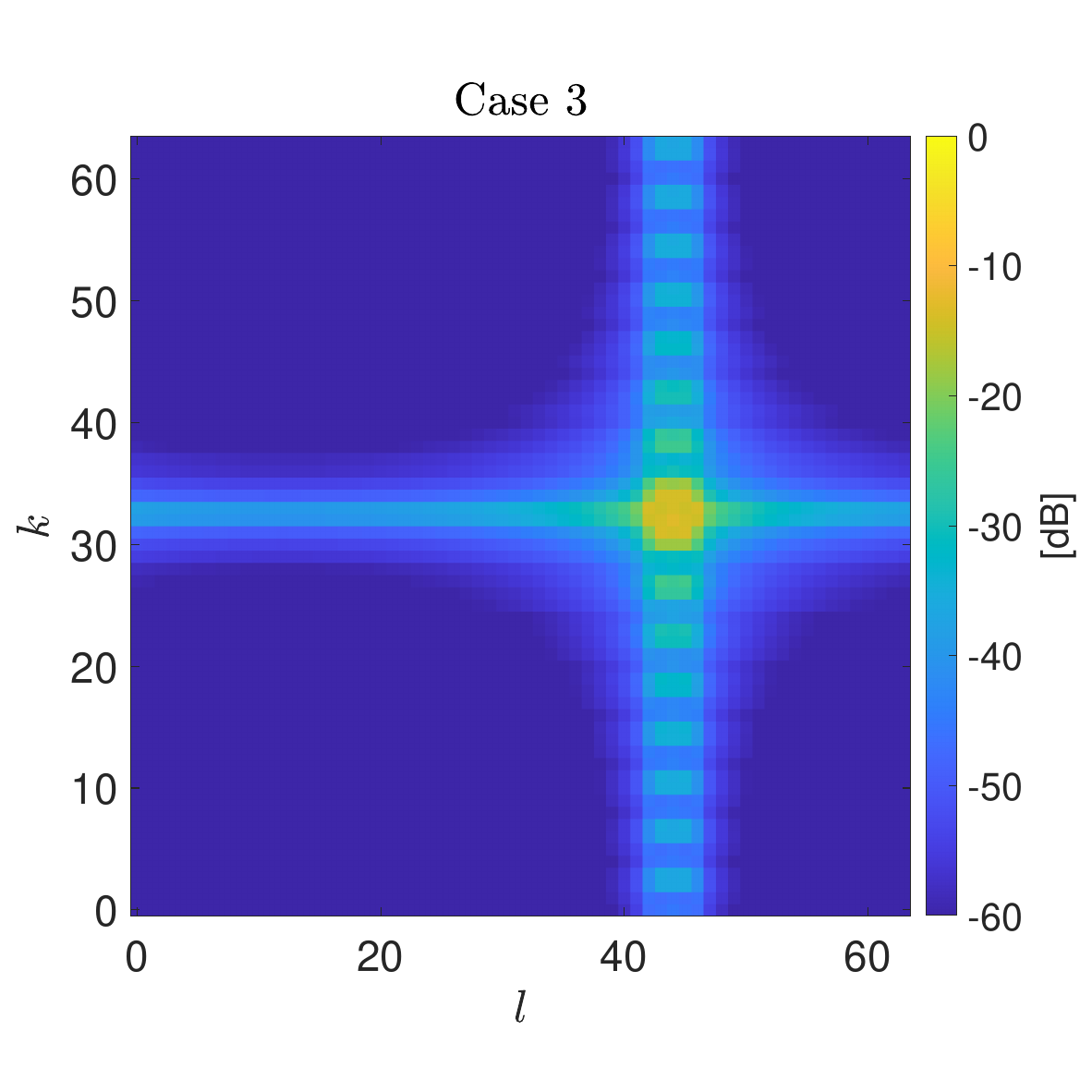}
    }
    \caption{$|\Phi[k,l]|$ vs $k=0,\ldots,N-1$ and $l=0,\ldots,M-1$, when $B=4$, $N=M=64$, $l_{0,{\rm int}}=42$, $k_{\rm int}=32$, and ideal shaping filters. Case~1 (left): $l_{1,{\rm int}}=l_{2,{\rm int}}=l_{3,{\rm int}}=42$ and $l_{0,{\rm fra}}=\cdots=l_{3,{\rm fra}}=k_{\rm fra}=0$. Case~2 (center): $l_{1,{\rm int}}=l_{2,{\rm int}}=l_{3,{\rm int}}=42$ and $l_{0,{\rm fra}}=\cdots=l_{3,{\rm fra}}=k_{\rm fra}=0.5$. Case~3 (right): $l_{1,{\rm int}}=43$, $l_{2,{\rm int}}=44$, $l_{3,{\rm int}}=45$, and $l_{0,{\rm fra}}=\cdots=l_{3,{\rm fra}}=k_{\rm fra}=0.5$.}
    \label{fig_hw}
\end{figure*}

\begin{example}\label{Example-ideal}
    To provide further insight into the effect of range migration on the target point-spread function in~\eqref{hw_fast}, we consider a toy example with $B=4$, $N=M=64$, $l_{0,{\rm int}}=42$, and $k_{\rm int}=32$. Fig.~\ref{fig_hw} reports $|\Phi[k,l]|$ versus $k=0,\ldots,N-1$ and $l=0,\ldots,M-1$ for different target kinematics. In Case~1, we assume no range migration (i.e., $l_{1,{\rm int}}=l_{2,{\rm int}}=l_{3,{\rm int}}=42$), integer range (i.e., $l_{0,{\rm fra}}=\cdots=l_{3,{\rm fra}}=0$), and integer range-rate (i.e., $k_{\rm fra}=0$). It is seen that $|\Phi[k,l]|$ contains a single non-zero entry at $(k,l)=(32,42)$, as predicted by~\eqref{rx_signal_DD_ideal_fast_07}. In Case~2, we assume no range migration, fractional range with $l_{0,{\rm fra}}=\cdots=l_{3,{\rm fra}}=0.5$, and fractional range-rate with $k_{\rm fra}=0.5$. By comparing Cases~1 and~2, it is seen that fractional range and range-rate induce a spreading of $|\Phi[k,l]|$ along both the delay and Doppler dimensions, as predicted by~\eqref{rx_signal_DD_ideal_fast_05}. In Case~3, we consider range migration across the four blocks, with $l_{1,{\rm int}}=43$, $l_{2,{\rm int}}=44$, $l_{3,{\rm int}}=45$ and $k_{\rm fra}=l_{0,{\rm fra}}=\cdots=l_{3,{\rm fra}}=0.5$. By comparing Cases~2 and~3, it is seen that range migration induces additional spreading of $|\Phi[k,l]|$ along the delay and Doppler dimensions, in agreement with Remark~\ref{Remark-ideal-sparse}. This additional spreading constitutes the novelty captured by Proposition~\ref{Proposition_DD_ideal}. Nevertheless, only a small fraction of the delay-Doppler coefficients have appreciable magnitude, so that the target response preserves a sparse structure despite its enlarged support.
\end{example}

\subsubsection{Vectorized model}\label{SEC:echo-ideal}
The samples $\{Y_{\rm DD}[k,l]\}$ can be arranged into the vector $\bm{y}\in\mathbb{C}^{NM}$, where $(\bm{y})_{lN+k+1}=Y_{\rm DD}[k,l]$, for $k=0,\ldots,N-1$ and $l=0,\ldots,M-1$. From~\eqref{rx_signal_DD_ideal_fast-circ1}, we obtain
\begin{equation}
    \bm{y}=\alpha\bm{\Phi}\bm{x}, \label{vec_model_ideal_1}
\end{equation}
where $\bm{\Phi}\in\mathbb{C}^{NM\times NM}$ and $\bm{x}\in\mathbb{C}^{NM}$ are defined by $(\bm{\Phi})_{lN+k+1,l'N+k'+1}= \Phi\big[[k-k']_{N},[l-l']_{M}\big]$ and $(\bm{x})_{lN+k+1}=X_{\rm DD}[k,l]$, respectively, for $k,k'=0,\ldots,N-1$ and $l,l'=0,\ldots,M-1$.

Alternatively, from~\eqref{rx_signal_DD_ideal_fast-circ2}, we have
\begin{equation}
    \bm{y}= \alpha \bm{X}\bm{\phi}, \label{vec_model_ideal_2}
\end{equation}
where $\bm{X}\in\mathbb{C}^{NM\times NM}$ and $\bm{\phi}\in\mathbb{C}^{NM}$ are defined by $(\bm{X})_{lN+k+1,l'N+k'+1}= X_{\rm DD}\big[[k-k']_{N},[l-l']_{M}\big]$ and $(\bm{\phi})_{lN+k+1}=\Phi[k,l]$, respectively, for $k,k'=0,\ldots,N-1$ and $l,l'=0,\ldots,M-1$.

\subsection{Rectangular shaping filters} \label{SEC_IO_rect}
Assume now $g_{\rm tx}(t)=g_{\rm rx}(t)=\sqrt{1/T}\,\mathrm{rect}(t/T)$. This implies that $\gamma(\tau,\nu)=0$ for $|\tau|\geq T$. Moreover, for large $M$, we have~\cite{Viterbo-2018}
\begin{multline}
    \gamma(\tau,\nu)= \frac{T}{M} \sum_{p=0}^{M-1} g_{\rm tx}\left(\frac{pT}{M}\right) g^{*}_{\rm rx}\left(\frac{pT}{M}-\tau\right)\e^{-\i 2\pi \nu \left(\frac{pT}{M}-\tau\right)}\\ =\begin{cases}\displaystyle
        \frac{1}{M} \sum_{p=0}^{\lceil M+\tau M /T\rceil-1}\!\!
        \e^{-\i 2\pi \nu T \left(\frac{p}{M}-\frac{\tau}{T}\right)}, & \tau\in (-T,0),\\
        \displaystyle
        \frac{1}{M} \sum_{p=\lceil \tau M /T\rceil}^{M-1}\!\!
        \e^{-\i 2\pi \nu T \left(\frac{p}{M}-\frac{\tau}{T}\right)}, & \tau\in [0,T).
    \end{cases}\label{cross-ambiguity-rect-approx}
\end{multline}
Consequently,~\eqref{rx_signal_TF_fast} simplifies as follows. For $n=1,\ldots,N-1$ and $m=0,\ldots,M-1$, we have
\begin{align}
    Y_{\rm TF}[n,m]&= H_{n,m}[n,m]X_{\rm TF}[n,m]\notag \\
    & \quad +\sum_{m'=0,\,m' \neq m}^{M-1} H_{n,m}[n,m']X_{\rm TF}[n,m']\notag \\
    & \quad +\sum_{m'=0}^{M-1} H_{n,m}[n-1,m']X_{\rm TF}[n-1,m'],\label{rx_signal_TF_fast-2}
\end{align}
where the second and third terms on the right-hand side correspond to ICI and ISI, respectively~\cite{Viterbo-2018}. For $n = 0$, instead, no ISI is present due to the guard interval between consecutive OTFS frames, and~\eqref{rx_signal_TF_fast} reduces to
\begin{align}
    Y_{\rm TF}[0,m]&= H_{0,m}[0,m]X_{\rm TF}[0,m]\notag \\
    & \quad +\sum_{m'=0,\, m'\neq m}^{M-1} H_{0,m}[0,m']X_{\rm TF}[0,m'],\label{rx_signal_TF_fast-2-0}
\end{align}
for $m=0,\ldots,M-1$. The following result extends~\cite[Theorem~2]{Viterbo-2018} to a non-integer target range varying every $N/B$ symbols. The proof is given in Appendix~\ref{Appendix_DD_rect}.
\begin{proposition}\label{Proposition_DD_rect}
    Assume that rectangular shaping filters are employed. Under the blockwise constant-range approximation, we have
    \begin{equation}
        Y_{\rm DD}[k,l]= \alpha \sum_{k'=0}^{N-1}\sum_{l'=0}^{M-1} \! X_{\rm DD}[k',l']\Psi_{k',l'}[k-k',l-l'], \label{rx_signal_DD_rect_fast}
    \end{equation}
    for $k=0,\ldots,N-1$ and $l=0,\ldots,M-1$, where
    \begin{align}
        \Psi_{k',l'}[k,l]&= \frac{1}{B}\sum_{b=0}^{B-1} \Psi_{b,k',l'}[k,l],\label{hw_rect_fast}
    \end{align}
    while
    \begin{align}
        &\Psi_{b,k',l'}[k,l]=\e^{\i 2 \pi \frac{v}{\lambda}\frac{r\left(bT_{B}\right)}{c}} \e^{\i 2 \pi \frac{v T}{\lambda}\frac{l'}{M}} \notag \\
        &\quad \times \e^{-\i 2 \pi \left(\frac{k}{N}-\frac{v T}{\lambda} \right)\frac{bN}{B}} \mathcal{D}_{\frac{N}{B}}\!\left(\frac{k}{N}-\frac{v T}{\lambda}\right) \notag \\
        &\quad \times\mathcal{D}_{M}\!\left(-\left(\frac{l}{M }-\frac{r\left(bT_{B}\right) \Delta}{c}\right)\right),\label{rx_signal_DD_rect_fast_A}
    \end{align}
    if $0\leq l'\leq \big\lceil M-\frac{r\left(bT_{B}\right)M}{cT}\big\rceil-1$, and
    \begin{align}
        &\Psi_{b,k',l'}[k,l]= \e^{-\i 2\pi\frac{k'}{N}} \e^{\i 2 \pi \frac{v}{\lambda}\frac{r\left(bT_{B}\right)}{c}} \e^{\i 2 \pi \frac{v T}{\lambda}\frac{l'-M}{M}} \notag \\
        &\quad \times \e^{-\i 2 \pi \left(\frac{k}{N}-\frac{v T}{\lambda} \right)\frac{bN}{B}} \left[\mathcal{D}_{\frac{N}{B}}\!\left(\frac{k}{N}-\frac{v T}{\lambda}\right)-\mathbbm 1_{\{b=0\}} \frac{B}{N}\right] \notag \\
        &\quad \times\mathcal{D}_{M}\!\left(-\left(\frac{l+M}{M }-\frac{r\left(bT_{B}\right) \Delta}{c}\right)\right),\label{rx_signal_DD_rect_fast_B}
    \end{align}
    if $ \big\lceil M-\frac{r\left(bT_{B}\right)M}{cT}\big\rceil\leq l'\leq M-1$.
\end{proposition}

Next, we provide further insight into Proposition~\ref{Proposition_DD_rect}.

\begin{remark}
    The right-hand side of~\eqref{rx_signal_DD_rect_fast} is not a two-dimensional convolution, since $\Psi_{k',l'}[k,l]$ depends on $k'$ and $l'$ due to ISI and ICI. This is in sharp contrast with the ideal shaping filter case. For fixed $k'$ and $l'$, $\Psi_{k',l'}[k,l]$ acts as a location-dependent kernel relating the input symbol $X_{\rm DD}[k',l']$ to the output symbols $Y_{\rm DD}[k,l]$, and is periodic in $k$ and $l$ with period $N$ and $M$, respectively. Accordingly, $\alpha \Psi_{k',l'}[k,l]$ represents the target response in the delay-Doppler domain to an input symbol located at $(k',l')$, and $\Psi_{k',l'}[k,l]$ can be interpreted as a location-dependent point-spread function, specified by the target initial-range and range-rate.
\end{remark}

\begin{remark}
    The result in~\eqref{rx_signal_DD_rect_fast} can be recast as
    \begin{align}
        &Y_{{\rm DD}}[k,l]=\frac{\alpha}{B} \sum_{b=0}^{B-1}\sum_{k'=0}^{N-1} \notag \\
        &\times
        \Bigg[
        \sum_{l'=0}^{M-l_{b,{\rm int}}-\lfloor l_{b,{\rm fra}}\rfloor-1}
        \!\!\! X_{\rm DD}[k',l'] \Psi_{b,k',l'}[k-k',l-l']\notag \\
        & +
        \sum_{l'=M-l_{b,{\rm int}}-\lfloor l_{b,{\rm fra}}\rfloor}^{M-1}
        \!\!\! X_{\rm DD}[k',l'] \Psi_{b,k',l'}[k-k',l-l']
        \Bigg].
    \end{align}
    Introducing $\bar{k}=k'-k+k_{\rm int}$, $\bar{l}_{b}=l'-l+l_{b,{\rm int}}$ in the first delay interval, and $\bar{l}_{b}=l'-l+l_{b,{\rm int}}-M$ in the second, using~\eqref{tau_nu_int_fra_fast} and
    \eqref{rx_signal_DD_rect_fast_A}--\eqref{rx_signal_DD_rect_fast_B}, and exploiting
    the $N$-periodicity of the resulting summand in $\bar{k}$, we obtain
    \begin{align}
        &Y_{{\rm DD}}[k,l]
        =\frac{\alpha}{B} \sum_{b=0}^{B-1}
        \sum_{\bar{k}=-\lfloor N/2\rfloor}^{\lceil N/2\rceil-1}
        \sum_{\bar{l}_{b}=-l-\lfloor l_{b,{\rm fra}}\rfloor}
        ^{M-l-1-\lfloor l_{b,{\rm fra}}\rfloor}
        \notag \\
        &\quad \times
        \e^{\i 2 \pi \frac{k_{\rm int}+k_{\rm fra}}{N}
        \frac{l_{b,{\rm int}}+l_{b,{\rm fra}}}{M}}
        \e^{\i 2 \pi \frac{k_{\rm int}+k_{\rm fra}}{N}
        \frac{l-l_{b,{\rm int}}+\bar{l}_{b}}{M}}
        \notag \\
        &\quad \times
        \e^{\i 2 \pi \left(\frac{\bar{k}+k_{\rm fra}}{B} \right)b}
        A_{b,\bar{k},\bar{l}_{b}}[k,l]
        \mathcal{D}_{M}\!\left(
        \frac{\bar{l}_{b}+l_{b,{\rm fra}}}{M }\right)
        \notag \\
        &\quad \times
        X_{\rm DD}\Big[
        [k-k_{\rm int}+\bar{k}]_{N},
        [l-l_{b,{\rm int}}+\bar{l}_{b}]_{M}
        \Big],
        \label{rx_signal_DD_rect_fast-2}
    \end{align}
    where $A_{b,\bar{k},\bar{l}_{b}}[k,l]$ is equal to
    \begin{equation}
        \mathcal{D}_{\frac{N}{B}}\!\left(
        -\frac{\bar{k}+k_{\rm fra}}{N}\right),
    \end{equation}
    if $l_{b,{\rm int}}-l \leq \bar{l}_{b}
    \leq M-l-1-\lfloor l_{b,{\rm fra}}\rfloor$, and to
    \begin{equation}
        \left[
        \mathcal{D}_{\frac{N}{B}}\!\left(
        -\frac{\bar{k}+k_{\rm fra}}{N}\right)
        -\mathbbm 1_{\{b=0\}} \frac{B}{N}
        \right]
        \e^{-\i 2\pi
        \frac{[k-k_{\rm int}+\bar{k}]_{N}}{N}},
    \end{equation}
    if $-l-\lfloor l_{b,{\rm fra}}\rfloor \leq \bar{l}_{b}\leq l_{b,{\rm int}}-l-1$. If $l_{b,{\rm int}}=l_{{\rm int}}$ and $l_{b,{\rm fra}}=0$ for $b=0,\ldots,B-1$, then~\eqref{rx_signal_DD_rect_fast-2} becomes
    \begin{multline}
        Y_{{\rm DD}}[k,l]=\alpha
        \sum_{\bar{k}=-\lfloor N/2\rfloor}^{\lceil N/2\rceil-1}
        \e^{\i 2 \pi \frac{k_{\rm int}+k_{\rm fra}}{N}
        \frac{l_{{\rm int}}}{M}}
        \e^{\i 2 \pi \frac{k_{\rm int}+k_{\rm fra}}{N}
        \frac{l-l_{{\rm int}}}{M}}
        \\
        \times A_{\bar{k}}[k,l]
        X_{\rm DD}\Big[
        [k-k_{\rm int}+\bar{k}]_{N},
        [l-l_{{\rm int}}]_{M}
        \Big],
        \label{rx_signal_DD_rect_fast-4}
    \end{multline}
    where $A_{\bar{k}}[k,l]$ is equal to
    \begin{equation}
        \mathcal{D}_{N}\!\left(
        -\frac{\bar{k}+k_{\rm fra}}{N}\right),
    \end{equation}
    if $l_{{\rm int}} \leq l \leq M-1$, and to
    \begin{equation}
        \left[
        \mathcal{D}_{N}\!\left(
        -\frac{\bar{k}+k_{\rm fra}}{N}\right)
        -\frac{1}{N}
        \right]
        \e^{-\i 2\pi
        \frac{[k-k_{\rm int}+\bar{k}]_{N}}{N}},
    \end{equation}
    if $0\leq l\leq l_{{\rm int}}-1$, which is consistent with~\cite[Theorem~2]{Viterbo-2018}.
\end{remark}

\begin{figure*}[!t]
    \centerline{
        \includegraphics[width=0.3\textwidth,trim=0cm 0.8cm 0cm 1cm, clip]{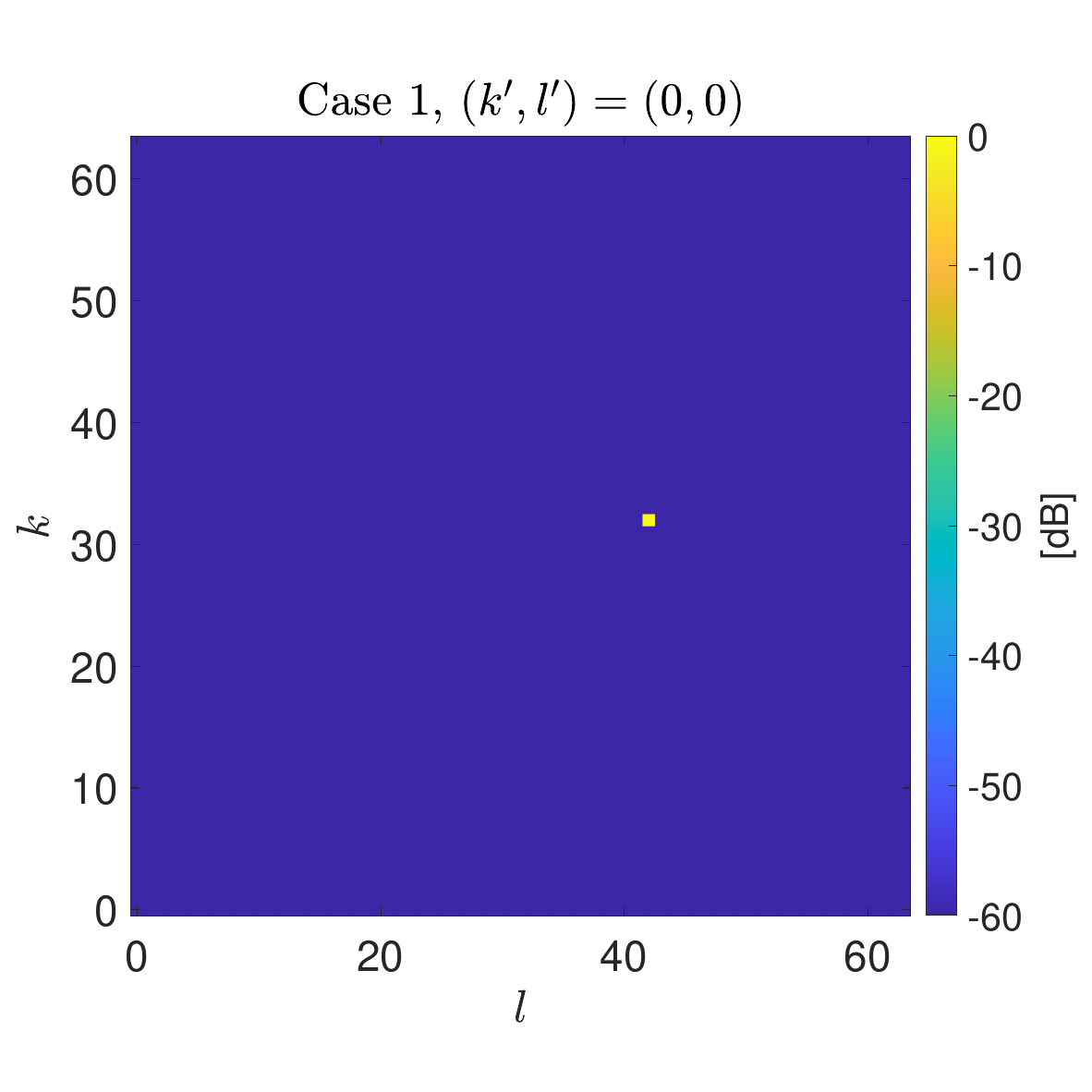}\hspace{0.5cm}
        \includegraphics[width=0.3\textwidth,trim=0cm 0.8cm 0cm 1cm, clip]{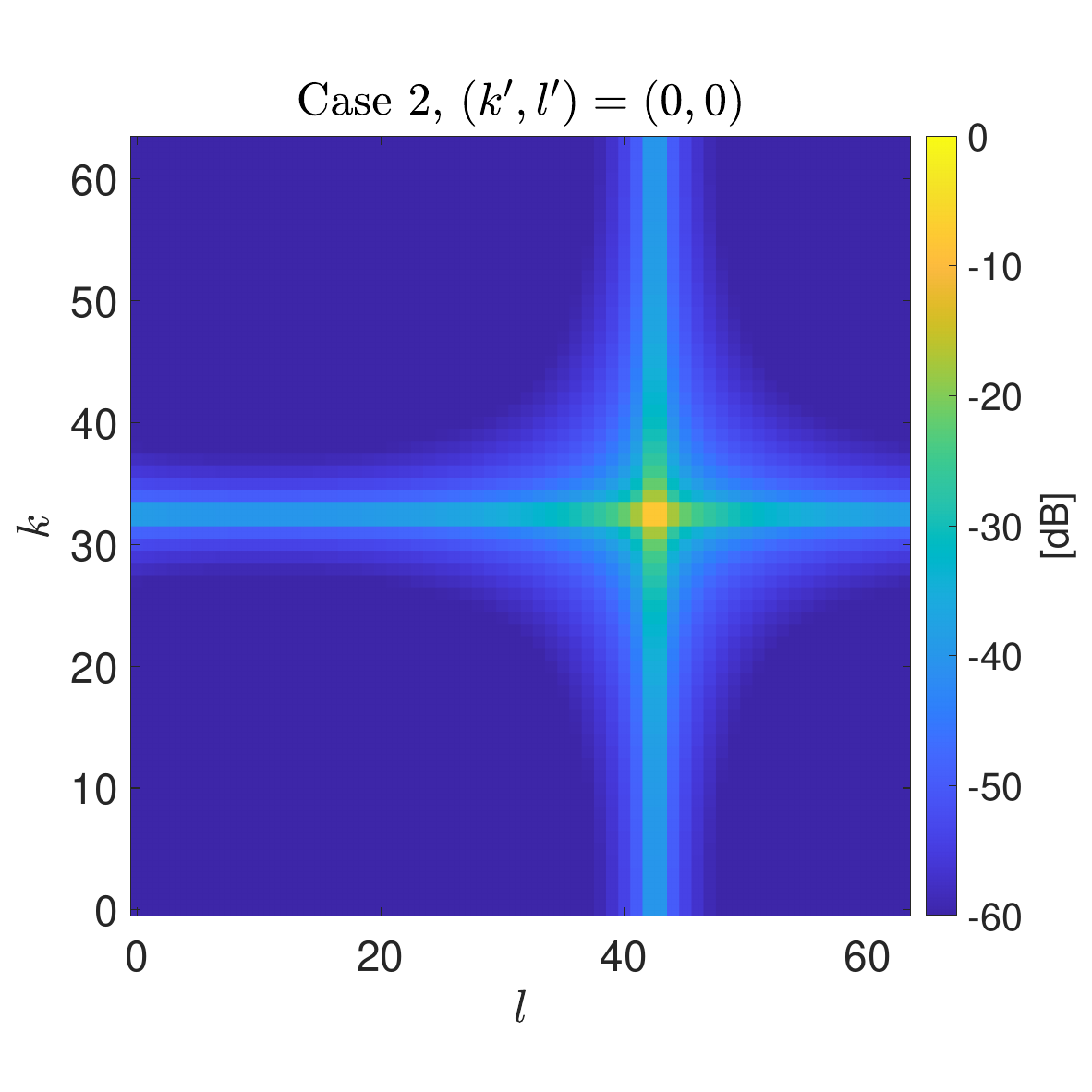}\hspace{0.5cm}
        \includegraphics[width=0.3\textwidth,trim=0cm 0.8cm 0cm 1cm, clip]{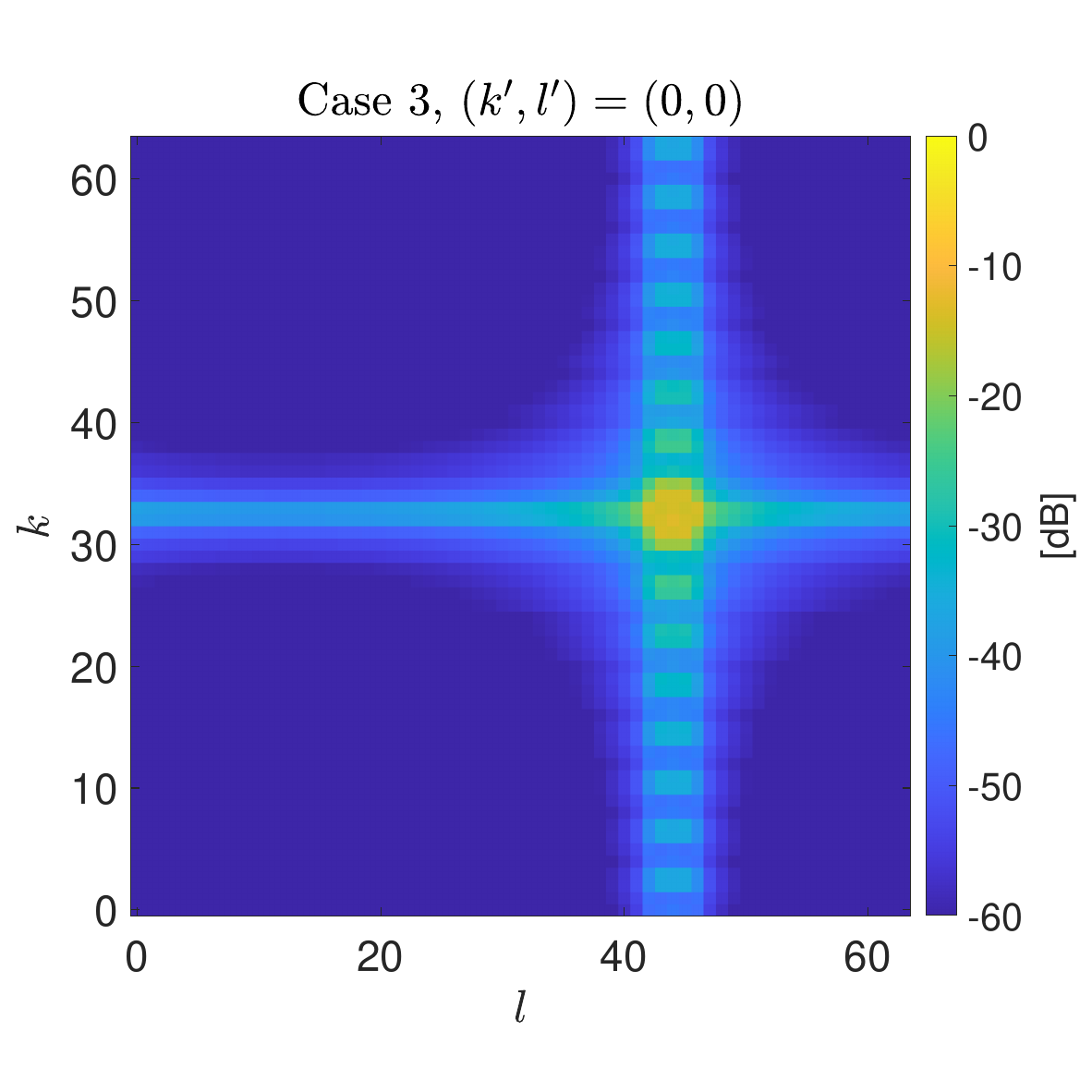}}
    \centerline{
        \includegraphics[width=0.3\textwidth,trim=0cm 0.8cm 0cm 1cm, clip]{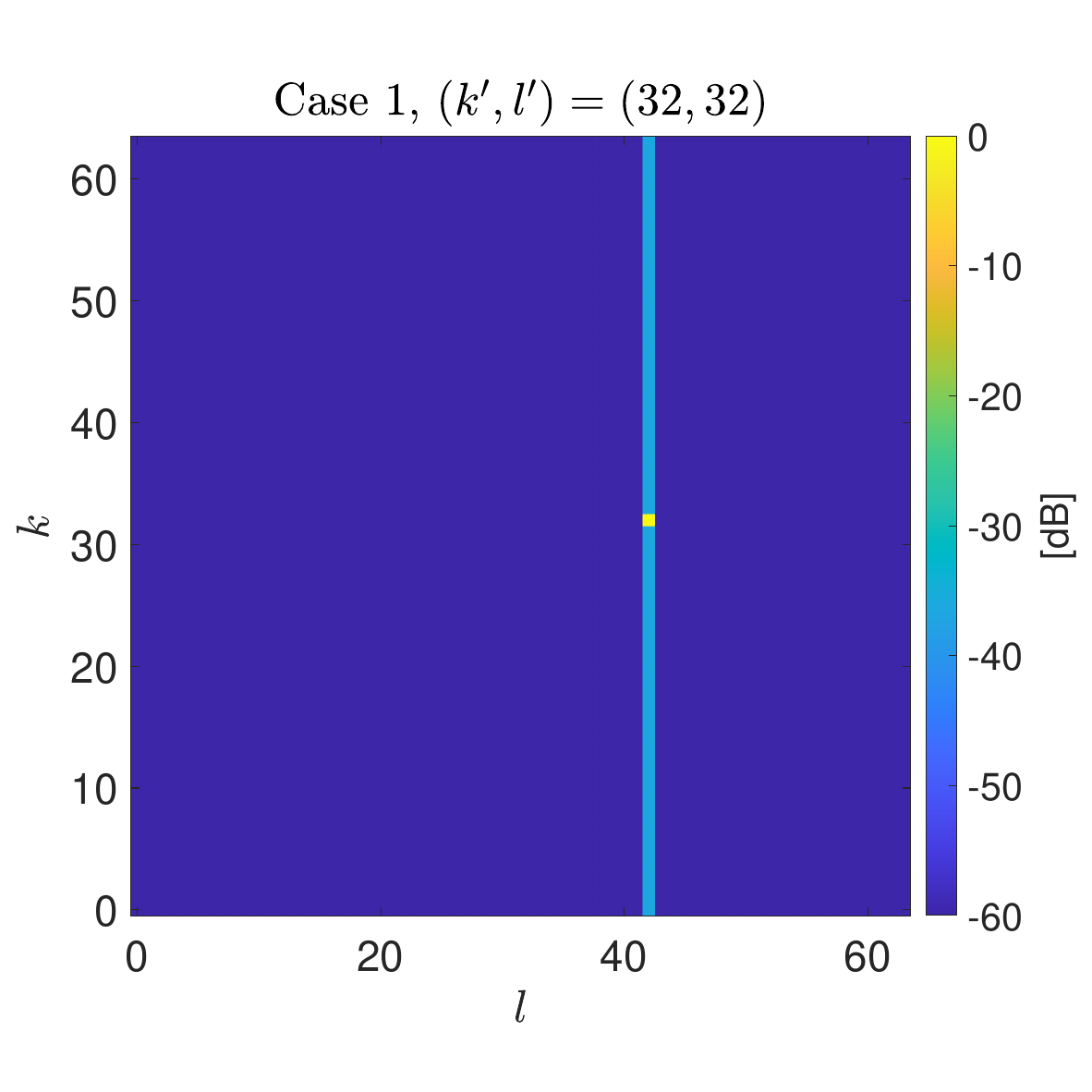}\hspace{0.5cm}
        \includegraphics[width=0.3\textwidth,trim=0cm 0.8cm 0cm 1cm, clip]{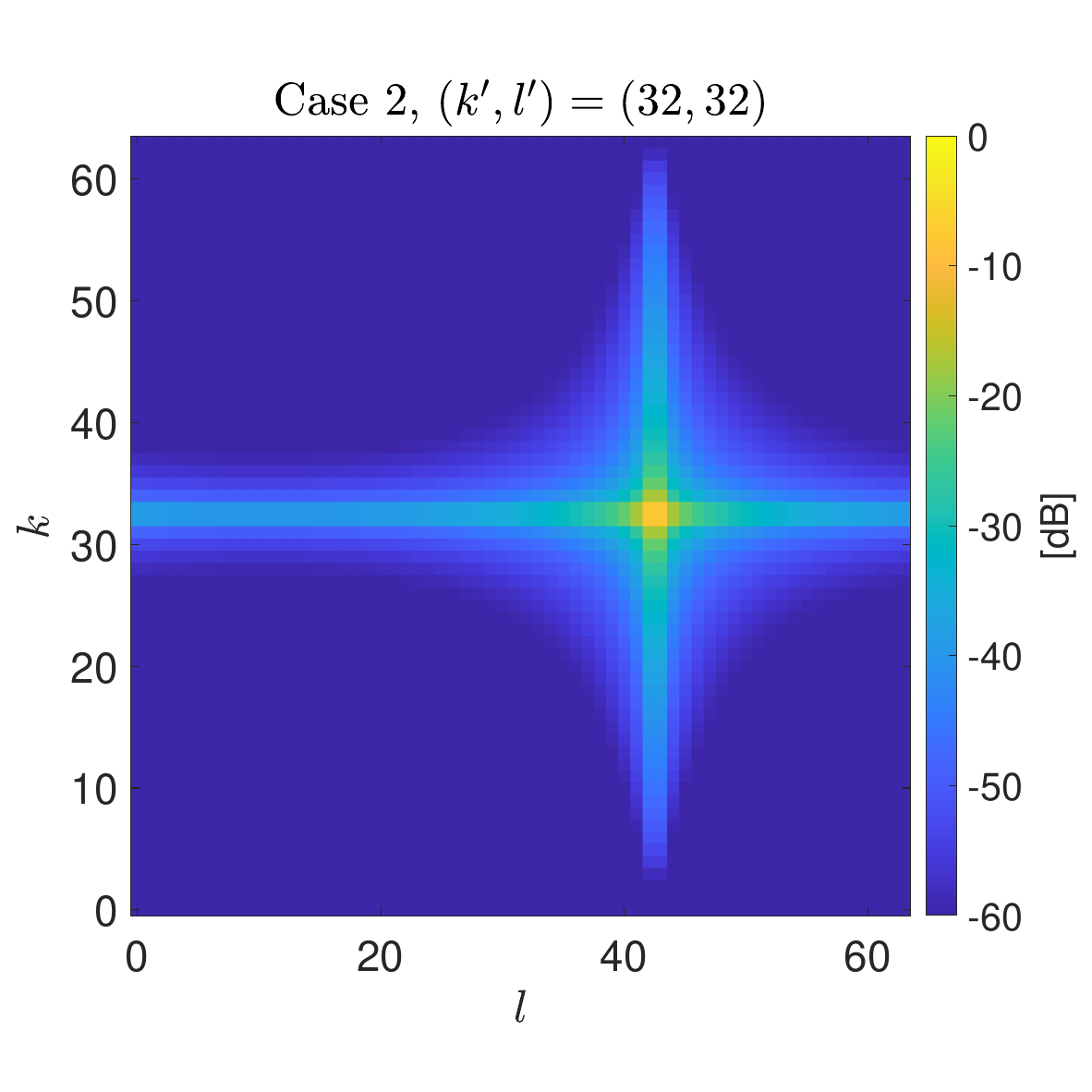}\hspace{0.5cm}
        \includegraphics[width=0.3\textwidth,trim=0cm 0.8cm 0cm 1cm, clip]{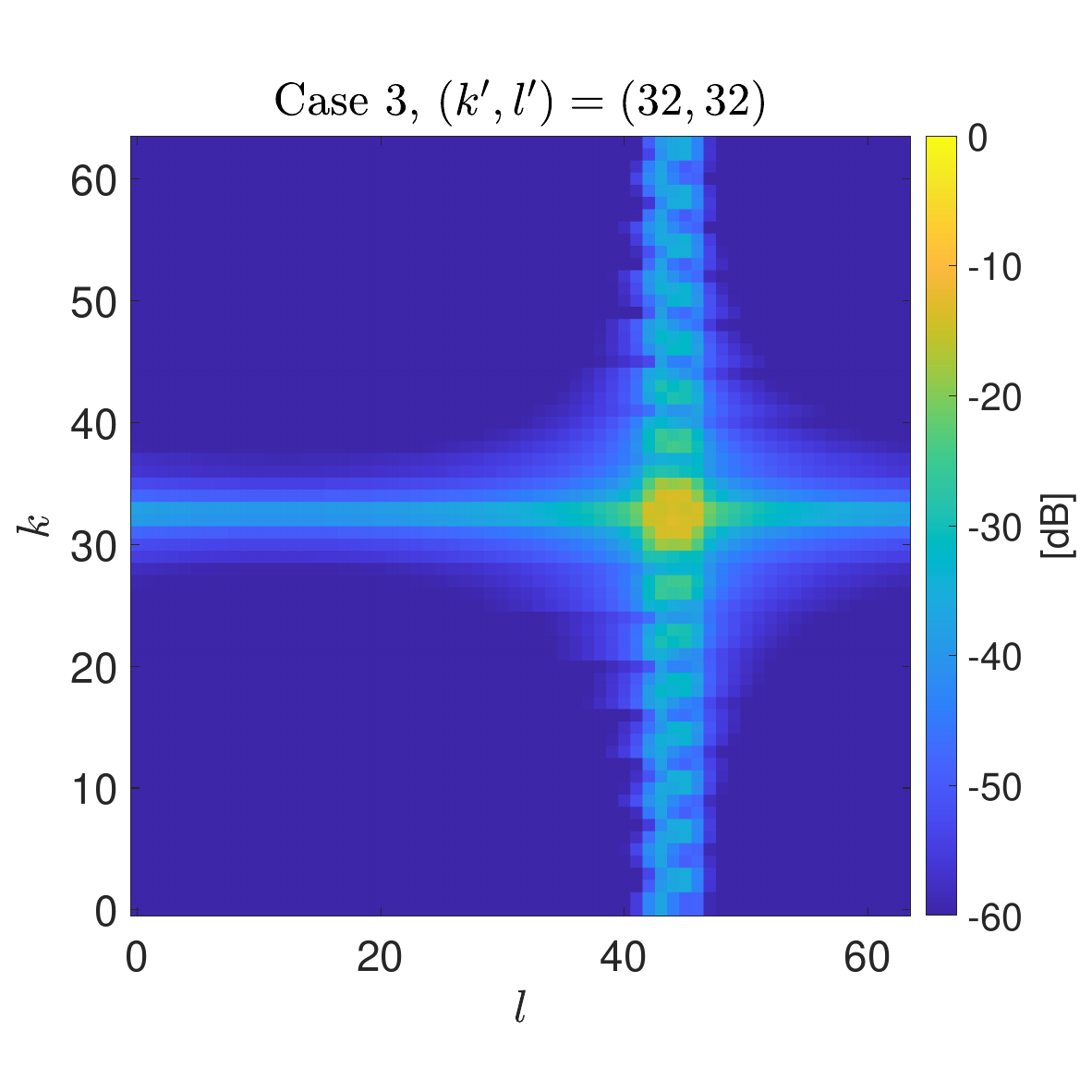}
    }
    \caption{$|\Psi_{k',l'}[k,l]|$ vs $k=0,\ldots,N-1$ and $l=0,\ldots,M-1$, when $B=4$, $N=M=64$, and rectangular shaping filters are employed. Top and bottom rows are for $(k',l')=(0,0)$ and $(32,32)$, respectively. Cases~1 (left), 2 (center), and~3 (right) are defined as in Fig.~\ref{fig_hw}.}
    \label{fig_hw_rect}
\end{figure*}

\begin{example}\label{Example-rect}
    Fig.~\ref{fig_hw_rect} reports $|\Psi_{k',l'}[k,l]|$ versus $k=0,\ldots,N-1$ and $l=0,\ldots,M-1$ under the same parameter settings considered in Example~\ref{Example-ideal}, for the input locations $(k',l')=(0,0)$ and $(32,32)$. Consistently with Remark~\ref{Remark-ideal-sparse} and Example~\ref{Example-ideal}, the three cases exhibit the same qualitative sparse-support behavior observed with ideal shaping filters, with range migration inducing additional spreading along both the delay and Doppler dimensions. With rectangular shaping filters, however, the shape of the point-spread function also depends on the input location because of ICI and ISI. In particular, for $(k',l')=(32,32)$, the Doppler profile is affected by the term $\mathbbm{1}_{\{b=0\}}B/N$ in~\eqref{rx_signal_DD_rect_fast_B}, which modifies the associated Dirichlet kernel due to ISI; this effect is not present for $(k',l')=(0,0)$.
\end{example}

\subsubsection{Vectorized model} \label{SEC:echo-rect}
The samples $\{Y_{\rm DD}[k,l]\}$ can again be arranged into the vector $\bm{y}\in\mathbb{C}^{NM}$. From~\eqref{rx_signal_DD_rect_fast}, we obtain
\begin{equation}\label{vec_model_rect}
    \bm{y}=\alpha\bm{\Psi}\bm{x},
\end{equation}
where $\bm{\Psi}\in\mathbb{C}^{NM\times NM}$ has entries $(\bm{\Psi})_{lN+k+1,l'N+k'+1}= \Psi_{k',l'}\big[k-k',l-l'\big]$. Note that the models in~\eqref{vec_model_ideal_1} and~\eqref{vec_model_rect} share the same structure; however, the matrices $\bm{\Phi}$ and $\bm{\Psi}$ differ.

The input-output relationship in~\eqref{rx_signal_DD_rect_fast}, or equivalently~\eqref{vec_model_rect}, captures the combined effects of range migration, ICI, and ISI. We next introduce a more tractable approximate model that preserves the dominant delay-Doppler spreading effects induced by range migration while simplifying the dependence of the target response on the transmitted-symbol location.

\subsubsection{Approximate model}
We follow the rationale adopted in~\cite{Viterbo-2018} to obtain an approximate input-output model. We first neglect the fractional part of the target range, i.e., we assume $l_{b,{\rm fra}}=0$, for $b=0,\ldots,B-1$. Accordingly, $\mathcal{D}_{M}(\bar l_b/M)$ in~\eqref{rx_signal_DD_rect_fast-2} is nonzero only for $\bar l_b=0$. Note next that the second case of $A_{b,\bar k,\bar l_b}[k,l]$ includes the boundary term $-\mathbbm 1_{\{b=0\}}B/N$ associated with the absence of ISI at the beginning of the OTFS frame that occurs only for $b=0$. Neglecting this term and reindexing the sums,~\eqref{rx_signal_DD_rect_fast-2} is approximated as
\begin{align}
    &Y_{{\rm DD}}[k,l]
    \approx\alpha\sum_{k'=0}^{N-1} \sum_{l'=0}^{M-1}
    \e^{\i 2 \pi \frac{k_{\rm int}+k_{\rm fra}}{NM}l}
    \Theta_{k',l'}[k,l]\notag \\
    &\quad \times \frac{1}{B} \sum_{b=0}^{B-1}
    \e^{-\i 2 \pi \left(\frac{k'-k_{\rm int}-k_{\rm fra}}{B} \right)b}
    \mathcal{D}_{\frac{N}{B}}\!\left(
    \frac{k'-k_{\rm int}-k_{\rm fra}}{N}\right) \notag \\
    &\quad \times
    \mathcal{D}_{M}\!\left(-\frac{l'-l_{b,{\rm int}}}{M}\right)
    X_{\rm DD}\Big[[k-k']_{N},[l-l']_{M}\Big],
    \label{rx_signal_DD_rect_fast-2-pprox}
\end{align}
where
\begin{align}
    \Theta_{k',l'}[k,l]=
    \left\{
    \begin{matrix}
        1, & l'\leq l,\\
        \e^{-\i 2\pi\frac{[k-k']_{N}}{N}}, & l'>l.
    \end{matrix}
    \right.
\end{align}
Based on~\eqref{rx_signal_DD_rect_fast-2-pprox}, we can now write
\begin{equation}\label{vec_model_rect_approx}
    \bm{y}\approx
    \alpha\big(\bm{X}\odot\bm{\Theta}\odot\bm{\Lambda}(v)\big)
    \bm{\phi},
\end{equation}
where $\bm{X}$ is the same quantity as in~\eqref{vec_model_ideal_2}, while $\bm{\phi}$ is evaluated here for $l_{b,{\rm fra}}=0$, for $b=0,\ldots,B-1$; $\bm{\Theta}$ and $\bm{\Lambda}(v)$ are $NM\times NM$ matrices with $(\bm{\Theta})_{lN+k+1,l'N+k'+1}=
\Theta_{k',l'}[k,l]$ and
\[
\big(\bm{\Lambda}(v)\big)_{lN+k+1,l'N+k'+1}
=
\e^{\i 2 \pi \frac{k_{\rm int}+k_{\rm fra}}{NM}l}
=
\e^{\i 2 \pi\frac{vT}{\lambda M}l},
\]
for $k,k'=0,\ldots,N-1$ and $l,l'=0,\ldots,M-1$.

To proceed, notice that $\{\e^{\i 2 \pi\frac{vT}{\lambda M}l}\}_{l=-\infty}^{\infty}$ is a complex exponential whose numerical frequency depends upon the target range-rate. Since $\frac{v_{\max}T}{\lambda}<1$ and the samples $\{\e^{\i 2 \pi\frac{vT}{\lambda M}l}\}_{l=0}^{M-1}$ present an overall phase variation less than $2\pi$, we propose to approximate $\bm{\Lambda}(v)$ by $\bm{\Lambda}(v_{\max}/2)$. Hence, we finally obtain
\begin{equation}\label{vec_model_rect_approx_2}
    \bm{y}\approx\alpha \bm{\Xi}\bm{\phi},
\end{equation}
where $\bm{\Xi}=\bm{X}\odot\bm{\Theta}\odot
\bm{\Lambda}(v_{\max}/2)$.
Thus, relative to~\eqref{vec_model_ideal_2}, the effects of rectangular shaping filters are approximately accounted for through the modified symbol matrix $\bm{\Xi}$.

\section{Estimation of the target parameters}\label{SEC_Detector_Estimator}
Upon accounting for the additive noise, after OTFS demodulation the received signal can be written as
\begin{equation}\label{vec_model}
    \bm{z}=\bm{y}+\bm{\omega}\in\mathbb{C}^{NM},
\end{equation}
where $\bm{\omega}$ is a circularly symmetric complex Gaussian vector with covariance matrix $\sigma_{\omega}^{2}\bm{I}_{NM}$. Based on $\bm{z}$ and the target echo model derived in Sec.~\ref{SEC_IO}, the radar receiver is faced with the problem of estimating the unknown target parameters, namely $d$, $v$, and $\alpha$. The ML estimator amounts here to a normalized matched-filter search over the initial-range/range-rate domain. Since the matched filtering is performed on the observations after OTFS demodulation, we hereafter refer to this estimator as the delay-Doppler-domain matched-filter (DD-MF) estimator. The corresponding estimates are~\cite{KayBook_vol1}
\begin{equation}\label{ML_estimator}
    \big(\hat{d};\,\hat{v}\big)=\argmax_{(d;\,v)\in \mathcal{S}} \frac{\big|\bm{e}(d,v)\herm\bm{z}\big|^2}{\|\bm{e}(d,v)\|^2}, \quad
    \hat{\alpha}=\frac{\bm{e}\big(\hat{d},\hat{v}\big)\herm\bm{z}}{\big\|\bm{e}\big(\hat{d},\hat{v}\big)\big\|^2},
\end{equation}
where $ \mathcal{S}=[0,r_{\max}]\times[0,v_{\max}]$, and
\begin{equation}\label{e_vector}
    \!\bm{e}(d,v)=
    \begin{cases}
        \bm{\Phi}(d,v)\bm{x}, & \text{for ideal shaping filters}\\
        \bm{\Psi}(d,v)\bm{x}, & \text{for rectangular shaping filters}
    \end{cases}
\end{equation}
is the echo template for the hypothesized initial-range $d$ and range-rate $v$. In practice, the continuous set $\mathcal{S}$ is replaced by a uniform rectangular grid $\mathcal{G}$ containing $G_{\rm r}=\lfloor r_{\max}/\Sigma_{\rm r}\rfloor$ and $G_{\rm rr}=\lfloor v_{\max}/\Sigma_{\rm rr}\rfloor$ points along the initial-range and range-rate domains, respectively, where $\Sigma_{\rm r}\leq R_{\rm r}$ and $\Sigma_{\rm rr}\leq R_{\rm rr}$ are the corresponding point spacings.

The implementation of~\eqref{ML_estimator} entails a complexity of
\begin{equation}\label{eq:complexity_ML}
    \mathcal{O}\big(G_{\rm r}G_{\rm rr}(MN)^2\big),
\end{equation}
which becomes computationally demanding when a fine search grid is required. To reduce complexity, we propose in Sec.~\ref{SEC_Two-step-estimator} a suboptimal approximation that exploits the sparse structure of the target echo. This procedure is further extended in Sec.~\ref{SEC_multitarget} to handle multiple targets.

\subsection{Suboptimal approximation of the DD-MF estimator}\label{SEC_Two-step-estimator}
We develop here a two-step approximation of the DD-MF estimator. In the first step, signal sparsity is exploited through BOMP to obtain a coarse estimate of the target response and, as a by-product, coarse estimates of the initial-range and range-rate, denoted as $\tilde{d}$ and $\tilde{v}$, respectively, as shown in Sec.~\ref{SEC_BOMP}. In the second step, the DD-MF search in~\eqref{ML_estimator} is restricted to a neighborhood of $\big(\tilde{d};\,\tilde{v}\big)$, thereby refining the initial-range and range-rate estimates and computing the amplitude estimate. In particular, we consider a uniform rectangular grid $\tilde{\mathcal{G}}$ centered at $\big(\tilde{d};\,\tilde{v}\big)$, with $\tilde G_{\rm r}$ and $\tilde G_{\rm rr}$ points and spacings $\Sigma_{\rm r}$ and $\Sigma_{\rm rr}$ along the initial-range and range-rate dimensions, respectively. We refer to this procedure as the DD-BOMP+MF estimator.

\subsubsection{Coarse estimation of the initial-range and range-rate}\label{SEC_BOMP}
Consider first ideal shaping filters. From~\eqref{vec_model_ideal_2}, the received signal is written as $\bm{z}= \bm{X}\bm{f}+\bm{\omega}$, where $\bm{f}=\alpha\bm{\phi}$ is the unknown target response vector. To proceed, $\bm{f}$ is partitioned into $M$ subvectors (blocks) of length $N$, namely, $\bm{f}=(\bm{f}_0;\cdots;\bm{f}_{M-1})$, where $\bm{f}_m=\big(\alpha\Phi[0,m];\ldots;\alpha\Phi[N-1,m]\big)$. Each block contains all Doppler coefficients associated with one delay index. As discussed in Remark~\ref{Remark-ideal-sparse} and Examples~\ref{Example-ideal} and~\ref{Example-rect}, the target response is sparse along the delay dimension; hence, $\bm f$ exhibits a block-sparse structure. Next, the dimension of $\bm{z}$ is reduced by means of an $NK\times NM$ selection matrix $\bm{S}$ whose rows are distinct rows of $\bm{I}_{NM}$. The parameter $K\in\{1,\ldots,M\}$ controls both the implementation complexity and the maximum number of delay blocks that can be recovered by BOMP, and it should therefore be selected according to the expected dominant delay support of the target response. The compressed measurement $\bm{z}_{s}=\bm{S}\bm{z}\in\mathbb{C}^{NK}$ is expressed as
\begin{equation}
    \bm{z}_{s}=\bm{X}_{s}\bm{f}+\bm{\omega}_{s}=\sum_{m=0}^{M-1}\bm{X}_{s,m}\bm{f}_{m}+\bm{\omega}_{s}, \label{vec_model_ideal_4}
\end{equation}
where $\bm{\omega}_{s}=\bm{S}\bm{\omega}$, $\bm{X}_{s}=(\bm{X}_{s,0}\,\cdots\,\bm{X}_{s,M-1})=\bm{S}\bm{X}$, and $\bm{X}_{s,0},\ldots,\bm{X}_{s,M-1}\in\mathbb{C}^{NK\times N}$.

Based on~\eqref{vec_model_ideal_4}, an estimate of the target response vector, say $\tilde{\bm{f}}$, can be computed by resorting to the BOMP algorithm~\cite{eldar2009block}.\footnote{\label{footnote-BOMP}Although $\bm{\phi}$ depends on $B$ through the blockwise echo model, BOMP does not exploit this dependence: the subvectors $\{\bm{f}_m\}$ are treated as unknown and estimated directly from the compressed measurements. Thus, $B$ does not explicitly enter the BOMP dictionary, block selection, or least-squares steps.} Let $\bm{z}^{(1)}_{s}=\bm{z}_{s}$ and $\mathcal{I}^{(1)}=\{0,\ldots,M-1\}$; then, at the $\ell$-th iteration, for $\ell\geq 1$, the index of the block in the dictionary matrix $\bm{X}_{s}$ best matched to $\bm{z}^{(\ell)}_{s}$ is first identified as follows
\begin{equation}
    i^{(\ell)}=\argmax_{i\in\mathcal{I}^{(\ell)}} \Big\|\bm{X}_{s,i}\herm \bm{z}^{(\ell)}_{s}\Big\|^2;
\end{equation}
then, an estimate of the subvectors $\bm{f}_{i^{(1)}},\ldots,\bm{f}_{i^{(\ell)}}$ corresponding to the currently selected blocks of $\bm{X}_{s}$ is computed by solving a least-squares problem, namely,
\begin{equation}
    \big(\tilde{\bm{f}}_{i^{(1)}}\cdots\tilde{\bm{f}}_{i^{(\ell)}}\big)=
    \argmin_{(\bm{f}_{i^{(1)}}\cdots\bm{f}_{i^{(\ell)}})\in \mathbb{C}^{N\times \ell}} \Big\|
    \bm{z}_{s}-\sum_{q=1}^{\ell}\bm{X}_{s,i^{(q)}}\bm{f}_{i^{(q)}}
    \Big\|^2;
\end{equation}
finally, the residual vector and the search set are updated as $
\bm{z}^{(\ell+1)}_{s}=\bm{z}_{s}-\sum_{q=1}^{\ell}\bm{X}_{s,i^{(q)}}\tilde{\bm{f}}_{i^{(q)}}$ and $\mathcal{I}^{(\ell+1)}=\mathcal{I}^{(\ell)}\setminus \{i^{(\ell)}\}$, respectively, and the procedure is stopped when $\ell=K$ or $\|\bm{z}^{(\ell+1)}_{s}\|^2<\epsilon NK\sigma_{\omega}^{2}$, with $\epsilon$ being a positive threshold. If the procedure ends in $\tilde \ell$ iterations, an estimate $\tilde{\bm{f}}$ of the target response vector is obtained from the indices $i^{(1)},\ldots,i^{(\tilde \ell)}$ and the subvectors $\tilde{\bm{f}}_{i^{(1)}},\ldots,\tilde{\bm{f}}_{i^{(\tilde \ell)}}$. At this point, the entries of $\tilde{\bm{f}}$ are arranged into the sequence $\tilde{F}[k,l]=(\tilde{\bm{f}})_{lN+k+1}$, for $k=0,\ldots,N-1$ and $l=0,\ldots,M-1$, which provides an estimate of $\alpha \Phi[k,l]$; finally, coarse estimates of the target initial-range and range-rate are obtained from the location of the maximum element of $|\tilde{F}[k,l]|$.

When a rectangular shaping filter is employed, from~\eqref{vec_model_rect_approx_2}, the received signal can be approximately expanded as $\bm{z}\approx\bm{\Xi}\bm{f}+\bm{\omega}$; accordingly, the BOMP-based coarse-estimation procedure can still be used after replacing $\bm{X}$ by the modified symbol matrix $\bm{\Xi}$. Note that the use of the approximate echo model in~\eqref{vec_model_rect_approx_2} is confined here to the coarse-estimation step, while the subsequent matched-filter step evaluates the statistic~\eqref{ML_estimator} using the full echo model in~\eqref{vec_model_rect}.

\subsubsection{Implementation complexity}
We now discuss the implementation complexity of the DD-BOMP+MF estimator. In the first step, BOMP is applied to the compressed model in~\eqref{vec_model_ideal_4}. At the $\ell$-th iteration, the block selection requires the computation of $\bm X_{s,i}^{\rm H}\bm z_s^{(\ell)}$ for all remaining blocks; since $\bm X_{s,i}\in\mathbb C^{NK\times N}$, this operation has complexity $\mathcal O((M-\ell+1)KN^2)$. The following least-squares update involves the matrix obtained by concatenating the $\ell$ selected blocks; a standard dense least-squares implementation therefore requires $\mathcal O(K\ell^2N^3+\ell^3N^3)$ operations. The residual updates have lower complexity and do not affect the dominant scaling. Summing over at most $K$ iterations yields a complexity $\mathcal O(MK^2N^2+K^4N^3)$ for the first step. In the second step, the metric in~\eqref{ML_estimator} is evaluated over the local grid $\tilde{\mathcal G}$, with complexity $\mathcal O\big(\tilde G_{\rm r}\tilde G_{\rm rr}(MN)^2\big)$. Therefore, the overall complexity of the DD-BOMP+MF estimator is
\begin{equation}
    \mathcal O\!\left(MK^2N^2+K^4N^3+\tilde G_{\rm r}\tilde G_{\rm rr}(MN)^2\right).
    \label{eq:complexity_two_step}
\end{equation}
The first two terms in~\eqref{eq:complexity_two_step} are controlled by $K$, which bounds the number of BOMP iterations, as discussed in Sec.~\ref{SEC_BOMP}. Importantly, the local grid $\tilde{\mathcal G}$ uses the same spacings $\Sigma_{\rm r}$ and $\Sigma_{\rm rr}$ as the exhaustive grid $\mathcal G$; hence, the reduction in matched-filter complexity is obtained by restricting the search region. When $K\ll M$ and $\tilde G_{\rm r}\tilde G_{\rm rr}\ll G_{\rm r}G_{\rm rr}$, the resulting complexity is substantially lower than that in~\eqref{eq:complexity_ML}.

\subsection{Extension to multiple targets}\label{SEC_multitarget}
Multiple targets can be handled through the CLEAN algorithm~\cite{Colone-2016,Bose-2011,Misiurewicz-2012,Bosse-2018}, which repeatedly estimates and removes the dominant target echo from the received signal. Let $\bm r^{(1)}=\bm z$ denote the initial residual. At the $p$-th iteration, the DD-BOMP+MF estimator is applied to $\bm r^{(p)}$ to extract the dominant remaining target and estimate its initial-range $\hat d^{(p)}$, range-rate $\hat v^{(p)}$, and amplitude $\hat \alpha^{(p)}$. The corresponding echo is then reconstructed as $\hat{\bm y}^{(p)}=\hat{\alpha}^{(p)}\bm e\big(\hat d^{(p)},\hat v^{(p)}\big)$, and the residual is updated as $\bm r^{(p+1)}=\bm r^{(p)}-\hat{\bm y}^{(p)}$ to mitigate the interference caused by previously extracted targets in the next iteration $p+1$. In this work, the number of targets is assumed known, say $P$, and the algorithm is therefore run for $P$ iterations, yielding the DD-BOMP+MF+CLEAN estimator.

Each iteration requires one execution of the DD-BOMP+MF estimator, with complexity given in~\eqref{eq:complexity_two_step}, together with a residual update of complexity $\mathcal O(MN)$. Hence, for a fixed value of $K$, the complexity of the DD-BOMP+MF+CLEAN estimator scales linearly with the number of extracted targets. If $P$ targets are processed, the resulting complexity is
\begin{equation}\mathcal O\left(P \big(MK^2N^2+K^4N^3+\tilde G_{\rm r}\tilde G_{\rm rr}(MN)^2\big)\right).
\end{equation}

\section{Numerical analysis}\label{SEC_Analysis}

To assess the system performance, we consider the parameters in Table~\ref{tab_1}. For these values, the maximum variation of the target range during one OTFS frame is $\bar{v}_{\rm pk}NT\approx 136$~m, which corresponds to about $3.5$ times the range resolution $R_{\rm r}$. The initial-range $\bar{d}$ and range-rate $\bar{v}$ are randomly generated within the intervals $[\bar{r}_{\min},\,\bar{r}_{\max}]$ and $[\bar{v}_{\min},\,\bar{v}_{\max}]$, respectively, while a Swerling~I fluctuation model is adopted for the amplitude $\alpha$. The information symbols $\{X_{\rm DD}[k,l]\}$ are independently drawn from a unit-energy $4$-QAM constellation.

The entries of the vector $\bm z$ observed in the delay-Doppler domain are generated according to~\eqref{baseband_echo}, \eqref{rx_signal_TF}, and \eqref{rx_signal_DD}. The radar receiver, instead, is designed according to the model developed in Secs.~\ref{SEC_IO_ideal} and~\ref{SEC_IO_rect} for ideal and rectangular shaping filters, respectively. Unless otherwise stated, we set $B=16$. In the implementation of the DD-BOMP+MF estimator, we use $K=4P$, $\Sigma_{\rm r}=R_{\rm r}/100$, $\Sigma_{\rm rr}=R_{\rm rr}/100$, $\tilde{G}_{\rm r}=401$, $\tilde{G}_{\rm rr}=401$, and $\epsilon=1$. The choice $K=4P$ provides a block-sparsity budget commensurate with the number of targets and the maximum range excursion of about $3.5R_{\rm r}$. With the adopted grid spacings, $G_{\rm r}\simeq4.86\times10^{4}$ and $G_{\rm rr}\simeq1.14\times10^{4}$, so that the exhaustive DD-MF grid contains $G_{\rm r}G_{\rm rr}\simeq5.53\times10^{8}$ points. In contrast, the local refinement grid contains $\tilde G_{\rm r}\tilde G_{\rm rr}=401^2\simeq1.61\times10^{5}$ points. Thus, BOMP localization reduces the number of matched-filter evaluations by a factor of approximately $3.4\times10^{3}$.

Finally, we define the target signal-to-noise ratio (SNR) as $\mathrm{SNR}=\mathrm{E}[|\alpha|^{2}]NM/\sigma_{\omega}^{2}$. For the true target parameters $(d,v,\alpha)$ and the corresponding estimates $(\hat d,\hat v,\hat\alpha)$, we also define the root mean square errors (RMSEs) of the initial-range and range-rate estimates and the normalized root mean square error (NRMSE) of the amplitude estimate as $\mathrm{RMSE}_{\rm r}=(\mathrm{E}[|d-\hat d|^2])^{1/2}$, $\mathrm{RMSE}_{\rm rr}=(\mathrm{E}[|v-\hat v|^2])^{1/2}$, and $\mathrm{NRMSE}_{\alpha}=(\mathrm{E}[|\alpha-\hat\alpha|^2]/\mathrm{E}[|\alpha|^2])^{1/2}$, respectively. We next consider the single- and multi-target cases.
\begin{table}[t]
    \caption{System parameters}
    \centering
    \begin{tabular}{lll}
        \toprule
        $c/\lambda$&Carrier frequency & 4 GHz\\
        $\Delta$&Subcarrier spacing & 15 kHz\\
        $M$&Number of subcarriers& 512\\
        $N$&Number of symbol intervals& 128\\
        $R_{\rm r}$&Range resolution& $\approx$ 39 m\\
        $R_{\rm rr}$&Range-rate resolution& $\approx$ 8.8 m/s\\
        $[\bar{r}_{\min},\,\bar{r}_{\max}]$&Inspected initial-range interval& [500,\,519] km\\
        $[\bar{v}_{\min},\,\bar{v}_{\max}]$&Inspected range-rate interval& [15,\,16] km/s\\
        \bottomrule
    \end{tabular}\label{tab_1}
\end{table}

\begin{figure*}[!tp]
    \centerline{
        \includegraphics[width=0.33\textwidth,trim=1cm 0cm 2cm 0.4cm, clip]{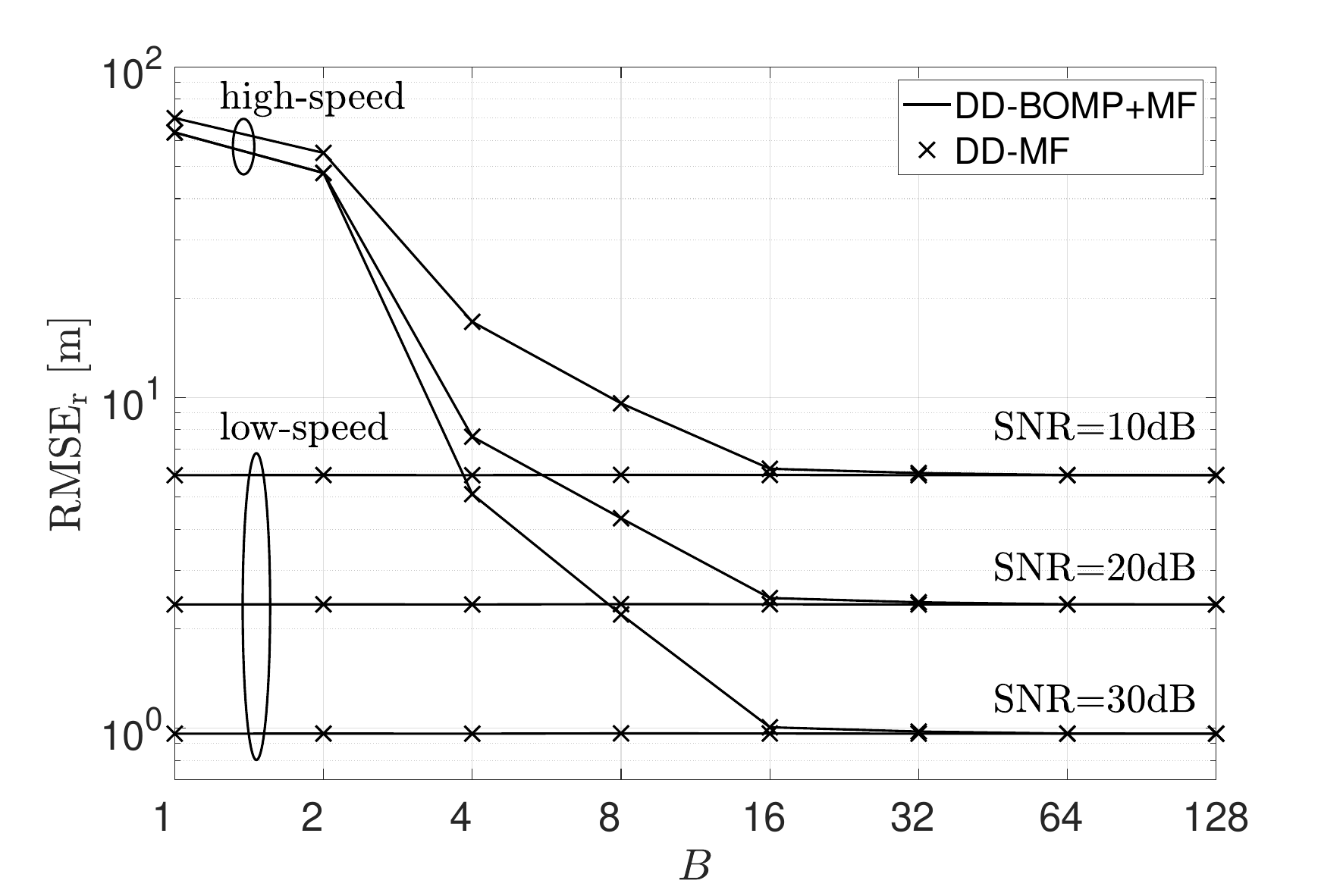}
        \includegraphics[width=0.33\textwidth,trim=0.5cm 0cm 2cm 0.4cm, clip]{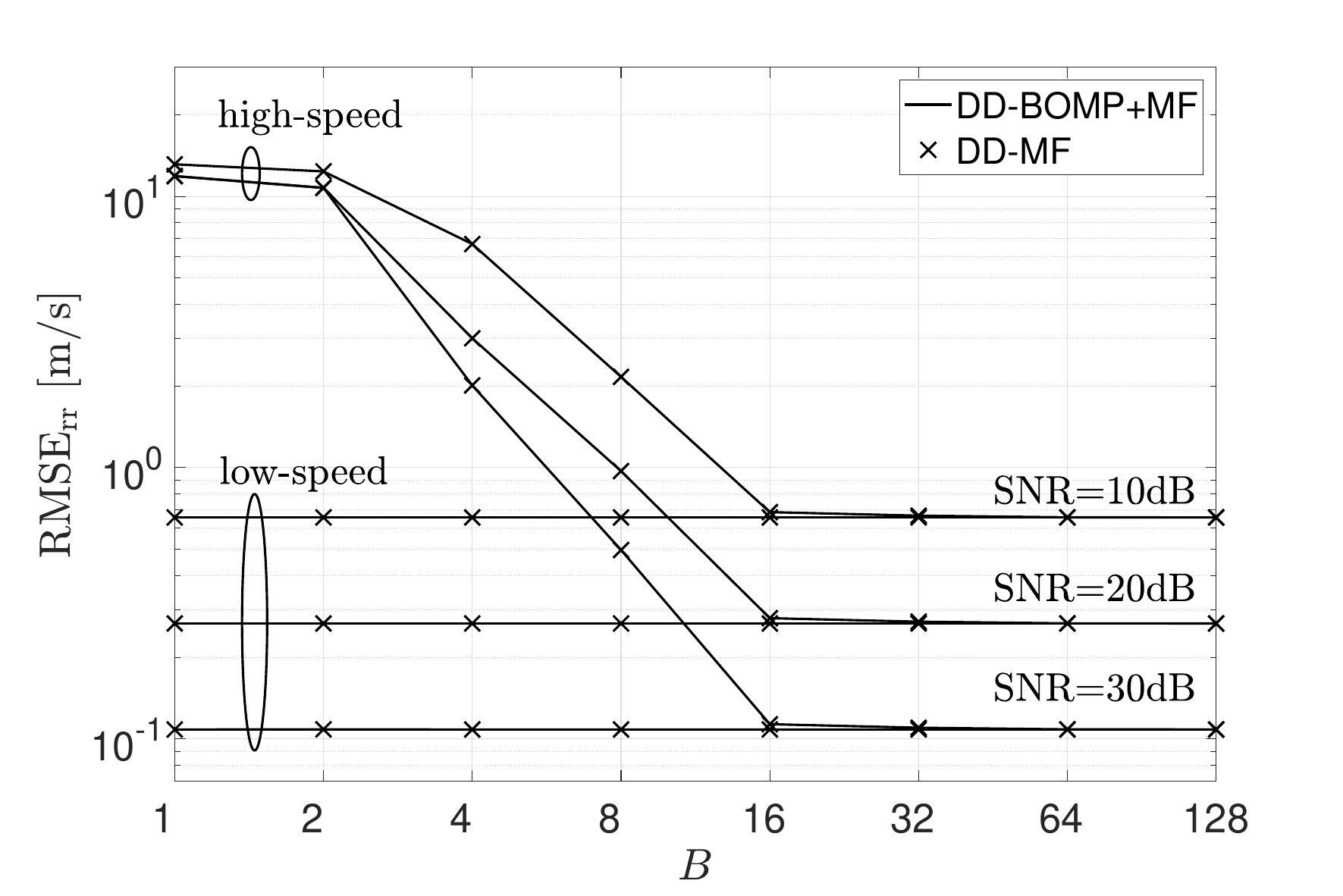}
        \includegraphics[width=0.33\textwidth,trim=0.5cm 0cm 2cm 0.4cm, clip]{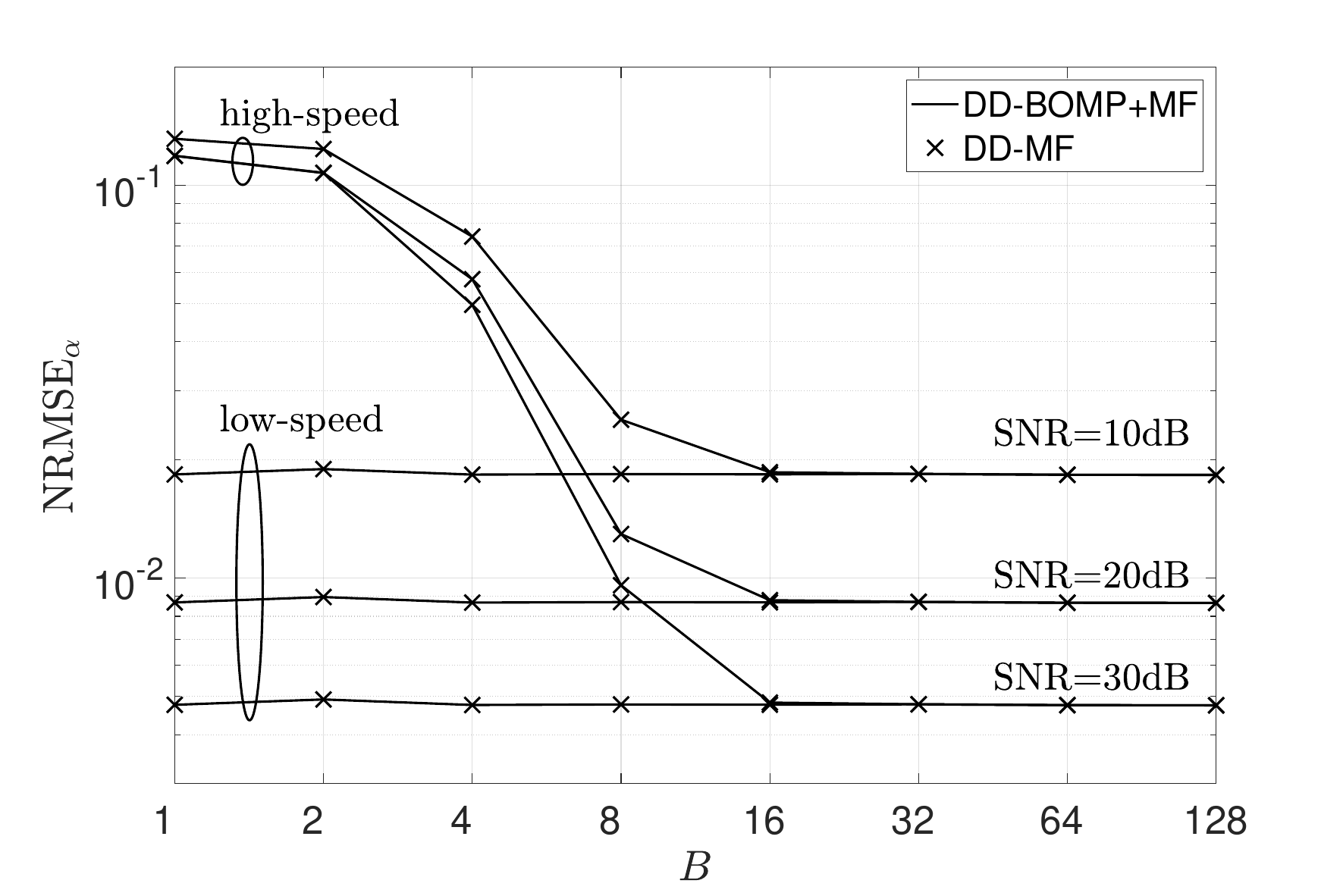}
    }
    \caption{$\mathrm{RMSE}_{\rm r}$ (left), $\mathrm{RMSE}_{\rm rr}$ (center), and $\mathrm{NRMSE}_{\alpha}$ (right) versus $B=1,2,4,8,16,32,64,128$ for $\mathrm{SNR}=10,20,30$~dB, when a high- or low-speed target is present and ideal shaping filters are employed. Both the DD-MF and DD-BOMP+MF estimators are considered.}
    \label{fig_single_1}
\end{figure*}

\begin{figure*}[!tp]
    \centerline{
        \includegraphics[width=0.33\textwidth,trim=1cm 0cm 2cm 0.4cm, clip]{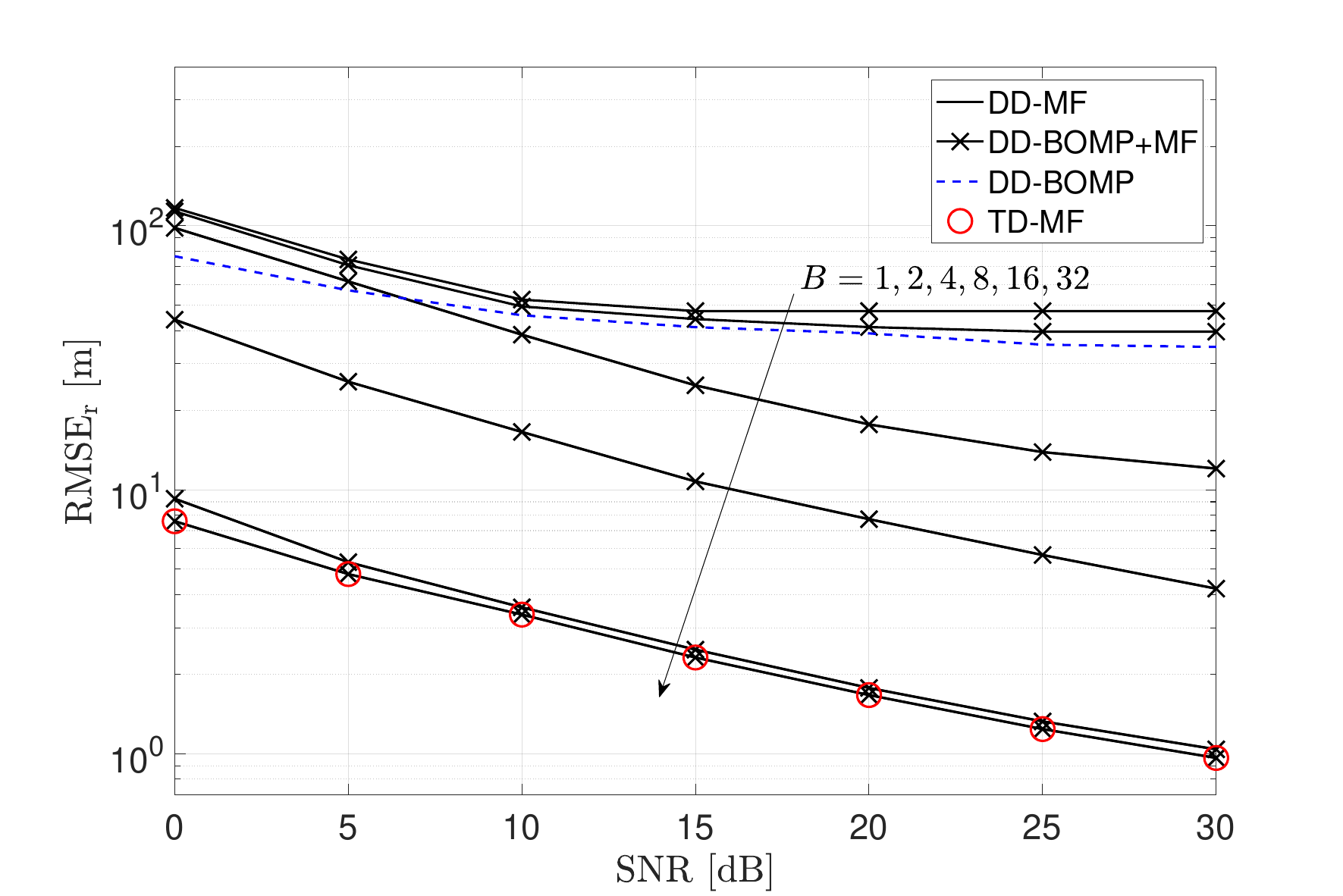}
        \includegraphics[width=0.33\textwidth,trim=0.5cm 0cm 2cm 0.4cm, clip]{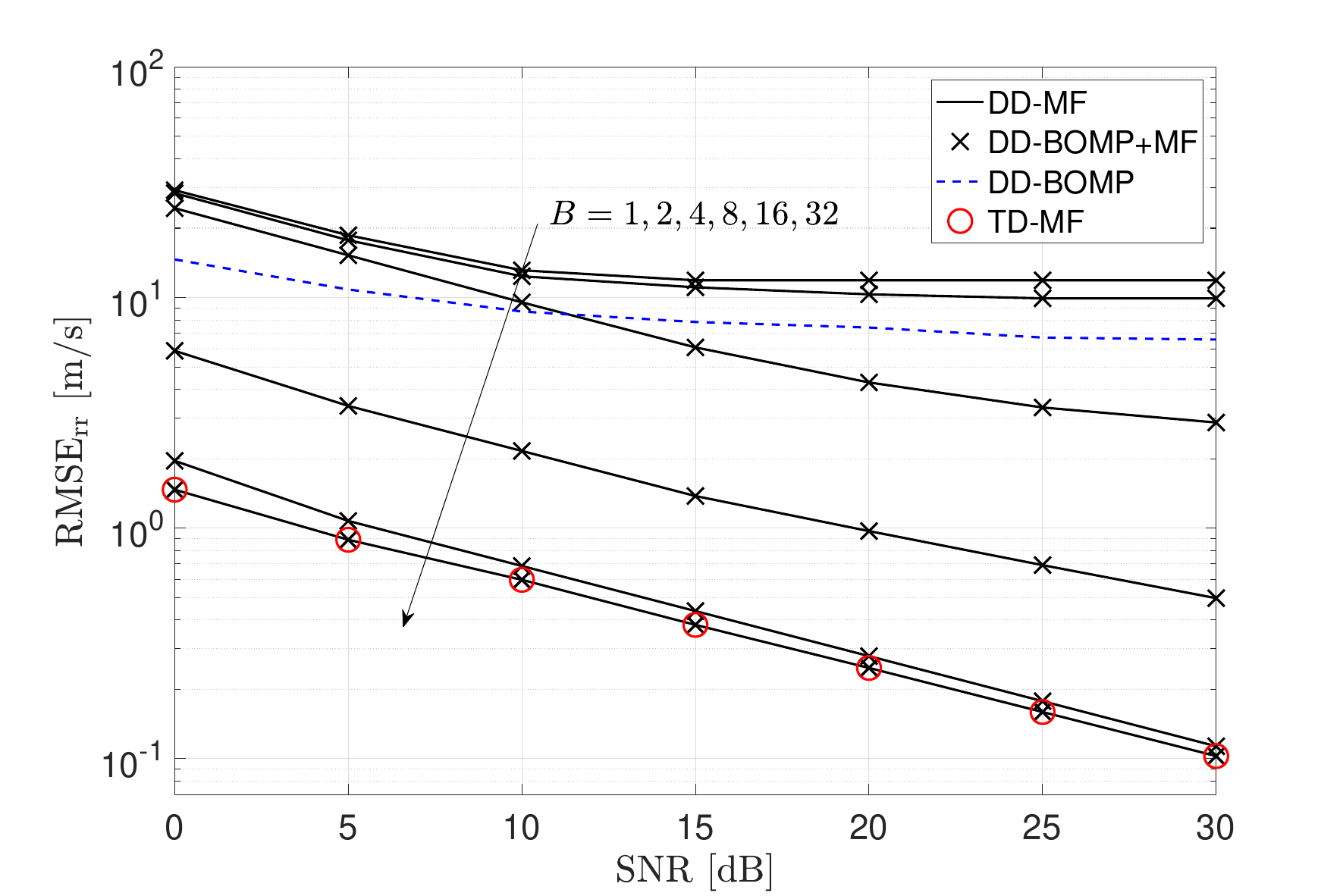}
        \includegraphics[width=0.33\textwidth,trim=0.5cm 0cm 2cm 0.4cm, clip]{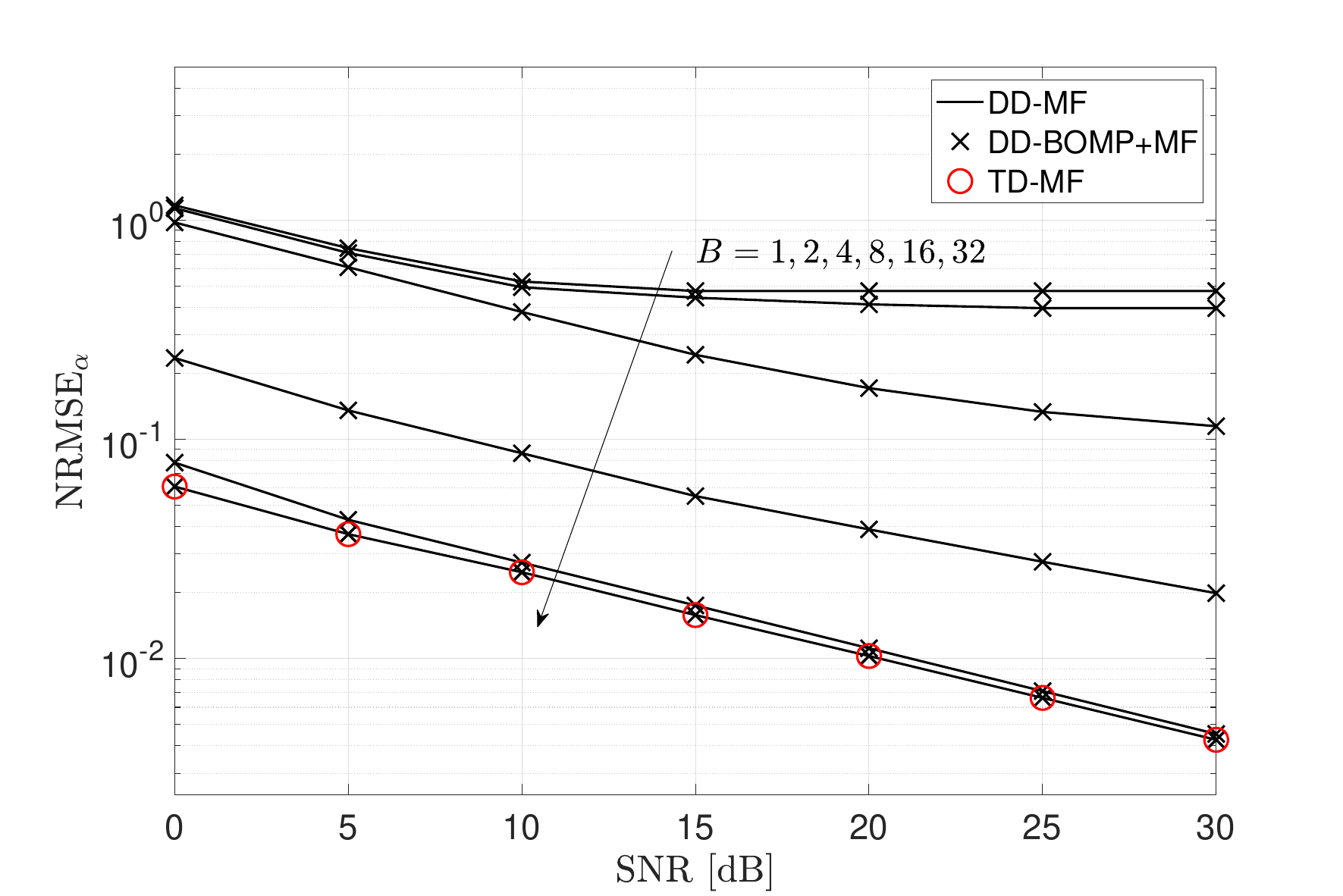}
    }
    \caption{$\mathrm{RMSE}_{\rm r}$ (left), $\mathrm{RMSE}_{\rm rr}$ (center), and $\mathrm{NRMSE}_{\alpha}$ (right) versus $\mathrm{SNR}$ when a high-speed target is present and rectangular shaping filters are employed. The DD-MF and DD-BOMP+MF estimators are implemented for $B=1,2,4,8,16,32$, while the coarse DD-BOMP initial-range and range-rate estimates and TD-MF are included for comparison.} \label{fig_single_2}
\end{figure*}

\begin{figure*}[!tp]
    \centerline{
        \includegraphics[width=0.33\textwidth,trim=1cm 0cm 2cm 0.4cm, clip]{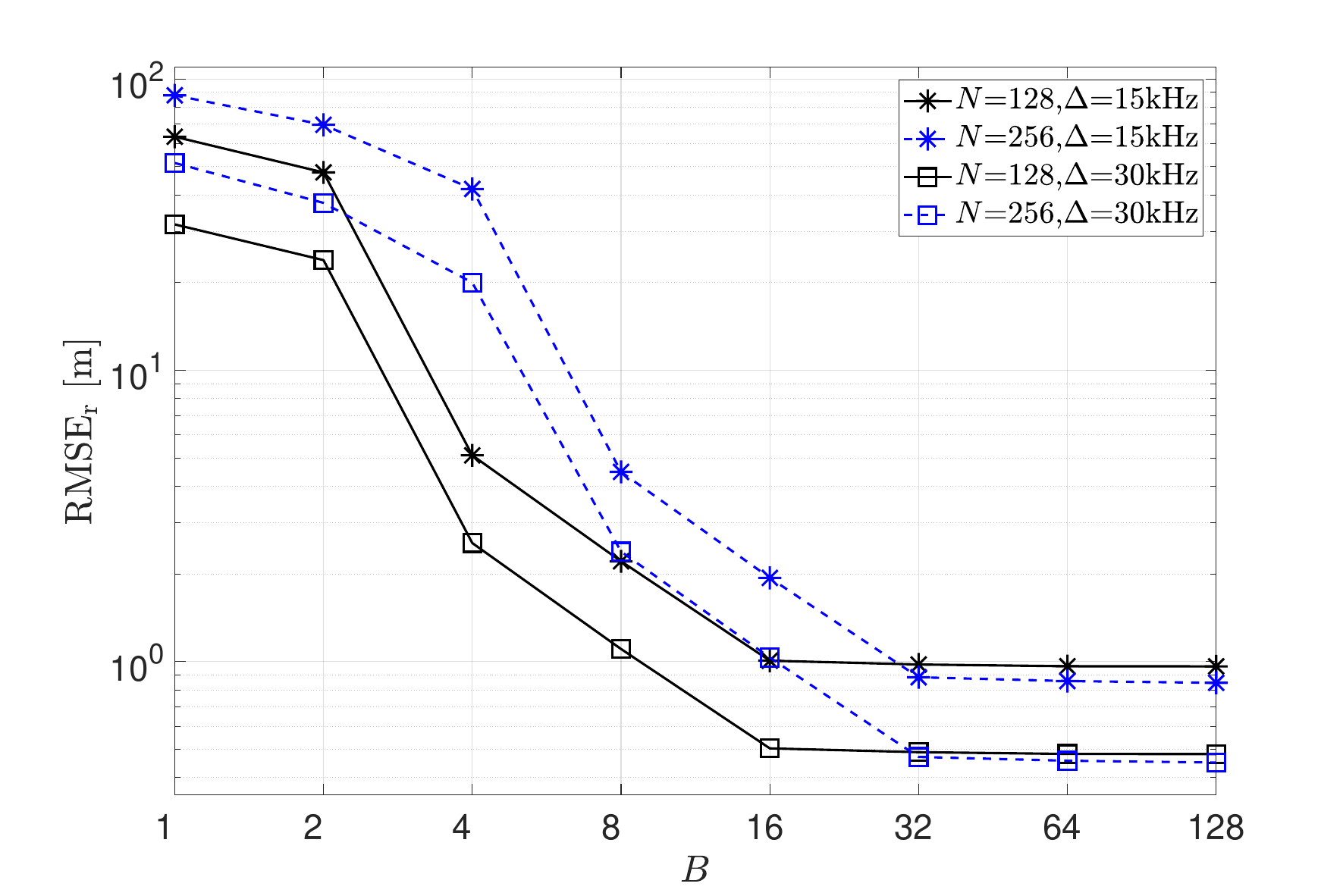}
        \includegraphics[width=0.33\textwidth,trim=0.5cm 0cm 2cm 0.4cm, clip]{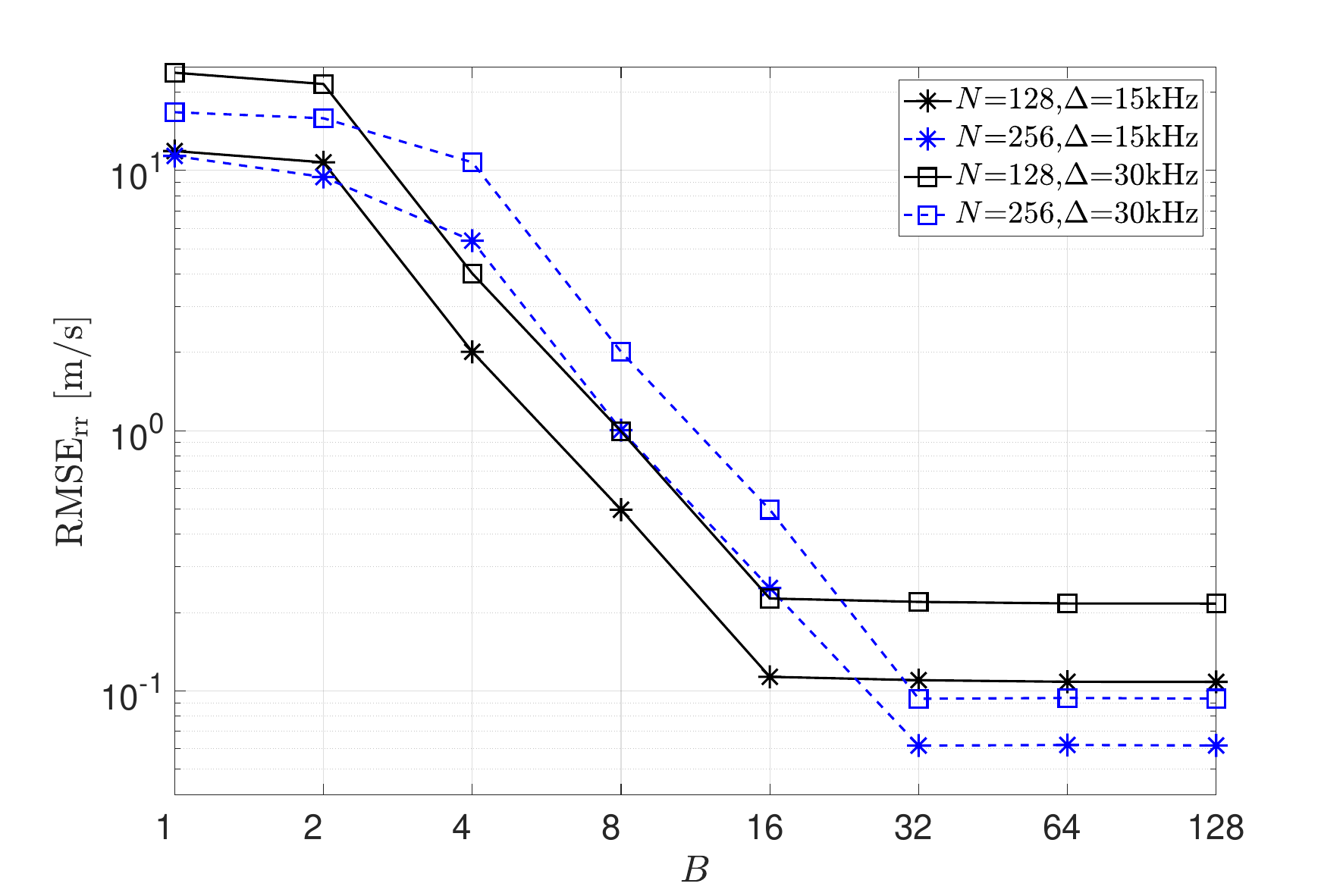}
        \includegraphics[width=0.33\textwidth,trim=0.5cm 0cm 2cm 0.4cm, clip]{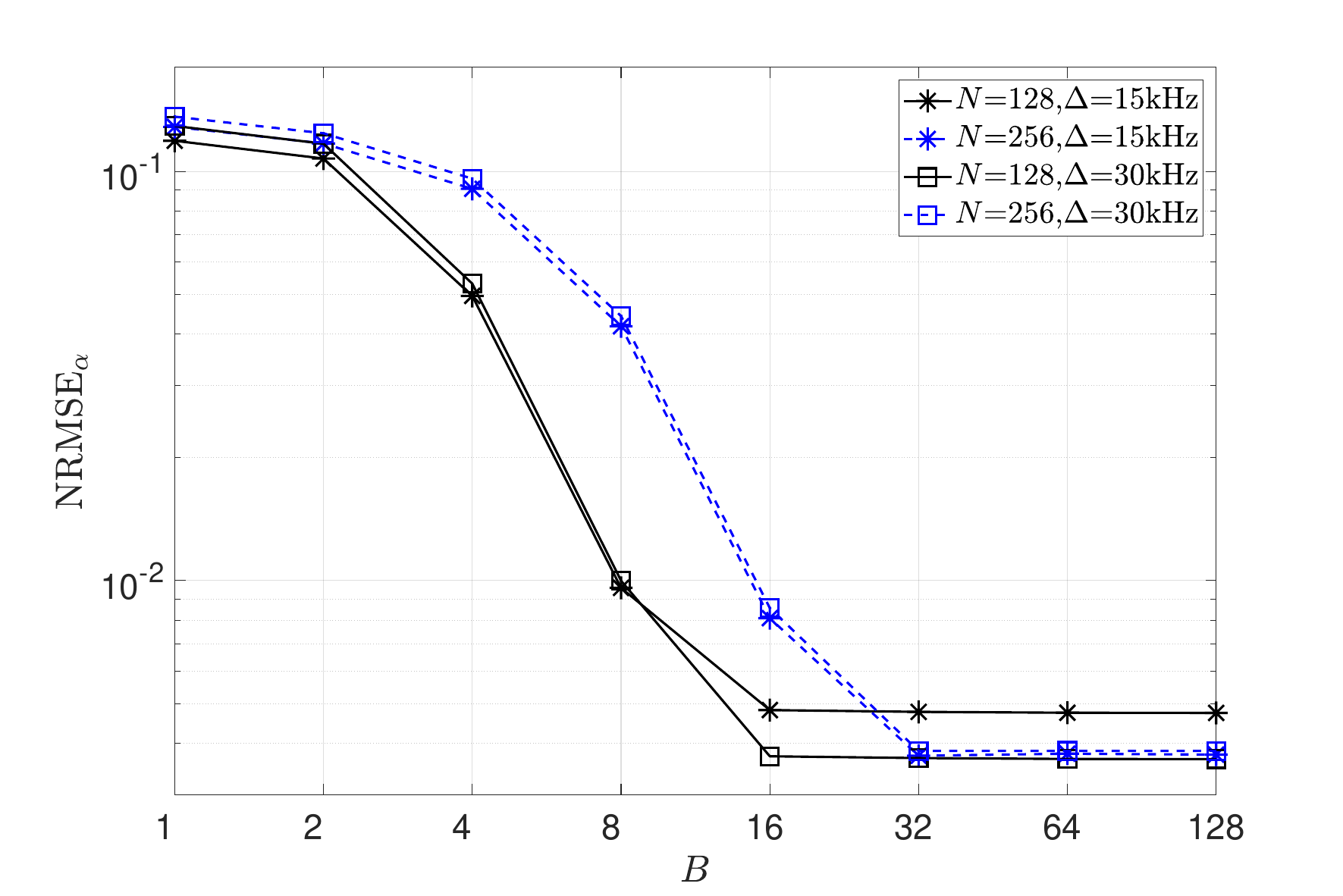}
    }
    \caption{$\mathrm{RMSE}_{\rm r}$ (left), $\mathrm{RMSE}_{\rm rr}$ (center), and $\mathrm{NRMSE}_{\alpha}$ (right) versus $B=1,2,4,8,16,32,64,128$ for different values of $N$ and $\Delta$, when a high-speed target is present, rectangular shaping filters are employed, $\mathrm{SNR}=30$~dB, and the DD-BOMP+MF estimator is considered.}
    \label{fig_single_3}
\end{figure*}

\begin{figure*}[!t]
    \centerline{
        \includegraphics[width=0.33\textwidth,trim=1cm 0cm 2cm 0.4cm, clip]{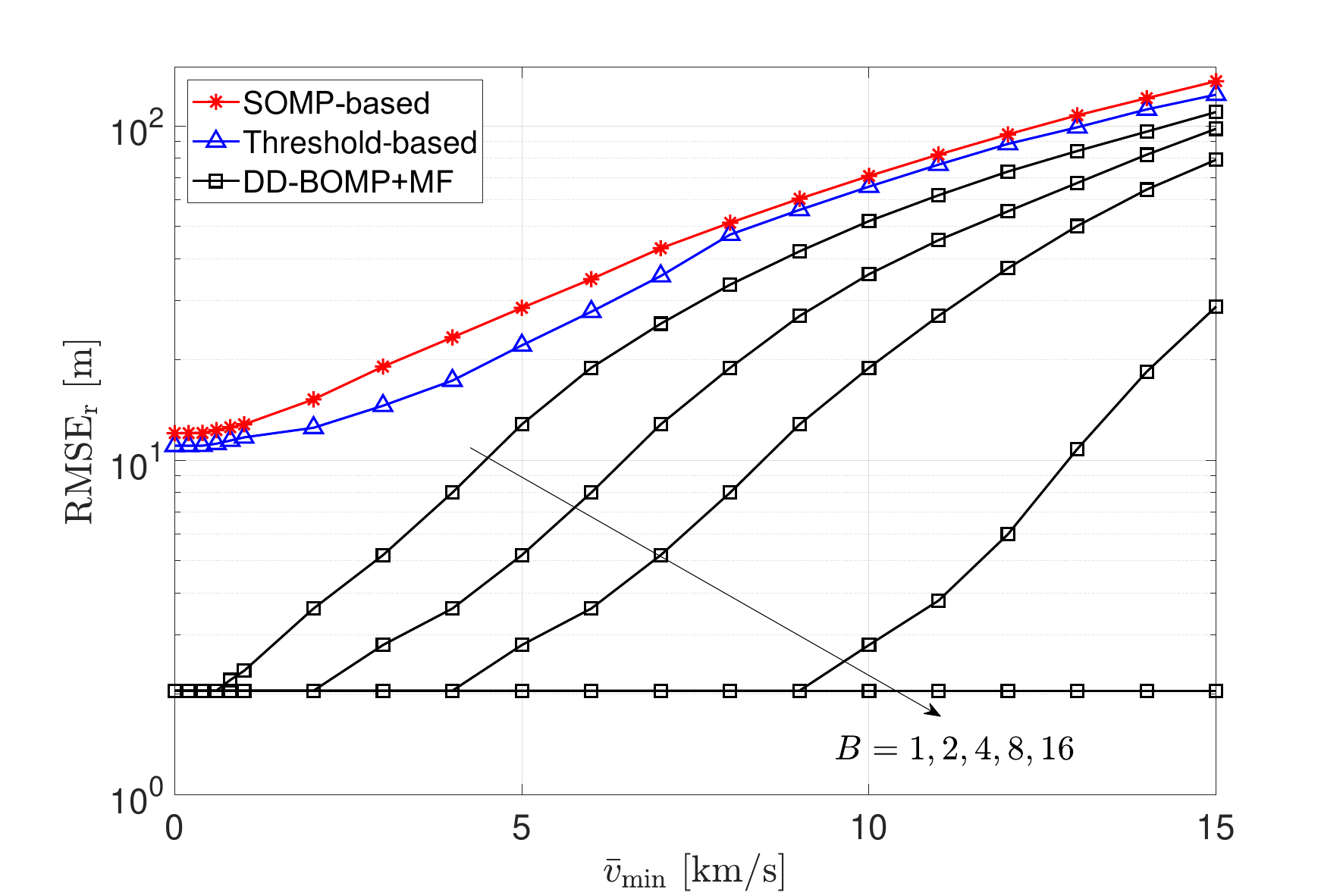}
        \includegraphics[width=0.33\textwidth,trim=0.5cm 0cm 2cm 0.4cm, clip]{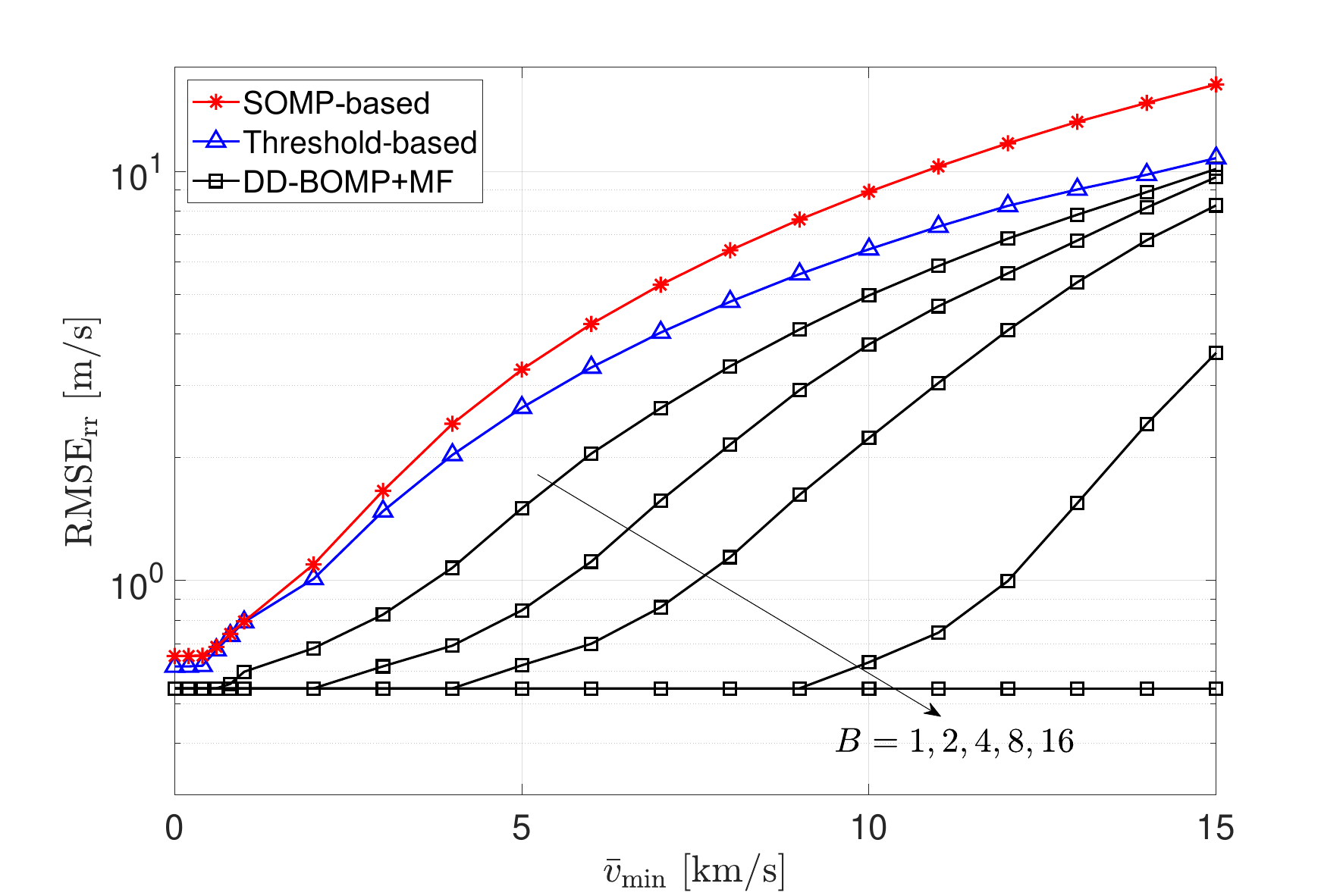}
        \includegraphics[width=0.33\textwidth,trim=0.5cm 0cm 2cm 0.4cm, clip]{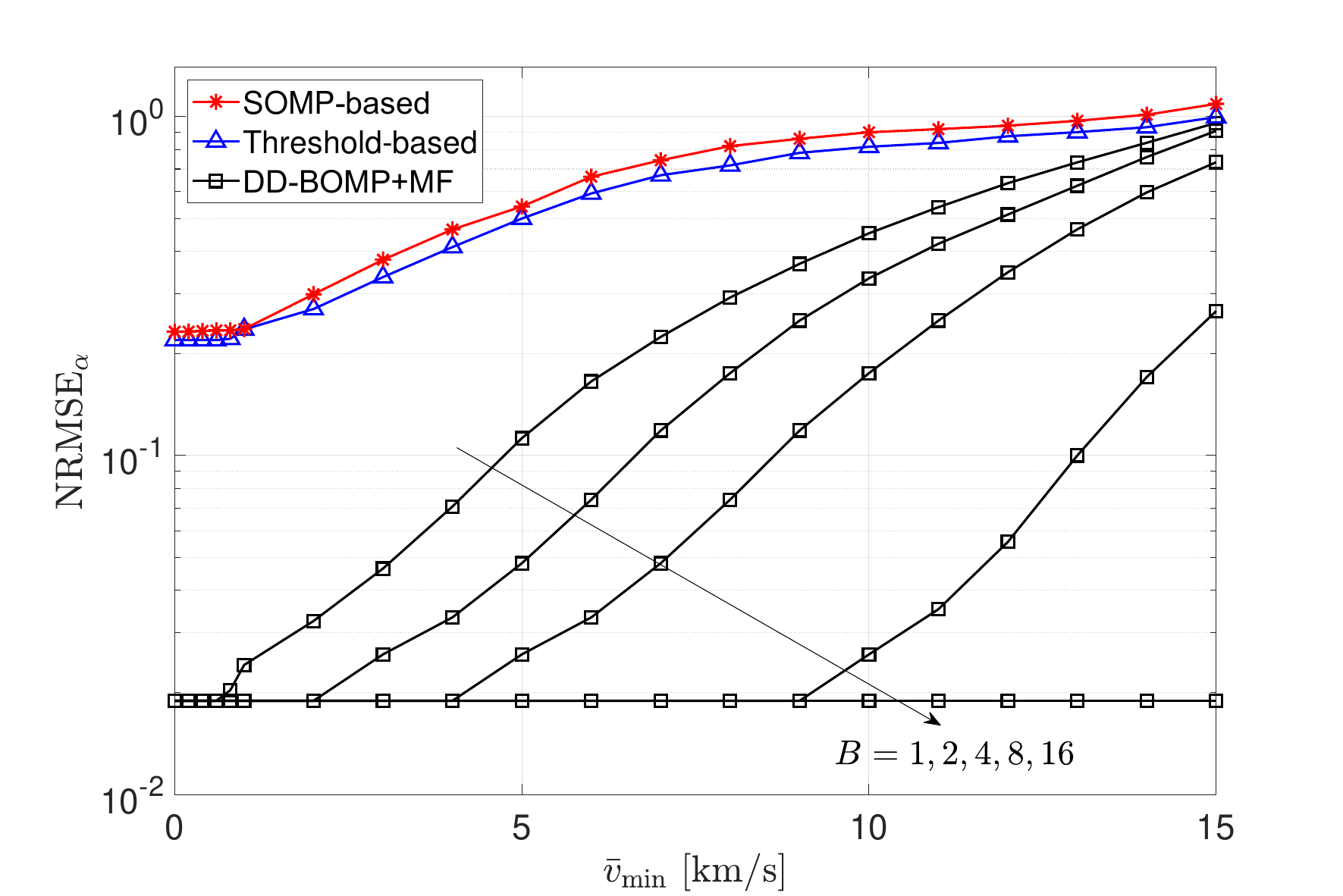}
    }
    \caption{$\mathrm{RMSE}_{\rm r}$ (left), $\mathrm{RMSE}_{\rm rr}$ (center), and $\mathrm{NRMSE}_{\alpha}$ (right) versus $\bar{v}_{\min}$, when rectangular shaping filters are used, $\mathrm{SNR}=15$~dB, and $\bar{v}_{\max}=\bar{v}_{\min}+1$~km/s. The DD-BOMP+MF estimator is implemented for $B=1, 2, 4, 8, 16$. For comparison, the threshold-based estimator in~\cite{Raviteja-2019-pilot} and the SOMP-based estimator in~\cite{Wenqian-2019} are included.}\label{fig_single_4}
\end{figure*}

\subsection{Single target}
We first examine the estimation performance with ideal shaping filters. Fig.~\ref{fig_single_1} presents the metrics $\mathrm{RMSE}_{\rm r}$, $\mathrm{RMSE}_{\rm rr}$, and $\mathrm{NRMSE}_{\alpha}$ as functions of $B$ for $\mathrm{SNR}=10,20,30$~dB. The parameters of a high-speed target cannot be accurately recovered when range variation during the OTFS frame is neglected at the system design (i.e., $B=1$). As $B$ increases from 1 to 16, estimation accuracy improves significantly, while further increases in $B$ yield diminishing returns. This behavior is consistent with the criterion $\eta_B\ll1$. Indeed, for the considered parameters, $\eta_B\simeq3.5/B$ from~\eqref{ratio_range_migration}, so that for $B=1$ the blockwise constant-range approximation is inaccurate. For $B=16$, the ratio reduces to about $0.22$, corresponding to a maximum range variation within one block of approximately $8.5$~m, at which point the estimation performance has essentially reached its saturation region.

For comparison, Fig.~\ref{fig_single_1} also includes results for a low-speed target (i.e., one for which range migration within the OTFS frame is negligible under the considered first-order range model) with $[\bar{v}_{\min}~\bar{v}_{\max}]=[0~1]$~km/s. In this case, the estimation accuracy remains unaffected by the choice of $B$; indeed, the maximum range variation during the OTFS frame is $\bar{v}_{\rm pk}NT\approx 8.5$~m, and $B = 1$ already provides an adequate model. Notably, the estimation accuracy for high-speed targets can match that of low-speed targets if range variation is properly accounted for in the echo model employed by the radar receiver (more on this in Fig.~\ref{fig_single_4}). Finally, note that the DD-BOMP+MF estimator achieves performance close to that of the DD-MF estimator across all considered scenarios, underscoring its effectiveness and practical utility.

Hereafter, rectangular shaping filters are employed. Fig.~\ref{fig_single_2} shows $\mathrm{RMSE}_{\rm r}$, $\mathrm{RMSE}_{\rm rr}$, and $\mathrm{NRMSE}_{\alpha}$ as functions of $\mathrm{SNR}$ when the DD-MF and DD-BOMP+MF estimators are implemented with $B=1,2,4,8,16,32$. For comparison, we also report the coarse initial-range and range-rate estimates provided by BOMP, denoted as DD-BOMP, and consider a motion-aware time-domain matched-filter (TD-MF) operating on the received noisy waveform before OTFS demodulation. For each hypothesized initial-range/range-rate pair, TD-MF reconstructs the corresponding data-dependent echo and evaluates a normalized matched-filter statistic. Unlike DD-MF, TD-MF does not rely on the additional blockwise constant-range approximation; however, both estimators are based on the same adopted first-order range model. Under a direct dense implementation, its exhaustive search has complexity $\mathcal{O}\big(G_{\rm r}G_{\rm rr}(MN)^2\big)$, i.e., the same dominant scaling as DD-MF.

The DD-BOMP performance is independent of $B$ (see Footnote~\ref{footnote-BOMP}) and is limited by the coarse delay-Doppler localization. Instead, the subsequent local matched-filter search in DD-BOMP+MF depends on $B$ through the full echo model, as does the global matched-filter search in DD-MF. For small $B$, the blockwise constant-range approximation is inaccurate and the resulting model mismatch may cause both DD-MF and DD-BOMP+MF to perform worse than the coarse DD-BOMP estimates, while DD-BOMP+MF still closely tracks DD-MF. As $B$ increases, the echo model becomes increasingly accurate, and the performance of both DD-MF and DD-BOMP+MF improves and approaches that of TD-MF. Altogether, these results show that the approximate model used by BOMP provides useful coarse localization independently of $B$, whereas accurate final estimation with the matched-filter step requires a sufficiently large $B$ to properly account for range migration.

Fig.~\ref{fig_single_3} examines the impact of the OTFS parameters $\Delta$ and $N$ on the performance of the DD-BOMP+MF estimator at $\mathrm{SNR}=30$~dB. The initial-range accuracy improves with $\Delta$ because the bandwidth $M\Delta$ increases, while it is only marginally affected by $N$ at fixed SNR. Conversely, the range-rate accuracy improves with larger $N$ and smaller $\Delta$, as both increase the coherent observation time $NT$ (with $T=1/\Delta$). The impact of $B$ is governed by $\eta_B$ in~\eqref{ratio_range_migration}, which is independent of $\Delta$ and scales linearly with $N$. Hence, the pairs $(N,\Delta)=(128,15~\text{kHz})$ and $(128,30~\text{kHz})$, shown by solid lines, have a similar validity region for the blockwise constant-range approximation, whereas the $N=256$ cases, shown by dashed lines, require approximately twice the value of $B$ to achieve comparable accuracy.

Finally, we assess the DD-BOMP+MF estimator across different mobility levels by setting $\bar{v}_{\max}=\bar{v}_{\min}+1$~km/s and varying $\bar{v}_{\min}$ from 0 to 15~km/s. For comparison, we include the threshold-based estimator in~\cite{Raviteja-2019-pilot} and the SOMP-based estimator in~\cite{Wenqian-2019}, both designed under a constant integer-range assumption over the OTFS frame. For a fair comparison, $X_{{\rm DD}}[k,l]$ has a single non-zero entry at $(k,l)=(64,256)$, with all remaining entries set to zero to form a guard region.\footnote{In this setup, the DD-BOMP+MF complexity reduces to $\mathcal{O}\big(MK^2N^2 + K^4N^3 + \tilde{G}_{\text{r}}\tilde{G}_{\text{rr}}MN\big)$ because $\bm{x}$ has a single non-zero entry at a known location, making the matrix-vector multiplication in~\eqref{e_vector} trivial during the second step. The benchmarks in~\cite{Raviteja-2019-pilot} and~\cite{Wenqian-2019} both have complexity $\mathcal{O}(MN)$.} Fig.~\ref{fig_single_4} reports $\mathrm{RMSE}_{\rm r}$, $\mathrm{RMSE}_{\rm rr}$, and $\mathrm{NRMSE}_{\alpha}$ versus $\bar{v}_{\min}$ for $\mathrm{SNR}=15$~dB. The estimators in~\cite{Raviteja-2019-pilot} and~\cite{Wenqian-2019} deteriorate rapidly beyond $\bar{v}_{\min}=1$~km/s. In this regime, the within-frame range excursion becomes a significant fraction of, and eventually exceeds, $R_{\rm r}$, causing a mismatch between the assumed and actual echo models. In contrast, the DD-BOMP+MF estimator remains robust over a much wider mobility range by increasing $B$, which improves the blockwise constant-range approximation in~\eqref{stop-and-go-approx}. In particular, $B=16$ provides stable performance over the entire considered range of $\bar{v}_{\min}$, consistently with Figs.~\ref{fig_single_1} and~\ref{fig_single_3}. Even as $\bar{v}_{\min}\rightarrow 0$, the DD-BOMP+MF estimator outperforms the existing methods owing to the second matched-filter step, at the cost of higher computational complexity.

\begin{figure*}[!t]
    \centerline{
        \includegraphics[width=0.33\textwidth,trim=1cm 0cm 2cm 0.4cm, clip]{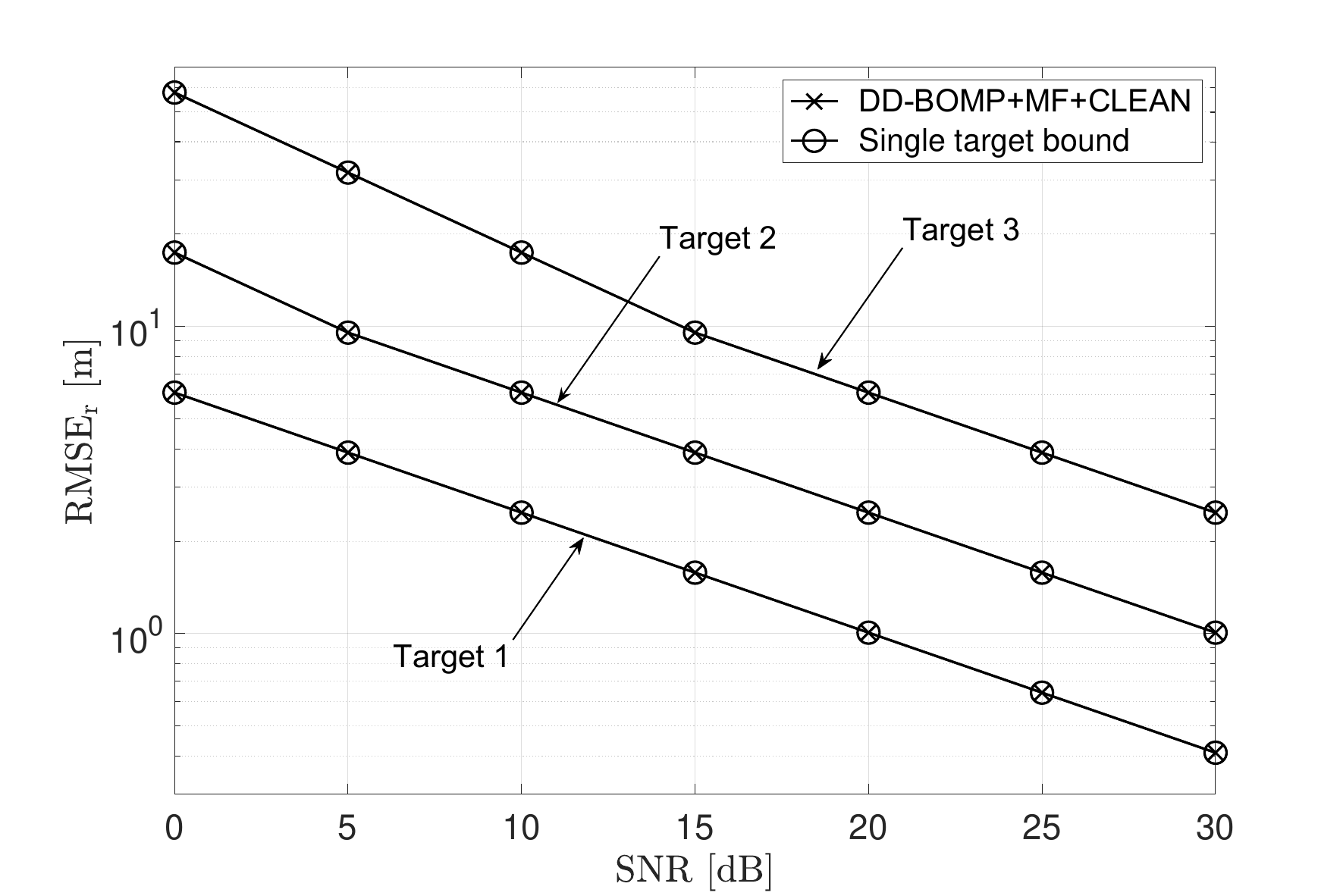}
        \includegraphics[width=0.33\textwidth,trim=0.5cm 0cm 2cm 0.4cm, clip]{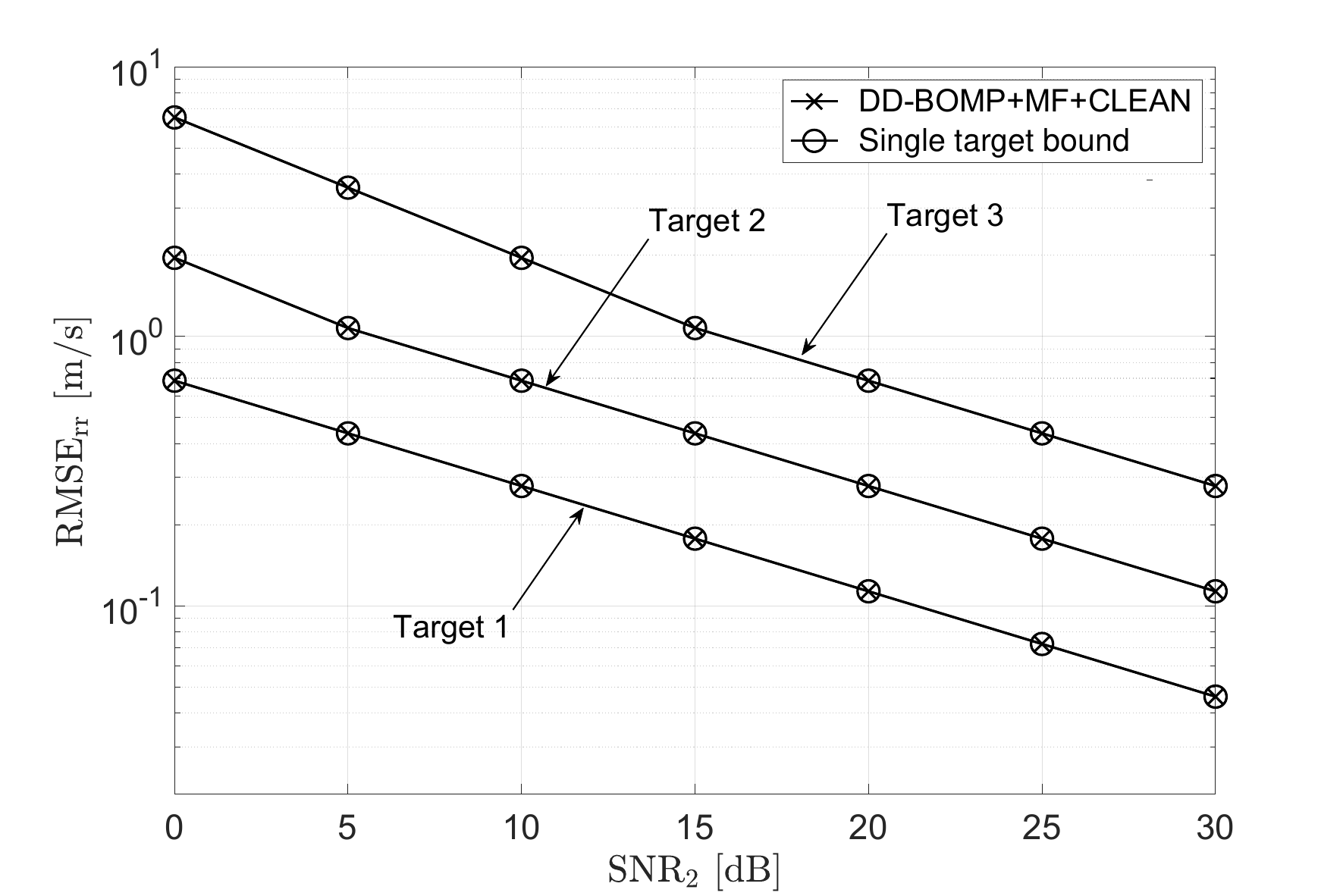}
        \includegraphics[width=0.33\textwidth,trim=0.5cm 0cm 2cm 0.4cm, clip]{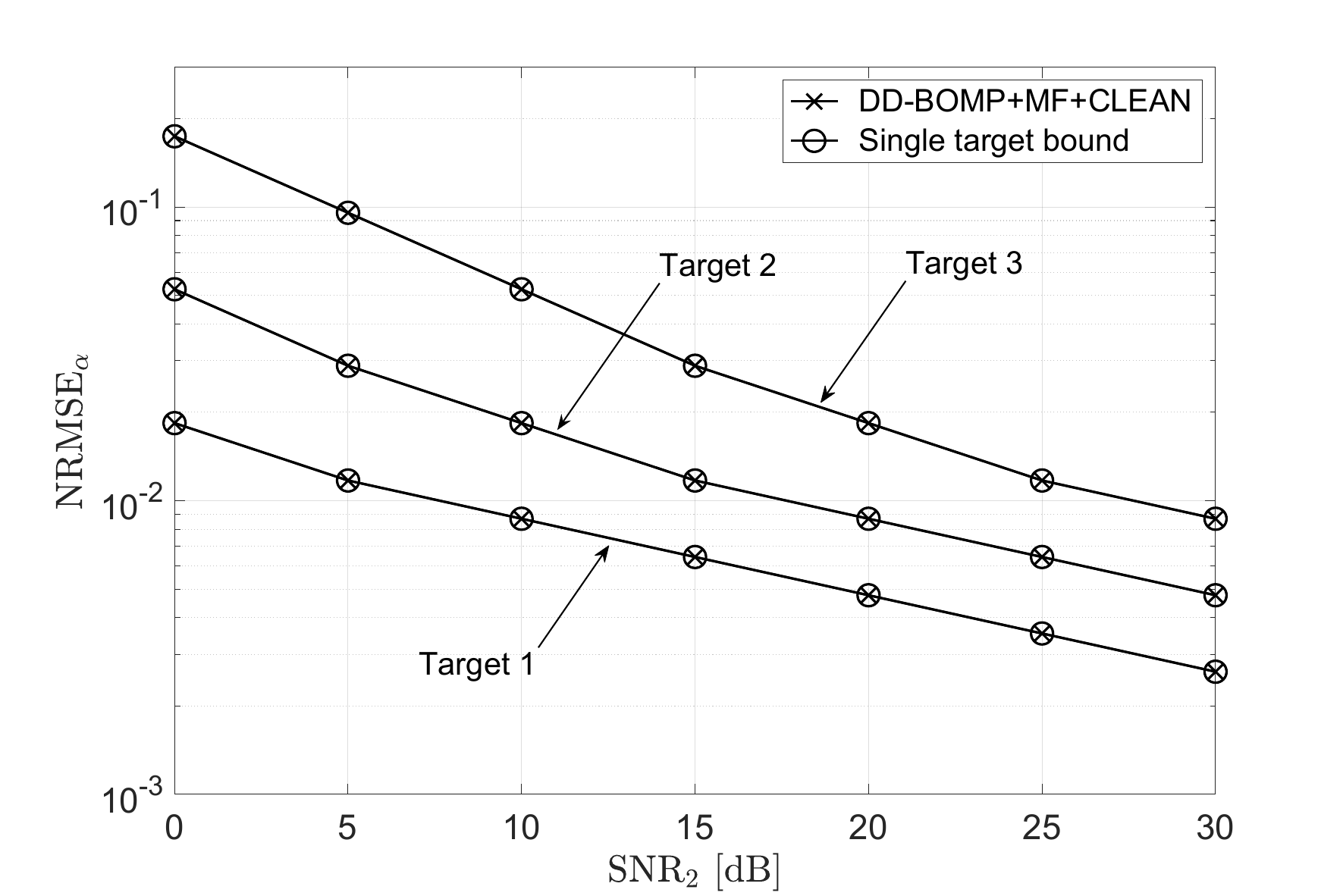}
    }
    \caption{$\mathrm{RMSE}_{\rm r}$ (left), $\mathrm{RMSE}_{\rm rr}$ (center), and $\mathrm{NRMSE}_{\alpha}$ (right) versus $\mathrm{SNR}_{2}$, when rectangular shaping filters are used, $P=3$ high-speed targets are present, $B=16$, and $\mathrm{SNR}_{1}/\mathrm{SNR}_{2}=\mathrm{SNR}_{2}/\mathrm{SNR}_{3}=10$. The performance of the DD-BOMP+MF+CLEAN estimator and the corresponding single-target performance bound are reported.}
    \label{fig_multi_1}
\end{figure*}

\begin{figure*}[!t]
    \centerline{
        \includegraphics[width=0.33\textwidth,trim=0.5cm 0cm 0.5cm 0.4cm, clip]{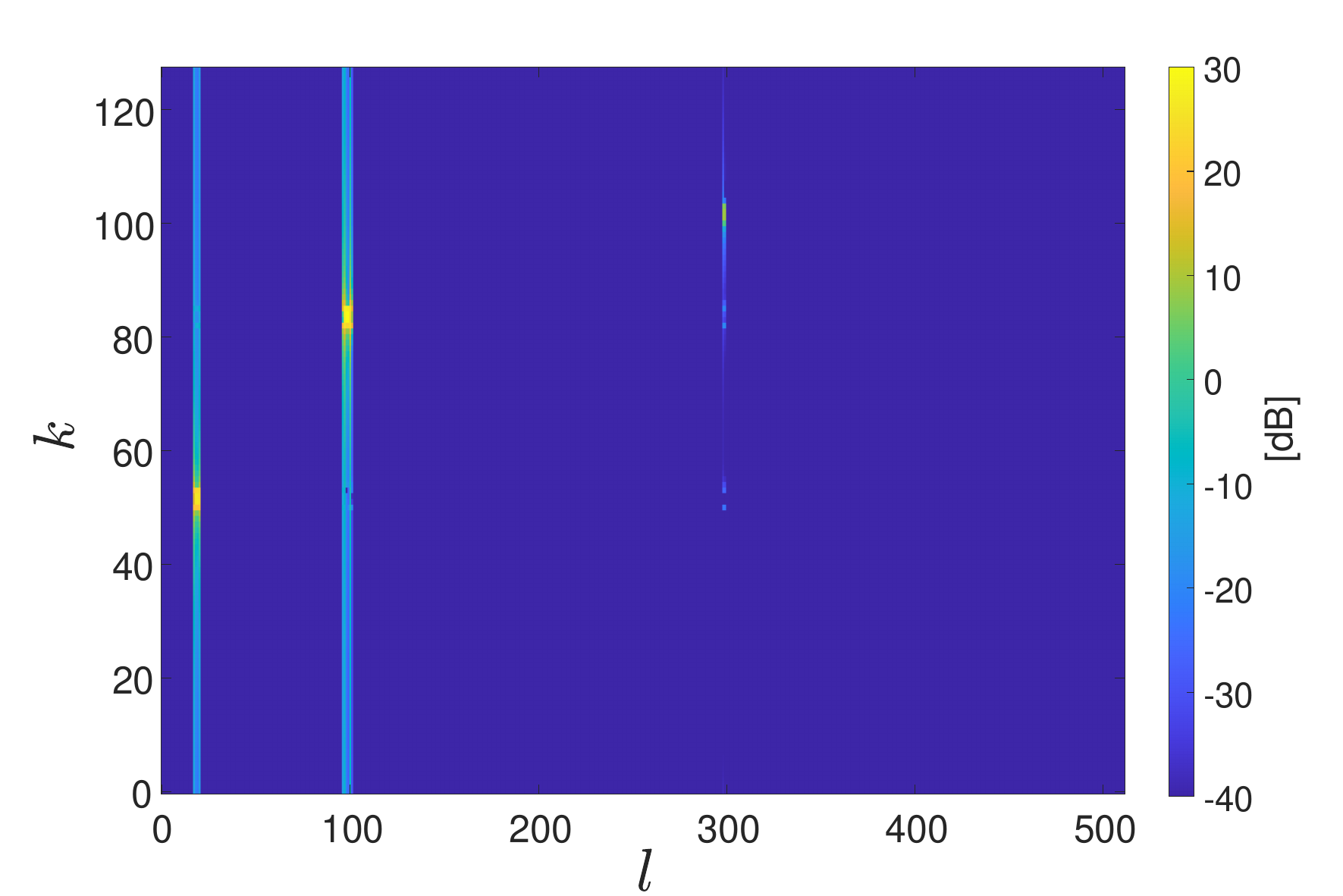}
        \includegraphics[width=0.33\textwidth,trim=0.5cm 0cm 0.5cm 0.4cm, clip]{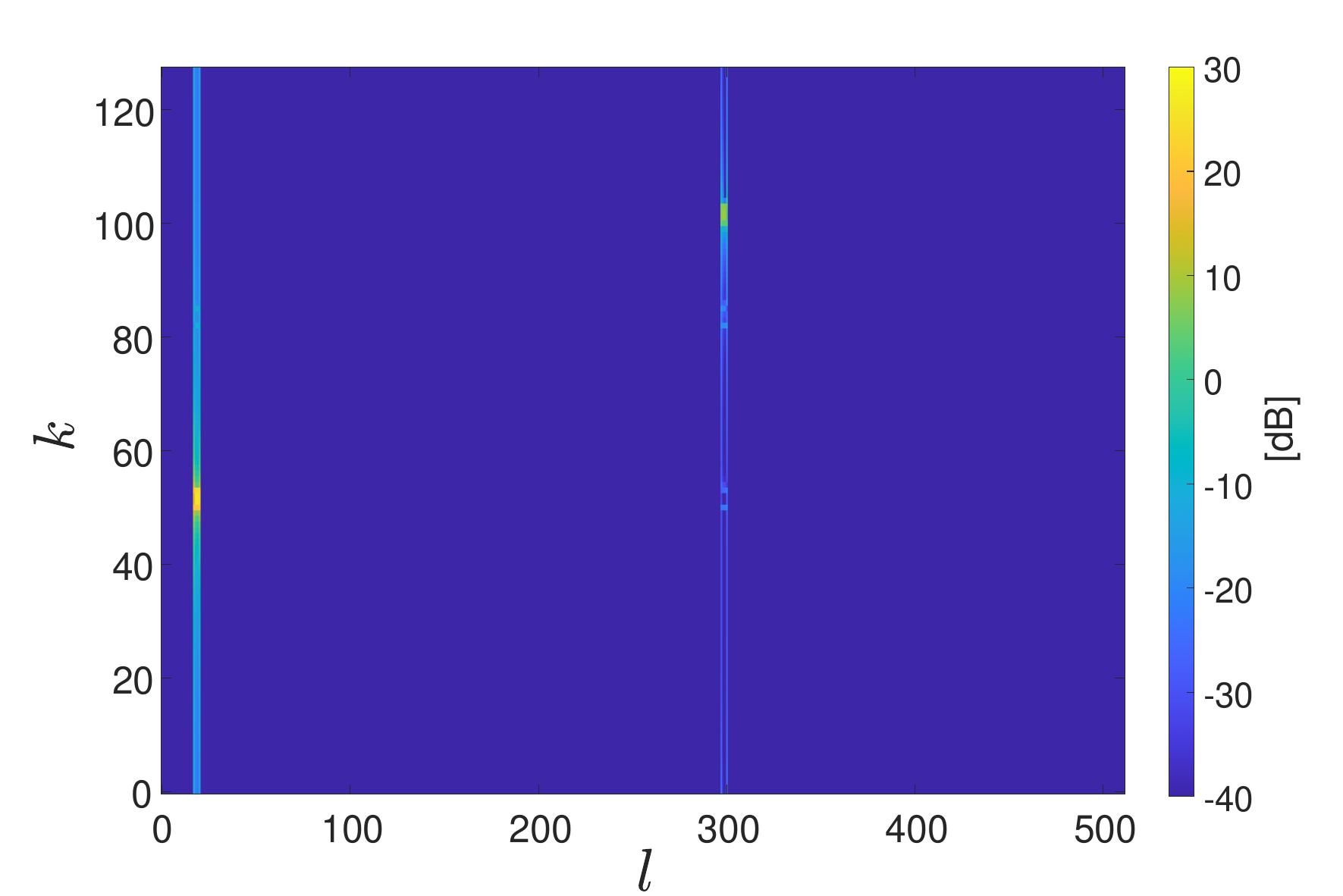}
        \includegraphics[width=0.33\textwidth,trim=0.5cm 0cm 0.5cm 0.4cm, clip]{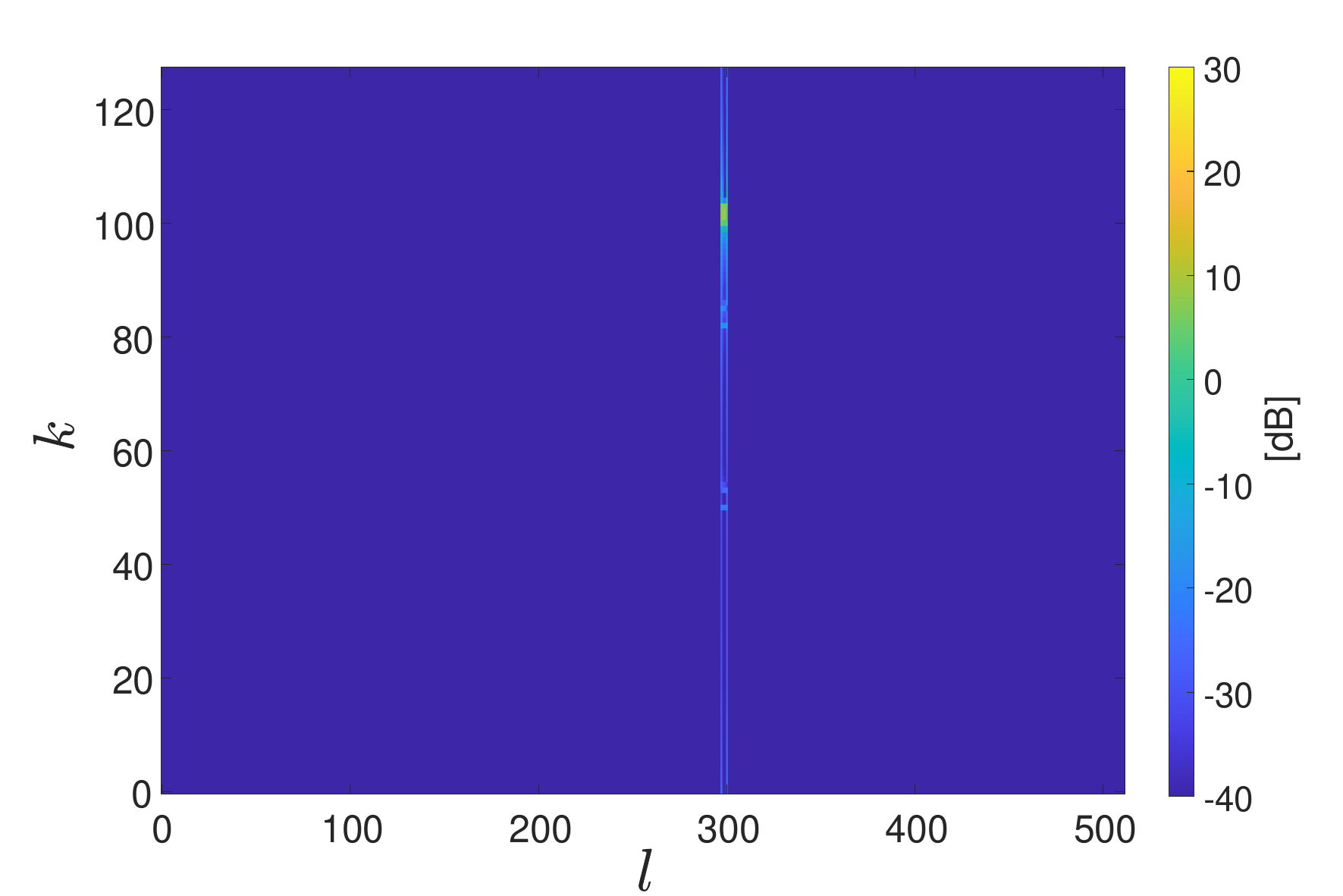}
    }
    \caption{$|\tilde{F}[k,l]|/\sigma_{\omega}$ vs $k=0,\ldots,N-1$ and $l=0,\ldots,M-1$ at the first (left), second (center), and third (right) iteration of the CLEAN algorithm, when rectangular shaping filters are employed, $P=3$ high-speed targets are present, $B=16$, $\mathrm{SNR}_{1}=30$~dB, $\mathrm{SNR}_{2}=20$~dB, and $\mathrm{SNR}_{3}=10$~dB. }
    \label{fig_multi_2}
\end{figure*}
\begin{figure*}[!t]
    \centerline{
        \includegraphics[width=0.33\textwidth,trim=1cm 0cm 2cm 0.4cm, clip]{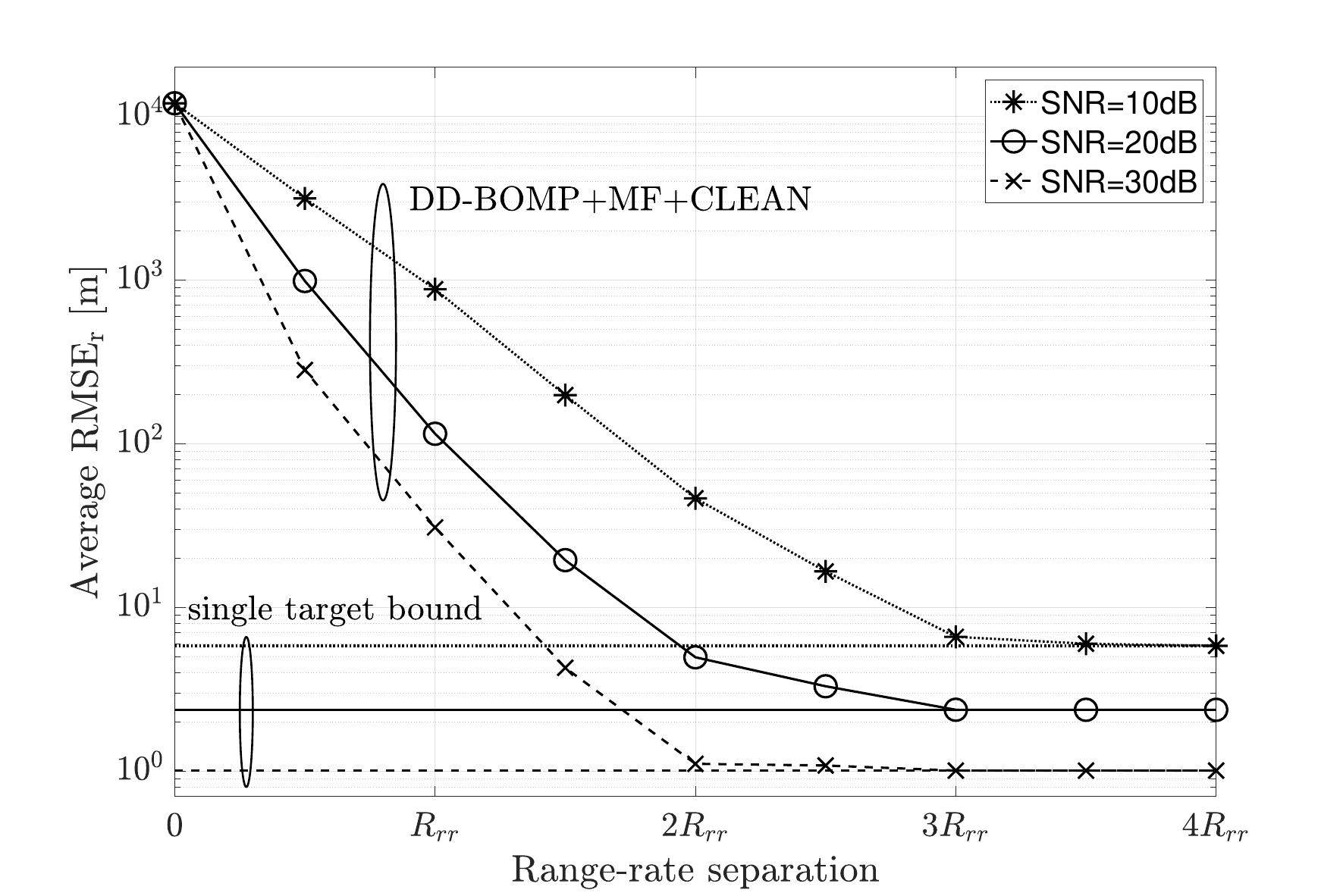}
        \includegraphics[width=0.33\textwidth,trim=0.5cm 0cm 2cm 0.4cm, clip]{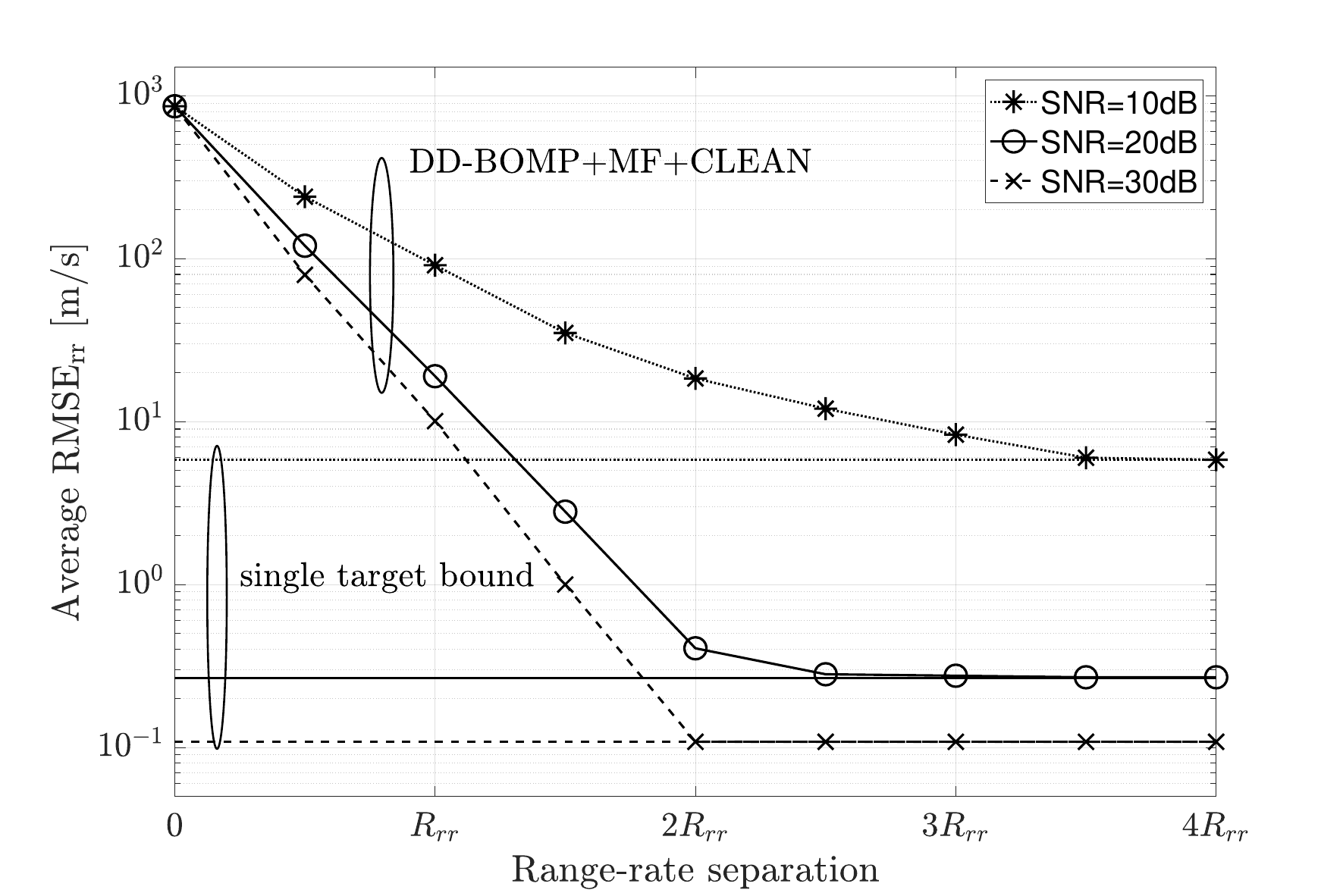}
        \includegraphics[width=0.33\textwidth,trim=0.5cm 0cm 2cm 0.4cm, clip]{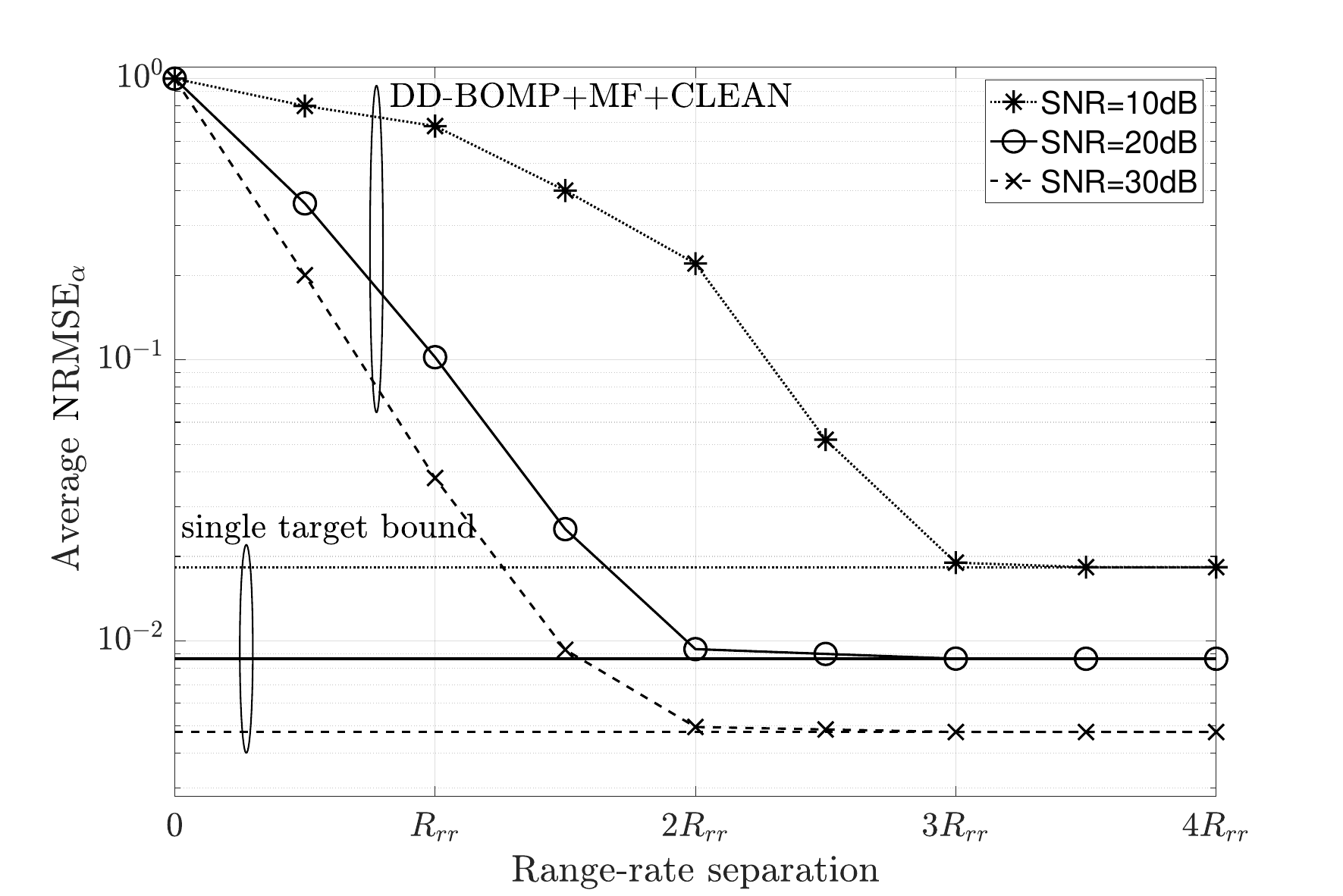}
    }
    \caption{Average $\mathrm{RMSE}_{\rm r}$ (left), $\mathrm{RMSE}_{\rm rr}$ (center), and $\mathrm{NRMSE}_{\alpha}$ (right) versus range-rate separation for $\mathrm{SNR}=10,20,30$~dB, when rectangular shaping filters are used, $P=2$ high-speed targets with the same initial-range and SNR are present, and $B=16$. The performance of the DD-BOMP+MF+CLEAN estimator and the corresponding single-target performance bound are reported.}
    \label{fig_multi_3}
\end{figure*}

\subsection{Multiple targets}
We first assume $P=3$ and denote by $\mathrm{SNR}_{p}$ the SNR of Target~$p$ for $p=1,\ldots,P$. For each target, Fig.~\ref{fig_multi_1} reports the metrics $\mathrm{RMSE}_{\rm r}$, $\mathrm{RMSE}_{\rm rr}$, and $\mathrm{NRMSE}_{\alpha}$ versus $\mathrm{SNR}_{2}$, when $B=16$, $\mathrm{SNR}_{1}/\mathrm{SNR}_{2}=\mathrm{SNR}_{2}/\mathrm{SNR}_{3}=10$, and the DD-BOMP+MF+CLEAN estimator is employed. For comparison, we also report the corresponding single-target performance obtained with the DD-MF estimator in~\eqref{ML_estimator}, which we refer to as the single-target bound. The proposed procedure achieves performance close to this bound for all three targets. For a single instance with $\mathrm{SNR}_{2}=20$~dB, Fig.~\ref{fig_multi_2} shows the BOMP-recovered sequence $|\tilde{F}[k,l]|/\sigma_{\omega}$ at each CLEAN iteration. It is seen that the target echoes are progressively recovered and removed from the observed data.

Recall that CLEAN is most effective when a dominant target can be accurately estimated before subtracting its reconstructed echo~\cite{Colone-2016,Bose-2011,Misiurewicz-2012,Bosse-2018}. For targets with similar powers and strongly overlapping delay-Doppler responses, cancellation errors may propagate across iterations. To illustrate this limitation, we consider $P=2$ targets with the same initial-range and SNR and vary their range-rate separation. Fig.~\ref{fig_multi_3} reports $\mathrm{RMSE}_{\rm r}$, $\mathrm{RMSE}_{\rm rr}$, and $\mathrm{NRMSE}_{\alpha}$ of the DD-BOMP+MF+CLEAN estimator, averaged over the two targets, for different SNR values. When the separation is below the nominal resolution $R_{\rm rr}$, the targets are difficult to distinguish and the estimation errors increase significantly. As it exceeds $R_{\rm rr}$, the targets become reliably separable and the performance approaches the single-target bound. This confirms that $R_{\rm rr}$ provides a meaningful reference for range-rate discrimination.

\section{Conclusions}\label{SEC_Conclusions}
In this work, we have investigated the use of OTFS communication signals for estimating the initial-range, range-rate, and amplitude of point-like space targets experiencing range migration under a first-order range model over the OTFS frame in LEO satellite sensing applications. Our main contribution is a novel delay-Doppler-domain echo model based on a blockwise constant-range approximation over the OTFS frame. We have shown that range migration spreads the target response while preserving its sparse structure. Exploiting this sparsity, we have proposed a two-step estimator of the target parameters that avoids an exhaustive delay-Doppler matched-filter search. When range migration is properly accounted for, the estimation accuracy for high-speed targets remains comparable to that for low-speed targets.

Future work may relax the assumption of perfectly known communication symbols and consider more general propagation and target models, including extended targets, rotational effects, multipath, and higher-order motion. Coordinated multi-satellite and multi-frame processing may also enable estimation of target position and velocity and help resolve initial-range and range-rate ambiguities.

\appendices
\section{Proof of Proposition~\ref{Proposition_TF}}
\label{Appendix_TF}
Upon plugging~\eqref{rx_signal-fast} into~\eqref{rx_signal_TF}, we have
\begin{align*}
    & Y_{\rm TF}[n,m]= \sum_{n'=0}^{N-1}\sum_{m'=0}^{M-1} \alpha X_{\rm TF}[n',m'] \notag \\
    &\quad \times \int_{-\infty}^{\infty}g_{\rm tx}\!\left(t-n'T -\frac{r\left(n'T\right)}{c}\right) g^{*}_{\rm rx}\!\left(t-nT\right) \notag \\
    &\quad \times \e^{\i 2\pi m' \Delta \left(t-n'T-\frac{r\left(n'T\right)}{c}\right)} \e^{-\i 2\pi m \Delta\left(t-nT\right)}\e^{\i 2\pi\frac{v}{\lambda}t}dt\notag \\
    &=\sum_{n'=0}^{N-1}\sum_{m'=0}^{M-1} X_{\rm TF}[n',m'] \notag \\
    &\quad \times \alpha \e^{\i 2\pi m \Delta\left((n-n')T-\frac{r\left(n'T\right)}{c}\right)}
    \e^{\i 2\pi\frac{v}{\lambda}\left(n'T+\frac{r\left(n'T\right)}{c}\right)}\notag \\
    &\quad \times
    \e^{-\i 2\pi \left( (m-m') \Delta -\frac{v}{\lambda}\right)\left((n-n')T-\frac{r\left(n'T\right)}{c}\right)}
    \notag \\
    &\quad \times
    \int_{-\infty}^{\infty}g_{\rm tx}\!\left(\beta\right) g^{*}_{\rm rx}\!\left(\beta-(n-n')T+\frac{r\left(n'T\right)}{c}\right) \notag \\
    &\quad \times
    \e^{-\i 2\pi \left( (m-m') \Delta -\frac{v}{\lambda}\right)\left(\beta-(n-n')T+\frac{r\left(n'T\right)}{c}\right)}d\beta\notag\\
    &=\sum_{n'=0}^{N-1}\sum_{m'=0}^{M-1} X_{\rm TF}[n',m']
    \alpha \e^{\i 2\pi\frac{v}{\lambda}nT}
    \e^{-\i 2\pi m' \Delta\frac{r\left(n'T\right)}{c}}
    \notag \\
    &\quad \times \gamma\left((n-n')T-\frac{r\left(n'T\right)}{c},(m-m') \Delta -\frac{v}{\lambda}\right),
\end{align*}
which completes the proof.

\section{Proof of Proposition~\ref{Proposition_DD_ideal}}
\label{Appendix_DD_ideal}
After exploiting~\eqref{ISFFT_TX} and~\eqref{rx_signal_TF_ideal_fast}, Eq.~\eqref{rx_signal_DD} becomes
\begin{multline}
    Y_{\rm DD}[k,l] =\frac{1}{NM}\!\sum_{k'=0}^{N-1}\sum_{l'=0}^{M-1} \! X_{\rm DD}[k',l'] \\
    \times \underbrace{\sum_{n=0}^{N-1}\sum_{m=0}^{M-1} H_{n,m}[n,m]\e^{-\i 2 \pi\left(\frac{k-k'}{N}\right) n}
    \e^{\i 2 \pi \left(\frac{l-l'}{M}\right)m}}_{(\star)}.
\end{multline}
Finally, notice that we have
\begin{align}
    (\star)&= M \alpha\sum_{n=0}^{N-1} \e^{-\i 2 \pi\left(\frac{k-k'}{N}-\frac{vT}{\lambda}\right) n}\notag \\ &\quad \times \left[\frac{1}{M} \sum_{m=0}^{M-1}
    \e^{\i 2 \pi \left(\frac{l-l'}{M}-\frac{r\left(nT\right)\Delta}{c}\right) m}\right]
    \notag \\
    &= M \alpha\sum_{n=0}^{N-1} \e^{-\i 2 \pi\left(\frac{k-k'}{N}-\frac{vT}{\lambda}\right) n}\notag \\ &\quad \times \mathcal{D}_{M}\!\left(-\left(\frac{l-l'}{M}-\frac{r\left(nT\right)\Delta}{c}\right)\right)
    \notag \\
    &= NM \alpha \sum_{b=0}^{B-1}\frac{1 }{B}\left[\frac{B}{N}\sum_{n=bN/B}^{(b+1)N/B-1}\e^{-\i 2 \pi\left(\frac{k-k'}{N}-\frac{vT}{\lambda}\right) n}\right] \notag \\ &\quad \times \mathcal{D}_{M}\!\left(-\left(\frac{l-l'}{M }-\frac{r\left(bT_{B}\right) \Delta}{c}\right)\right)
    \label{hw_fast_simeq}\\
    &=NM \alpha \Phi[k-k',l-l'],\notag
\end{align}
where~\eqref{hw_fast_simeq} follows from~\eqref{stop-and-go-approx}.

\section{Proof of Proposition~\ref{Proposition_DD_rect}}
\label{Appendix_DD_rect}
Upon plugging~\eqref{rx_signal_TF_fast-2} and~\eqref{rx_signal_TF_fast-2-0} in~\eqref{rx_signal_DD}, we have
\begin{align}
    Y_{\rm DD}[k,l]&=\frac{1}{\sqrt{NM}}\sum_{n=0}^{N-1}\sum_{m=0}^{M-1}
    \e^{-\i 2\pi\left(\frac{n k}{N}-\frac{m l}{M}\right)} \notag \\
    &\quad \times \sum_{m'=0}^{M-1} H_{n,m}[n,m']X_{\rm TF}[n,m']
    \notag \\
    &\quad + \frac{1}{\sqrt{NM}}\sum_{n=1}^{N-1}\sum_{m=0}^{M-1}
    \e^{-\i 2\pi\left(\frac{n k}{N}-\frac{m l}{M}\right)} \notag \\
    &\quad \times \sum_{m'=0}^{M-1} H_{n,m}[n-1,m']X_{\rm TF}[n-1,m'] \notag \\
    &=Y_{\rm DD}^{\rm ici}[k,l]+Y_{\rm DD}^{\rm isi}[k,l]. \label{rx_signal_DD_fast_rect}
\end{align}

\begin{figure*}[!t]
    \centering
    \small
    \begin{align}
        &\Pi_{k,l}^{\rm ici}[k',l']=\frac{\alpha}{NM^{2}}\sum_{n=0}^{N-1}\sum_{m=0}^{M-1}\sum_{m'=0}^{M-1}\left[
        \e^{\i 2\pi\frac{v}{\lambda}nT} \e^{-\i 2\pi m' \Delta\frac{r\left(nT\right)}{c}} \sum_{p=0}^{\big\lceil M-\frac{r\left(nT\right)M}{cT}\big\rceil-1} \e^{-\i 2 \pi \left((m-m')\Delta- \frac{v}{\lambda}\right)\left(\frac{pT}{M}+\frac{r\left(nT\right)}{c}\right)}\right]
        \e^{\i2\pi n\left(\frac{k'-k}{N}\right)} \e^{\i 2\pi \left(\frac{ml-m'l'}{M}\right)}
        \notag \\
        &= \frac{\alpha}{N} \sum_{n=0}^{N-1}\e^{-\i 2 \pi n\left(\frac{k-k'}{N}-\frac{v T}{\lambda} \right)}\e^{\i 2 \pi \frac{v}{\lambda}\frac{r\left(nT\right)}{c}}\sum_{p=0}^{\big\lceil M-\frac{r\left(nT\right)M}{cT}\big\rceil-1}
        \left[\frac{1}{M}\sum_{m=0}^{M-1}\e^{-\i 2 \pi \frac{m}{M}\left(p+\frac{r\left(nT\right)}{c}\frac{M}{T}-l\right)}\right]
        \left[\frac{1}{M}\sum_{m'=0}^{M-1} \e^{\i 2 \pi \frac{m'}{M}(p-l')}\right]
        \e^{\i 2 \pi \frac{v T}{\lambda}\frac{p}{M}}
        \notag \\
        &= \sum_{b=0}^{B-1}\frac{\alpha}{N}\sum_{n=bN/B}^{(b+1)N/B-1} \e^{-\i 2 \pi n\left(\frac{k-k'}{N}-\frac{v T}{\lambda} \right)}\e^{\i 2 \pi \frac{v}{\lambda}\frac{r\left(nT\right)}{c}} \sum_{p=0}^{\big\lceil M-\frac{r\left(nT\right)M}{cT}\big\rceil-1}\mathcal{D}_{M}\!\left(\frac{p+\frac{r\left(nT\right)}{c}\frac{M}{T}-l}{M}\right)\delta\big[[p-l']_{M}\big]\e^{\i 2 \pi \frac{v T}{\lambda}\frac{p}{M}}
        \notag \\
        &= \frac{\alpha}{B} \sum_{b=0}^{B-1}\e^{\i 2 \pi \frac{v}{\lambda}\frac{r\left(bT_{B} \right)}{c}} \e^{-\i 2 \pi \left(\frac{k-k'}{N}-\frac{v T}{\lambda} \right)\frac{bN}{B}}
        \mathcal{D}_{\frac{N}{B}}\!\left(\tfrac{k-k'}{N}-\tfrac{v T}{\lambda}\right) \sum_{p=0}^{\big\lceil M-\frac{r\left(bT_{B}\right)M}{cT}\big\rceil-1}\mathcal{D}_{M}\!\left(\tfrac{p+\frac{r\left(b T_{B}\right)}{c}\frac{M}{T}-l}{M}\right)\delta\big[[p-l']_{M}\big]\e^{\i 2 \pi \frac{v T}{\lambda}\frac{p}{M}}. \label{h_fast_rect_ici}
    \end{align}
    \hrulefill
\end{figure*}
\begin{figure*}[!t]
    \centering \small
    \begin{align}
        &\Pi_{k,l}^{\rm isi}[k',l']=\frac{\alpha}{NM^{2}}\!\sum_{n=1}^{N-1}\sum_{m=0}^{M-1}\sum_{m'=0}^{M-1}\!\left[
        \e^{\i 2\pi\frac{v}{\lambda}nT} \e^{-\i 2\pi m' \Delta\frac{r\left((n-1)T\right)}{c}}\!\!\!\!
        \sum_{p=\left\lceil M-\frac{r\left((n-1)T\right)M}{cT}\right\rceil}^{M-1}\!\!\!\!\!\!\e^{-\i 2 \pi \left((m-m')\Delta- \frac{v}{\lambda}\right)\left(\frac{pT}{M}-T+\frac{r\left((n-1)T\right)}{c}\right)}\!\right]\e^{-\i 2\pi n\left(\frac{k-k'}{N}\right)} \!\e^{\i 2\pi \left(\frac{ml-m'l'}{M}\right)}
        \notag \\
        &= \frac{\alpha}{N} \sum_{n=1}^{N-1} \e^{-\i 2 \pi n\left(\frac{k-k'}{N}-\frac{v T}{\lambda} \right)}\e^{\i 2 \pi \frac{v}{\lambda}\frac{r\left((n-1)T\right)}{c}} \!\!\sum_{p=\left\lceil M-\frac{r\left((n-1)T\right)}{c}\frac{M}{T} \right\rceil}^{M-1}\!\!
        \left[\frac{1}{M}\sum_{m=0}^{M-1}\e^{-\i 2 \pi \frac{m}{M}\left(p-M+\frac{r\left((n-1)T\right)}{c}\frac{M}{T}-l\right)}\right]
        \left[\frac{1}{M}\sum_{m'=0}^{M-1} \e^{\i 2 \pi \frac{m'}{M}(p-M-l')}\right]\e^{\i 2 \pi \frac{v T}{\lambda}\frac{p-M}{M}}
        \notag \\
        &= \sum_{b=0}^{B-1}\frac{\alpha}{N}\sum_{\substack{n=bN/B \\ n \neq 0}}^{(b+1)N/B-1} \e^{-\i 2 \pi n\left(\frac{k-k'}{N}-\frac{v T}{\lambda} \right)}\e^{\i 2 \pi \frac{v}{\lambda}\frac{r\left((n-1)T\right)}{c}}\sum_{p=\left\lceil M-\frac{r\left((n-1)T\right)}{c}\frac{M}{T} \right\rceil}^{M-1}\mathcal{D}_{M}\!\left(\frac{p-M+\frac{r\left((n-1)T\right)}{c}\frac{M}{T}-l}{M}\right)\delta\big[[p-l']_{M}\big] \e^{\i 2 \pi \frac{v T}{\lambda}\frac{p-M}{M}}
        \notag \\
        &= \frac{\alpha}{B}\sum_{b=0}^{B-1} \e^{\i 2 \pi \frac{v}{\lambda}\frac{r\left(bT_{B}\right)}{c}} \e^{-\i 2 \pi \left(\frac{k-k'}{N}-\frac{v T}{\lambda} \right)\frac{bN}{B}} \left[\mathcal{D}_{\frac{N}{B}}\!\left(\frac{k-k'}{N}-\frac{v T}{\lambda}\right)-\mathbbm 1_{\{b=0\}} \frac{B}{N}\right]
        \notag\\
        &\quad \times \sum_{p=\left\lceil M-\frac{r\left(b T_{B}\right)}{c}\frac{M}{T} \right\rceil}^{M-1}\mathcal{D}_{M}\!\left(\frac{p-M+\frac{r\left(b T_{B}\right)}{c}\frac{M}{T}-l}{M}\right) \delta\big[[p-M-l']_{M}\big] \e^{\i 2 \pi \frac{v T}{\lambda}\frac{p-M}{M}}. \label{h_fast_rect_isi}
    \end{align}
    \hrulefill
\end{figure*}

As to the term $Y_{\rm DD}^{\rm ici}[k,l]$ in~\eqref{rx_signal_DD_fast_rect}, we have:
\begin{align}
    Y_{\rm DD}^{\rm ici}[k,l]&=\frac{1}{NM}\sum_{n=0}^{N-1}\sum_{m=0}^{M-1}\e^{-\i 2\pi\left(\frac{n k}{N}-\frac{m l}{M}\right)} \sum_{m'=0}^{M-1} H_{n,m}[n,m']\notag\\
    &\quad \times \sum_{k'=0}^{N-1}\sum_{l'=0}^{M-1} X_{\rm DD}[k',l']\e^{\i 2\pi\left(\frac{n k'}{N}-\frac{m' l'}{M}\right)} \notag\\
    &=\sum_{k'=0}^{N-1}\sum_{l'=0}^{M-1} X_{\rm DD}[k',l']\Pi_{k,l}^{\rm ici}[k',l'] ,
    \label{Y_DD_fast_rect_ici}
\end{align}
where
\begin{align}
    \Pi_{k,l}^{\rm ici}[k',l']&=\frac{1}{NM}\sum_{n=0}^{N-1}\sum_{m=0}^{M-1}\sum_{m'=0}^{M-1}H_{n,m}[n,m']\notag \\ &\quad \times \e^{-\i 2\pi n\left(\frac{k-k'}{N}\right)}\e^{\i 2\pi \left(\frac{ml-m'l'}{M}\right)}.
\end{align}
Upon expanding $\Pi_{k,l}^{\rm ici}[k',l']$ as shown in~\eqref{h_fast_rect_ici}, we obtain
\begin{align}
    Y_{{\rm DD}}^{\rm ici}[k,l]&=\frac{\alpha}{B} \sum_{b=0}^{B-1}\sum_{k'=0}^{N-1}
    \sum_{l'=0}^{\big\lceil M-\frac{r\left(bT_{B}\right)M}{cT}\big\rceil-1}
    X_{\rm DD}[k',l'] \notag \\
    &\quad \times \e^{\i 2 \pi \frac{v}{\lambda}\frac{r\left(bT_{B}\right)}{c}} \e^{\i 2 \pi \frac{v T}{\lambda}\frac{l'}{M}}\notag \\
    &\quad \times \e^{-\i 2 \pi \left(\frac{k-k'}{N}-\frac{v T}{\lambda} \right)\frac{bN}{B}} \mathcal{D}_{\frac{N}{B}}\!\left(\frac{k-k'}{N}-\frac{v T}{\lambda}\right) \notag \\
    &\quad \times\mathcal{D}_{M}\!\left(-\left(\frac{l-l'}{M }-\frac{r\left(bT_{B}\right) \Delta}{c}\right)\right).
    \label{Y_DD_ICI}
\end{align}

As to the term $Y_{\rm DD}^{\rm isi}[k,l]$ in~\eqref{rx_signal_DD_fast_rect}, we have:
\begin{align}
    Y_{\rm DD}^{\rm isi}[k,l]&=\frac{1}{NM}\sum_{n=1 }^{N-1}\sum_{m=0}^{M-1}\e^{-\i 2\pi\left(\frac{n k}{N}-\frac{m l}{M}\right)} \sum_{m'=0}^{M-1} H_{n,m}[n-1,m']\notag\\
    &\quad \times \sum_{k'=0}^{N-1}\sum_{l'=0}^{M-1} X_{\rm DD}[k',l']\e^{\i 2\pi\left(\frac{(n-1) k'}{N}-\frac{m' l'}{M}\right)} \notag\\
    &=\sum_{k'=0}^{N-1}\sum_{l'=0}^{M-1} X_{\rm DD}[k',l']\Pi_{k,l}^{\rm isi}[k',l'] \e^{-\i 2\pi\frac{k'}{N}} ,
    \label{Y_DD_fast_rect_isi}
\end{align}
where
\begin{align}
    \Pi_{k,l}^{\rm isi}[k',l']&=\frac{1}{NM}\sum_{n=1}^{N-1}\sum_{m=0}^{M-1}\sum_{m'=0}^{M-1}H_{n,m}[n-1,m']\notag \\ &\quad \times \e^{-\i 2\pi n\left(\frac{k-k'}{N}\right)}\e^{\i 2\pi \left(\frac{ml-m'l'}{M}\right)}.
\end{align}
Upon expanding $\Pi_{k,l}^{\rm isi}[k',l']$ as shown in~\eqref{h_fast_rect_isi}, we obtain
\begin{align}
    Y_{{\rm DD}}^{\rm isi}[k,l]&=\frac{\alpha}{B} \sum_{b=0}^{B-1}\sum_{k'=0}^{N-1}
    \sum_{l'=\big\lceil M-\frac{r\left(bT_{B}\right)M}{cT}\big\rceil}^{M-1}
    X_{\rm DD}[k',l'] \e^{-\i 2\pi\frac{k'}{N}} \notag \\
    &\quad \times \e^{\i 2 \pi \frac{v}{\lambda}\frac{r\left(bT_{B}\right)}{c}} \e^{\i 2 \pi \frac{v T}{\lambda}\frac{l'-M}{M}} \e^{-\i 2 \pi \left(\frac{k-k'}{N}-\frac{v T}{\lambda} \right)\frac{bN}{B}}\notag \\
    &\quad \times \left[\mathcal{D}_{\frac{N}{B}}\!\left(\frac{k-k'}{N}-\frac{v T}{\lambda}\right)-\mathbbm 1_{\{b=0\}} \frac{B}{N}\right] \notag \\
    &\quad \times\mathcal{D}_{M}\!\left(-\left(\frac{l-l'+M}{M }-\frac{r\left(bT_{B}\right) \Delta}{c}\right)\right).
    \label{Y_DD_ISI}
\end{align}
In the above derivation, we have used $r((n-1)T)\simeq r(nT)$, consistently with the assumption of negligible target-range variation within one symbol interval.

Finally,~\eqref{rx_signal_DD_rect_fast} follows from~\eqref{Y_DD_ICI} and~\eqref{Y_DD_ISI}.

\balance

\end{document}